\documentclass{emulateapj}
\usepackage{natbib}

\def\aj{{AJ}}

\def\apj{{ApJ}}
\def\apjs{{ApJS}}
\def\mnras{{MNRAS}}

\def\pasp{{PASP}}

\def\s4g{{S$^4$G}}

\slugcomment{Accepted for publication in the Astrophysical Journal Supplement Series}

\shorttitle{Classical Morphology of S$^4$G Galaxies }
\shortauthors{Buta et al.}

\begin{document}

\title{A Classical Morphological Analysis of Galaxies in the {\it Spitzer} Survey of Stellar Structure in Galaxies ({S$^4$G})}

\author{Ronald J. Buta\altaffilmark{1},
Kartik Sheth\altaffilmark{2},
E. Athanassoula\altaffilmark{3},
A. Bosma\altaffilmark{3},
Johan H. Knapen\altaffilmark{4,5},
Eija Laurikainen\altaffilmark{6,7},
Heikki Salo\altaffilmark{6},
Debra Elmegreen\altaffilmark{8},
Luis C. Ho\altaffilmark{9,10,11},
Dennis Zaritsky\altaffilmark{12},
Helene Courtois\altaffilmark{13,14},
Joannah L. Hinz\altaffilmark{12},
Juan-Carlos Mu\~noz-Mateos\altaffilmark{2,15},
Taehyun Kim\altaffilmark{2,15,16},
Michael W. Regan\altaffilmark{17},
Dimitri A. Gadotti\altaffilmark{15},
Armando Gil de Paz\altaffilmark{18},
Jarkko Laine\altaffilmark{6},
Kar\'in Men\'endez-Delmestre\altaffilmark{19},
S\'ebastien Comer\'on\altaffilmark{6,7},
Santiago Erroz Ferrer\altaffilmark{4,5},
Mark Seibert\altaffilmark{20},
Trisha Mizusawa\altaffilmark{2,21},
Benne Holwerda\altaffilmark{22},
Barry F. Madore\altaffilmark{20}
}

\altaffiltext{1}{Department of Physics \& Astronomy, University of Alabama, Box 870324, Tuscaloosa, AL 35487-0324}
\altaffiltext{2}{National Radio Astronomy Observatory / NAASC, 520 Edgemont Road, Charlottesville, VA 22903}
\altaffiltext{3}{Aix Marseille Universite, CNRS, LAM (Laboratoire d'Astrophysique de Marseille) UMR 7326, 13388, Marseille, France}
\altaffiltext{4}{Departamento de Astrof\'isica, Universidad de La Laguna, 38206 La Laguna, Spain}
\altaffiltext{5}{Instituto de Astrof\'isica de Canarias, V\'ia L\'actea s/n 38205 La Laguna, Spain}
\altaffiltext{6}{Division of Astronomy, Department of Physical Sciences, University of Oulu, Oulu, FIN-90014, Finland}
\altaffiltext{7}{Finnish Centre of Astronomy with ESO (FINCA), University of Turku, Vaisalantie 20, FI-21500, Piikio, Finland}
\altaffiltext{8}{Vassar College, Deparment of Physics \& Astronomy, Poughkeepsie, NY 12604}
\altaffiltext{9}{Kavli Institute for Astronomy and Astrophysics, Peking University, Beijing 100871, China}
\altaffiltext{10}{Department of Astronomy, Peking University, Beijing 100871, China}
\altaffiltext{11}{The Observatories of the Carnegie Institution for Science, 813 Santa Barbara Street, Pasadena, CA 91101, USA}
\altaffiltext{12}{Steward Observatory, University of Arizona, 933 North Cherry Avenue, Tucson, Arizona 85721}
\altaffiltext{13}{Universit\'e Lyon 1, CNRS/IN2P3, Institut de Physique Nucl\'eaire, Lyon, France}
\altaffiltext{14}{Institute for Astronomy, University of Hawaii, 2680 Woodlawn Drive, Honolulu, HI 26822}
\altaffiltext{15}{European Southern Observatory, Casilla 19001, Santiago 19, Chile}
\altaffiltext{16}{Astronomy Program, Department of Physics and Astronomy, Seoul National University, Seoul 151-742, Korea}
\altaffiltext{17}{Space Telescope Science Institute, 3700 San Martin Drive, Baltimore, MD 21218}
\altaffiltext{18}{Departmento de Astrofisica, Universidad Complutense de Madrid, 28040 Madrid, Spain}
\altaffiltext{19}{University of Rio de Janeiro, Observatorio de Valongo, Ladeira Pedro Antonio, 43, CEP 20080-090, Rio de Janeiro, Brazil}
\altaffiltext{20}{The Observatories of the Carnegie Institution for Science, 813 Santa Barbara Street, Pasadena, CA 91101}
\altaffiltext{21}{Department of Physics and Space Sciences, Florida Institute of Technology, 150 W. University Boulevard, Melbourne, FL 32901}
\altaffiltext{22}{University of Leiden, Sterrenwacht Leiden, Niels Bohrweg 2, NL-2333 CA Leiden, The Netherlands}

\begin{abstract}
The {\it Spitzer} Survey of Stellar Structure in Galaxies (S$^4$G) is
the largest available database of deep, homogeneous middle-infrared
(mid-IR) images of galaxies of all types. The survey, which includes
2352 nearby galaxies, reveals galaxy morphology only minimally affected
by interstellar extinction. This paper presents an atlas and
classifications of S$^4$G galaxies in the Comprehensive de Vaucouleurs
revised Hubble-Sandage (CVRHS) system. The CVRHS system follows the
precepts of classical de Vaucouleurs (1959) morphology, modified to
include recognition of other features such as inner, outer, and nuclear
lenses, nuclear rings, bars, and disks, spheroidal galaxies, X patterns
and box/peanut structures, OLR subclass outer rings and pseudorings,
bar ansae and barlenses, parallel sequence late-types, thick disks, and
embedded disks in 3D early-type systems. We show that our CVRHS
classifications are internally consistent, and that nearly half of the
S$^4$G sample consists of extreme late-type systems (mostly bulgeless,
pure disk galaxies) in the range Scd-Im. The most common family
classification for mid-IR types S0/a to Sc is SA while that for types
Scd to Sm is SB. The bars in these two type domains are very different
in mid-IR structure and morphology. This paper examines the bar, ring,
and type classification fractions in the sample, and also includes
several montages of images highlighting the various kinds of ``stellar
structures" seen in mid-IR galaxy morphology.

\end{abstract}

\section{Introduction}

Galaxy morphology and classification are an essential step in
understanding how galaxies form and evolve. Morphology is rich in
clues to the internal and external physical processes that have
molded galactic shapes. It is, however, non-trivial to determine
exactly what a given morphology actually implies about the history
of a galaxy, because we only see the $z$$\approx$0 end-product of
all of these processes, whether secular in nature or not. Only by
examining the collective morphology of galaxies, both nearby and
very distant, in conjunction with physical data (such as luminosities,
diameters, inclinations, and bulge properties) and numerical
simulations of galaxy evolution, can we hope to piece together the
general evolutionary paths of different classes of galaxies.

The {\it Spitzer Space Telescope} (Werner et al. 2004) opened a new
window on galaxy structure at middle infrared (mid-IR) wavelengths.
With the Infrared Array Camera (IRAC; Fazio et al. 2004), {\it Spitzer}
provided four major IR bands for direct imaging: 3.6, 4.5, 5.8, and
8.0$\mu$m.\footnote{Pahre et al. (2004) considered all four IRAC bands
to be mid-IR, while the Infrared Processing and Analysis Center (IPAC)
considers the 3.6 and 4.5$\mu$m filters as near-IR and the 5.8 and
8.0$\mu$m filters as mid-IR
(www.ipac.caltech.edu/outreach/Edu/Regions/irregions.html). Here we use
mid-IR for all of the IRAC filters to distinguish them from the
ground-based near-IR studies of the past that were in the $IJHK$ bands
(ranging from 0.8 to 2.2$\mu$m).} These bands cover a unique part of
the galactic spectrum: the 3.6 and 4.5$\mu$m bands largely sample the
photospheric light of old stars (Pahre et al. 2004), while the 5.8 and
8.0$\mu$m bands reveal the dusty interstellar medium (Helou et al.
2004). In all of these bands, star formation and the interstellar
medium are evident to various degrees in the form of either emission
lines or thermal emission from dust heated by massive stars. Most
importantly, these mid-IR bands show galaxies mostly free of the
effects of extinction and reddening, revealing previously hidden
structures (e.g., rings in edge-on galaxies, or nuclear rings in the
dusty central areas of some barred galaxies).

The {\it Spitzer} Survey of Stellar Structure in Galaxies (S$^4$G;
Sheth et al. 2010) is the largest database of high-quality mid-IR
images of nearby galaxies available. Publicly released in 2013,
the S$^4$G includes 2352 galaxies imaged in the 3.6$\mu$m
and 4.5$\mu$m bands, selected according to redshift, distance,
apparent brightness, and galactic latitude. Because these filters
are predominantly sensitive to the light of old stars, they trace
the distribution of stellar mass. The primary goal of the S$^4$G
was to ``obtain a complete census of the stellar structures in
galaxies in the local volume." For this purpose, the images have
been used for studies of structures in the faint outskirts of
galaxies and of tidal debris (S. Laine et al.  2014; Kim et al.
2012), the properties of thick disks seen in edge-on galaxies of
types Sb to Sdm and profile breaks (Comer\'on et al.  2011a,b,c;
2012), mid-IR flocculent and grand design spiral structure and
star-forming regions (Elmegreen et al. 2011, 2014), conversion of
3.6 and 4.5$\mu$m light into stellar mass maps (Meidt et al. 2012,
2014; Querajeta et al. 2014), properties of stellar mass galactic
rings (Comer\'on et al. 2014), bar brightness profiles (Kim et al.
2014), outer disk brightness profiles (Munoz-Mateos et al.  2013,
J. Laine et al. 2014), quantitative morphology using cosmologically
relevant parameters (Holwerda et al.  2014), mid-IR asymmetries and
the mid-IR Tully-Fisher relation (Zaritsky et al. 2013, 2014),
examination of a possible relation between nuclear activity and bar
strength (Cisternas et al. 2013), and for the photometric decompositions
of bulges, disks, and bars in the mid-IR (Salo et al.  2014). Knapen
et al.  (2014) have also gathered high-quality optical images for
about 60\% of the survey galaxies.

The S$^4$G also provides an opportunity to examine the mid-IR
structure of a large number of galaxies from the point of view of
{\it classical morphological analysis}, meaning the classification
of galaxies in the well-known Hubble (1926, 1936) system and its
later offshoots (Sandage 1961; Sandage \& Bedke 1994; de Vaucouleurs
1959). This is worth doing for several reasons: (1) the value of
visual classification has increased over the past 20 years owing
to the explosion in imaging data available (e. g., the Sloan Digital
Sky Survey) as well as to advances in numerical simulations and
theoretical understanding of the processes that impact galactic
shapes (Kormendy 2012; Athanassoula 2012); (2) Because the mid-IR
provides the clearest view of galactic stellar mass morphology, the
symbolism of classical morphological analysis [i.e., SA(r)a, SB(s)bc,
etc.] has more meaning than it did when the $B$-band, the waveband
in which galaxy classification has traditionally been performed (as
in, e.g., Sandage 1961, Sandage \& Bedke 1994; de Vaucouleurs 1959;
Buta et al. 2007), was the only band used for such analysis; (3)
the high-quality of S$^4$G images with respect to uniformity, depth
of exposure, and resolution in the IR, and the detailed information
the images provide on both nuclear and outer structure allows us
to improve on the morphological types listed in published catalogues
like the Third Reference Catalogue of Bright Galaxies (RC3, de
Vaucouleurs et al.  1991); (4) examining such a high-quality database
at the level of detail needed for classical morphological analysis
can draw attention to special cases of interest; (5) classical
morphological types in the mid-IR complement the quantitative
analyses that are a major part of the S$^4$G project (Sheth et al.
2010); (6) specialized visual classifications are still essential
to the automated and crowd-sourced classifications that are a common
practice in astronomy today, especially for high redshift studies
(e.g., Coe et al.  2006; Huertas-Company et al.  2008; Lee et al.
2013); and (7) the large number of images that are homogeneous in
sensitivity, coverage, and spatial resolution avoid the problems
that would plague heterogeneous datasets.

Buta et al. (2010a, hereafter paper I) presented a preliminary
morphological analysis of nearly 200 S$^4$G galaxies from the {\it
Spitzer} archives, and showed that the old $B$-band classification
systems could be effectively applied in the mid-IR. This did not
mean that there were no problems in the actual application of a
$B$-band system in the mid-IR, only that on the whole the classical
systems could still be used for the majority of mid-IR galaxy types.
Eskridge et al. (2002) came to the same conclusion using near-IR
$H$-band (1.65$\mu$m) images. $H$-band types are compared with our
mid-IR classifications in Section 3.3.

In this paper, we present a similar analysis to paper I of the
entire S$^4$G sample. We use the notation of the ``Comprehensive
de Vaucouleurs revised Hubble-Sandage" (CVRHS) system (e.g., Buta
2014) to provide classifications similar to, but more extensive
than, those provided in the RC3. Much of the background for the
survey is already described in paper I; only a brief summary will
be provided here. In addition to an atlas of images of the 2168
S$^4$G galaxies not included in the paper I analysis, we highlight
specific aspects of mid-IR morphology (as well as interesting
individual cases). Because the CVRHS has ``evolved" since 2010 to
include more features (such as ansae bars and barlenses; Section
4.3) and also uses aspects of the van den Bergh (1976) parallel
sequence classification (following developments summarized by
Kormendy \& Bender 2012), the present study includes a re-examination
of the paper I galaxies.

\section{Galaxy Sample} 

The sample selection for the S$^4$G is described by Sheth et al.
(2010). All galaxies in the Hyperleda database (Paturel et al. 2003)
having an HI radial velocity ($V_{radio}$) $<$ 3000 km s$^{-1}$
(corresponding to a distance $D <$ 40 Mpc for $H_o$ = 75 km s$^{-1}$
Mpc$^{-1}$), a blue light isophotal diameter $D_{25}$ $>$
1\rlap{.}$^{\prime}$ 0), a blue photographic magnitude $m_B$ $<$
15.5 (corrected for internal extinction), and a Galactic latitude
of $b$ $>$ 30$^o$ were selected for the survey. The original sample
had 2331 galaxies (named in the on-line table accompanying Sheth
et al. 2010), but the final sample has 2352 galaxies owing to {\it
Spitzer's} improved efficiency which allowed 21 galaxies satisfying
the original criteria (except for HI detection) to be added. The
additional galaxies are primarily HI-poor ellipticals and S0s.

Of the selected galaxies, $\approx$600 had already been observed
by {\it Spitzer} for other projects, and thus images were already
available. New data were collected for the $\approx$1750 remaining
sample galaxies. The pixel size for each image is
0\rlap{.}$^{\prime\prime}$75, achieved using the drizzle technique
(Williams et al. 1996) on original images having a pixel size
of 1\rlap{.}$^{\prime\prime}$2. The point spread function for the
3.6 $\mu$m images has a mean full width at half maximum of
1\rlap{.}$^{\prime\prime}$66 (1\rlap{.}$^{\prime\prime}$72 for the
4.5$\mu$m filter; IRAC Instrument Handbook), which limits the
accuracy of the classifications of some of the smaller or more
distant galaxies in the sample. The processing of all S$^4$G images
followed a pipeline with a number of steps (P1-P4) outlined by Sheth
et al. (2010).

The use of 21-cm radial velocities to select S$^4$G sample galaxies
introduced a bias against inclusion of gas-poor early-type galaxies,
for which an HI radial velocity would not be available. The bias
is being rectified in a supplementary survey (Sheth et al. 2013)
of 465 early-type galaxies that satisfy the same selection criteria
as the original survey but using an optical radial velocity for the
distance limit. With these galaxies, the full S$^4$G will include
2817 galaxies. A comparable morphological analysis of these additional
galaxies will be provided in a later study.

The images used for our morphological analysis are the Pipeline 1
(P1) images. These are the final, ``science ready" mosaics where
individual sub-images have been matched with regard to background
levels and drizzled to get the final pixel scale. S$^4$G images are
generally much more sensitive to low light levels than are typical
ground-based near-IR images, owing to the greatly reduced and much
more stable background levels that space observations have compared
to ground-based images.

Our approach to galaxy classification from the P1 images is the
same as was used in paper I. The final P1 images were background-subtracted
and then converted into units of mag arcsec$^{-2}$ using a common
(Vega) zero point of 17.6935. This type of ``classification-ready"
image has the advantage that all of the galaxies can be displayed
in a homogeneous way; with the Vega zero point, the range 11.5 -
26.5 mag arcsec$^{-2}$ covered the full range of surface brightnesses
for the sample. The same faint limit was used for most of the
galaxies, but the bright limit was adjusted for individual objects.

\begin{figure}
\figurenum{1.0001}
\plotone{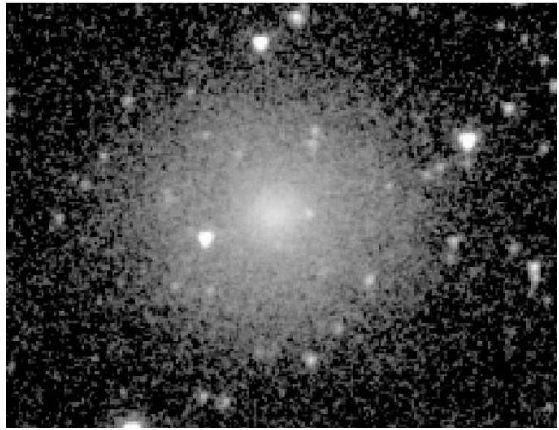}
%\vspace{-2.0truein}
\caption{
UGC12893; filter: 3.6$\mu$m; mean (Phase1,2) CVRHS type: dSA(l)0$^o$ /
Sph; north up, east left; field: 2\rlap{.}$^{\prime}$73 $\times$
2\rlap{.}$^{\prime}$10; surface brightness range: 18.0-26.5 mag
arcsec$^{-2}$. (Figure 1 is published in its entirety in the electronic
edition of the Astrophysical Journal Supplement Series. A portion is
shown here for guidance regarding its form and content.) }
\label{atlas}
\end{figure}

Our final list has 2412 galaxies (NGC 4038 and 4039 are counted as
one), 60 more than the extended S$^4$G sample. All of the additional
galaxies are companions or in the same area as an S$^4$G sample
galaxy.  As in paper I, we include the full set of classification-ready
images as an atlas (Figure~\ref{atlas}). Each image was displayed on a
24-bit monitor within an area that is recorded in the caption to the
image.  Many, but not all, of the additional galaxies are covered in
the atlas images.

Smaller or more distant galaxies in the sample are not well-resolved
in the IRAC image. Even so, S$^4$G images are of far higher depth
than could ever have been achieved from the ground at near-IR
wavelengths, especially for low luminosity, low surface brightness
galaxies.

Also as in paper I, only the 3.6$\mu$m images were used for the visual
classifications presented in this paper. The reason is that these
images tend to have a greater depth of exposure (and thus greater
sensitivity to stellar mass) than the 4.5$\mu$m dataset. With azimuthal
averaging of the luminosity distribution, Sheth et al. (2010) showed
that these images can detect surface mass densities as low as
$\approx$1$M_{\odot}$ pc$^{-2}$.

Although 3.6$\mu$m is an excellent wavelength for seeing the
distribution of stellar mass in galaxies, it is not perfect, and indeed
no IR band perfectly traces such mass. The main drawbacks of the
3.6$\mu$m filter are contamination by ``hot dust" and the inclusion of
a 3.3$\mu$m emission feature due to a polycyclic aromatic hydrocarbon
associated with star-forming regions (e.g., Meidt et al. 2012).  As
shown by Kendall et al. (2008), hot dust emission at 3.6$\mu$m can be
removed using an IRAC 8.0$\mu$m image if available.  However, most
S$^4$G galaxies do not have an 8.0$\mu$m image. Meidt et al. (2012,
2014) and Querajeta et al. (2014) use [3.6]$-$[4.5] colors and a
technique known as ``independent component analysis" to locate and
remove young contaminants and derive stellar mass maps. Our
morphological analysis is based on the original 3.6$\mu$m images and not
on the corrected stellar mass maps. The main reason for this is that
galaxy classification has traditionally been based on the distribution
of luminosity and not the distribution of mass.

\section{S$^4$G Morphology}

\subsection{Classification System}

\subsubsection{The VRHS classification}

The CVRHS system is a modified version of the de Vaucouleurs (1959)
revised Hubble-Sandage (VRHS) system that is described in the de
Vaucouleurs Atlas of Galaxies (dVA, Buta et al.  2007). More detail
on the application of the system, and extensive illustrations of
different CVRHS morphological features, is provided in the complementary
reviews of Buta (2012; IAC Winter school lectures on morphology and
secular evolution) and Buta (2013; phenomenology of galaxy morphology
and classification). Table 1 provides a summary of the meaning of
the notations of CVRHS morphology used in this paper.

 \begin{deluxetable*}{ll}
 \tablewidth{0pc}
 \tablenum{1}
 \tablecaption{Explanation of CVRHS 
Symbols\tablenotemark{a}
 }
 \tablehead{
 \colhead{Symbol} &
 \colhead{Description} 
 \\
 \colhead{1} &
 \colhead{2} 
 }
 \startdata
\noalign{\vspace{3truemm}}                                                                                                                                 
\noalign{\centerline{\bf General Terms\phantom{0000000}}}                                                                                                  
\noalign{\vspace{3truemm}}                                                                                                                                 
ETG                       & An early-type galaxy, collectively referring to a  galaxy in the range of \\                                                   
                          & \ \ \ types E to Sa \\                                                                                                         
ITG                       & An intermediate-type galaxy, taken to be in the range Sab to Sbc \\                                                            
LTG                       & A late-type galaxy, collectively referring to a galaxy in the range of \\                                                      
                          & \ \ \ types Sc to Im \\                                                                                                        
\noalign{\vspace{3truemm}}                                                                                                                                 
ETS                       & An early-type spiral, taken to be in the range S0/a to Sa \\                                                                   
ITS                       & An intermediate-type spiral, taken to be in the range Sab to Sbc \\                                                            
LTS                       & A late-type spiral, taken to be in the range Sc to Scd \\                                                                      
XLTS                      & An extreme late-type spiral, taken to be in the range Sd to Sm \\                                                              
classical bulge           & A galaxy bulge that likely formed from early mergers of smaller galaxies \\                                                    
                          & \ \ \ (Kormendy \& Kennicutt 2004; Athanassoula 2005) \\                                                                       
pseudobulge               & A galaxy bulge made of disk material that has secularly collected into \\                                                      
                          & \ \ \ the central regions of a barred galaxy (Kormendy 2012) \\                                                                
PDG                       & A pure disk galaxy, a galaxy lacking a classical bulge and often also \\                                                       
                          & \ \ \ lacking a pseudobulge \\                                                                                                 
\noalign{\vspace{3truemm}}                                                                                                                                 
\noalign{\centerline{\bf Stage\phantom{000000}}}                                                                                                           
\noalign{\vspace{3truemm}}                                                                                                                                 
stage                     & The characteristic of galaxy morphology that recognizes development of \\                                                      
                          & \ \ \ structure, the widespread distribution of star formation, and the \\                                                     
                          & \ \ \ relative importance of a bulge component along a sequence that \\                                                        
                          & \ \ \ correlates well with basic characteristics such as integrated color, \\                                                  
                          & \ \ \ average surface brightness, and HI mass-to-blue luminosity ratio \\                                                      
\noalign{\vspace{3truemm}}                                                                                                                                 
\noalign{\centerline{\bf Elliptical Galaxies\phantom{000000}}}                                                                                             
\noalign{\vspace{3truemm}}                                                                                                                                 
E galaxy                  & A galaxy having a smoothly declining brightness distribution with little \\                                                    
                          & \ \ \ or no evidence of a disk component and no inflections (such as lenses) \\                                                
                          & \ \ \ in the luminosity distribution (examples: NGC 1052, 3193, 4472) \\                                                       
En                        & An elliptical galaxy of visual flattening n = $10(1-b/a)$, where $b/a$ is the \\                                               
                          & \ \ \ visual isophotal axis ratio (Hubble 1926) \\                                                                             
E$^+$n                    & A ``late" elliptical of visual flattening n, a transition stage to the S0 class \\                                             
                          & \ \ \ (de Vaucouleurs 1959); show slight traces of differentiated structure, \\                                                
                          & \ \ \ usually subtle evidence of lenses (example: NGC 5846) or a faint \\                                                      
                          & \ \ \ outer envelope; also has been used as a ``home" for Morgan cD galaxies \\                                                
                          & \ \ \ in RC3 \\                                                                                                                
E/E$^+$                   & An E galaxy that in our Phase 1,2 analysis averages between E and E$^+$ \\                                                     
                          & \ \ \ (example: NGC 3226)\\                                                                                                    
E(d)n                     & A disky elliptical galaxy of visual flattening n (Kormendy \& Bender \\                                                        
                          & \ \ \ 1996); a subclassification of ellipticals having pointy outer \\                                                         
                          & \ \ \ isophotes that is visually detectable only for the most obvious or \\                                                    
 \enddata
 \end{deluxetable*}
 %\clearpage
 \begin{deluxetable*}{ll}
 \tablewidth{0pc}
 \tablenum{1 (cont.)}
 \tablecaption{Explanation of CVRHS Symbols (cont.)\tablenotemark{a}
 }
 \tablehead{
 \colhead{Symbol} &
 \colhead{Description} 
 \\
 \colhead{1} &
 \colhead{2} 
 }
 \startdata
                          & \ \ \ most favorably oriented cases (example: NGC 3377)\\                                                                      
E(b)n                     & A boxy elliptical galaxy of visual flattening n (Kormendy \& Bender 1996); \\                                                  
                          & \ \ \ a subclassification of ellipticals having boxy outer isophotes that is \\                                                
                          & \ \ \ visually detectable only for the most obvious cases \\                                                                   
E(b,nd)                   & A boxy E galaxy with an inner disk (example: NGC 4370) \\                                                                      
\noalign{\vspace{3truemm}}                                                                                                                                 
\noalign{\centerline{\bf S0 Galaxies\phantom{000000}}}                                                                                                     
\noalign{\vspace{3truemm}}                                                                                                                                 
S0 galaxy                 & A disk-shaped galaxy lacking strong or obvious spiral structure; at \\                                                         
                          & \ \ \ minimum, a two-component system with a bulge and a disk \\                                                               
S0$^-$$\rightarrow$S0$^o$$\rightarrow$S0$^+$     & a stage sequence of S0 galaxies based on increasing \\                                                  
                          & \ \ \ development of structure, such as bars, lenses, and rings \\                                                             
E$^+$/S0$^-$              & An ETG that in our Phase 1,2 analysis averages between E$^+$ and \\                                                            
                          & \ \ \ S0$^-$ (example: NGC 4649)\\                                                                                             
S0$^-$                    & An early S0 showing clear evidence for a disk (envelope) but little  \\                                                        
                          & \ \ \ structure; all features are subtle (examples: NGC 4442, 5507)\\                                                          
S0$^{-/o}$                & An S0 galaxy that in our Phase 1,2 analysis averages between S0$^-$ \\                                                         
                          & \ \ \ and S0$^o$ \\                                                                                                            
S0$^o$                    & An intermediate stage S0 showing clear lenses or traces of  \\                                                                 
                          & \ \ \ rings/pseudorings (examples: NGC 1411, 1533) \\                                                                          
S0$^o$[d]                 & One example of a notation used for an apparently early-type galaxy in \\                                                       
                          & \ \ \ the catalogue having little or no apparent bulge; other examples \\                                                      
                          & \ \ \ include [c], [cd], [m]. These are related to the van den Bergh \\                                                        
                          & \ \ \ (1976) parallel-sequence Hubble classification system, although \\                                                       
                          & \ \ \ that system is not fully built into the catalogue (example: \\                                                           
                          & \ \ \ NGC 693) \\                                                                                                              
S0$^{o/+}$                & An S0 galaxy that in our Phase 1,2 analysis averages between S0$^o$\\                                                          
                          & \ \ \ and S0$^+$ \\                                                                                                            
S0$^+$                    & A late S0 stage showing strong rings and bars, and in some cases trace  \\                                                     
                          & \ \ \ spiral structure (examples: NGC 1291, 1326, 4138)\\                                                                      
S0$^+$[c]                 & A late S0 with an Sc-like central concentration (examples: NGC 4344, \\                                                        
                          & \ \ \ 4451) \\                                                                                                                 
\noalign{\vspace{3truemm}}                                                                                                                                 
\noalign{\centerline{\bf Spiral and Irregular Galaxies\phantom{000000}}}                                                                                   
\noalign{\vspace{3truemm}}                                                                                                                                 
Spiral galaxy             & A galaxy where a spiral pattern is a major part of the morphology \\                                                           
S0/a-Sa-Sb-Sc-Sd-Sm       & A stage sequence for spirals (with intermediate types Sab, Sbc, Scd, and \\                                                    
                          & \ \ \ Sdm) based on Hubble's three criteria: relative prominence of the\\                                                      
                          & \ \ \ bulge, the degree of openness of the arms, and the degree of \\                                                          
                          & \ \ \ resolution of the arms into star clusters or very luminous stars. \\                                                     
Irregular galaxy          & A complex system characterized by an irregular distribution of star \\                                                         
                          & \ \ \ formation; this irregular distribution can, however, be embedded in a \\                                                 
                          & \ \ \ more regular background \\                                                                                               
S$\underline{\rm 0}$/a    & An S0/a galaxy that is closer to S0$^+$ than to Sa (examples: NGC 522,\\                                                       
                          & \ \ \ 2681, 3626) \\                                                                                                           
 \enddata
 \end{deluxetable*}
 %\clearpage
 \begin{deluxetable*}{ll}
 \tablewidth{0pc}
 \tablenum{1 (cont.)}
 \tablecaption{Explanation of CVRHS Symbols (cont.)\tablenotemark{a}
 }
 \tablehead{
 \colhead{Symbol} &
 \colhead{Description} 
 \\
 \colhead{1} &
 \colhead{2} 
 }
 \startdata
S0/a                      & A transition stage showing clear but subtle tightly-wrapped spiral  \\                                                         
                          & \ \ \ arms; structure is generally smooth but trace star formation is seen \\                                                  
                          & \ \ \ in nearby examples (examples: NGC 1350, 1452, 4394, 4454, 4984, 5701, \\                                                 
                          & \ \ \ 6340, 7098)\\                                                                                                            
S0/$\underline{\rm a}$    & An S0/a galaxy that is closer to Sa than to S0$^+$ (examples: NGC 3185, 3900)\\                                                
Sa                        & An early-type spiral, usually defined by relatively smooth, tightly-wrapped spiral \\                                          
                          & \ \ \ arms and a significant bulge; standard interpretation may be violated in a \\                                            
                          & \ \ \ cluster environment or in presence of a bar (examples: NGC 1433, 1512, 3031, \\                                          
                          & \ \ \ 3788, 4260, 4450, 4548, 4800, 7513)\\                                                                                    
S$\underline{\rm a}$b     & An Sab galaxy that is closer to Sa than to Sb (example: NGC 2985)\\                                                            
Sab                       & An intermediate-type spiral galaxy similar to Sa but with a more knotty  \\                                                    
                          & \ \ \ structure (examples: NGC 210, 1097, 3992, 4995; IC 1993)\\                                                               
Sa$\underline{\rm b}$     & An Sab galaxy that is closer to Sb than to Sa (examples: NGC 3177, 4902)\\                                                     
Sb                        & An intermediate-type spiral having relatively more open, knotty arms and a  \\                                                 
                          & \ \ \ smaller bulge than Sa or Sab galaxies (examples: NGC 908, 3433, 3689, \\                                                 
                          & \ \ \ 3705, 4237, 7479)\\                                                                                                      
S$\underline{\rm b}$c     & An Sbc galaxy that is closer to Sb than to Sc (examples: NGC 3344, 3512; \\                                                    
                          & \ \ \ IC 769) \\                                                                                                               
Sbc                       & Typically, an intermediate-type galaxy having well-developed, open, knotty  \\                                                 
                          & \ \ \ spiral arms like an Sc galaxy but with a more significant bulge (examples: \\                                            
                          & \ \ \ NGC 1365, 3184, 3198, 3338, 3726) \\                                                                                     
Sb$\underline{\rm c}$     & An Sbc galaxy that is closer to Sc than to Sb (examples: NGC 2715, 4303)\\                                                     
Sc                        & A late-type spiral having well-developed, open, and knotty spiral arms with a  \\                                              
                          & \ \ \ small but significant bulge (examples: NGC 1042, 1084, 3486, 3810, 3893, \\                                              
                          & 4411B, 5457, 5970, 7448; IC 1953)\\                                                                                            
S$\underline{\rm c}$d     & An Scd galaxy that is closer to Sc than to Sd (examples: NGC 1073, 5033, 5468) \\                                              
Scd                       & Similar to an Sc but with little or no bulge; typically a small central object  \\                                             
                          & \ \ \ (nuclear star cluster or pseudobulge) may be seen; arms can be more open than  \\                                        
                          & \ \ \ for an Sc; these are generally pure disk galaxies (PDGs) (examples: NGC 1255, \\                                         
                          & \ \ \ 1559, 3346, 5334, 5595, 7741) \\                                                                                         
Sc$\underline{\rm d}$     & An Scd galaxy that is closer to Sd than to Sc (examples: NGC 255, 1253, 3359, \\                                               
                          &\ \ \ 4411A, 5597, 5668)\\                                                                                                      
Sd                        & An extreme late-type spiral, similar to Scd but with less apparent central \\                                                  
                          & \ \ \ concentration than an Scd; among the most common PDGs; asymmetry is often \\                                             
                          & \ \ \ present but less extreme than in Sdm or Sm,Im types (examples: NGC 3003, \\                                              
                          & \ \ \ 4294, 4731, 5068, 5669)\\                                                                                                
S$\underline{\rm d}$m     & An Sdm galaxy that is closer to Sd than to Sm (examples: NGC 247, 3556, 7151, \\                                               
                          & \ \ \ 7497)\\                                                                                                                  
Sdm                       & An extreme late-type spiral showing considerable asymmetry, usually with one  \\                                               
                          & \ \ \ arm longer and better defined than the other; considerable star formation also  \\                                       
                          & \ \ \ characterizes these PDGs (examples: NGC 300, 3906, 4395, 4630, 7154)\\                                                   
Sd$\underline{\rm m}$     & An Sdm galaxy that is closer to Sm than to Sd (example: NGC 2552)\\                                                            
Sm                        & A magellanic spiral, usually characterized by a single spiral arm emerging from  \\                                            
                          & \ \ \ a bar or central region and little or no bulge or central concentration (de  \\                                          
                          & \ \ \ Vaucouleurs \& Freeman 1972) (examples: NGC 55, 5474, 7091)  \\                                                          
 \enddata
 \end{deluxetable*}
 %\clearpage
 \begin{deluxetable*}{ll}
 \tablewidth{0pc}
 \tablenum{1 (cont.)}
 \tablecaption{Explanation of CVRHS Symbols (cont.)\tablenotemark{a}
 }
 \tablehead{
 \colhead{Symbol} &
 \colhead{Description} 
 \\
 \colhead{1} &
 \colhead{2} 
 }
 \startdata
S/Im                      & A magellanic galaxy that is closer to Im than to Sm (example: NGC 4353)\\                                                      
Im                        & A magellanic irregular galaxy, characterized by an irregular shape sometimes  \\                                               
                          & \ \ \ within a smooth background of starlight; often show considerable star formation  \\                                      
                          & \ \ \ and a wide range of luminosities (examples: NGC 4214, 4242, 4449, 4605)\\                                                
Im (cc)                   & A clump cluster, usually referring to an irregular galaxy with a large number of \\                                            
                          & \ \ \ scattered star forming regions. The term was originally applied to high \\                                               
                          & \ \ \ redshift clumpy galaxies (e.g., Elmegreen et al. 2009) (examples: IC 1826, \\                                            
                          & \ \ \ 2040; UGC 1945)\\                                                                                                        
I0                        & A type of galaxy seen mainly in blue light images where a highly irregular dust \\                                             
                          & \ \ \ distribution is seen within an S0 or E-like background; at 3.6$\mu$m, I0 \\                                              
                          & \ \ \ galaxies are relatively normal-looking ETGs (examples: NGC 2968, 3077, 5195, \\                                          
                          & \ \ \ 5253, 5363) \\                                                                                                           
\noalign{\vspace{3truemm}}                                                                                                                                 
\noalign{\centerline{\bf Dwarf and Spheroidal Galaxies\phantom{000000}}}                                                                                   
\noalign{\vspace{3truemm}}                                                                                                                                 
dE                        & A ``dwarf elliptical" galaxy, defined to have a ``smooth intensity distribution \\                                             
                          & \ \ \ over the face and ... low surface brightness" (Sandage \& Binggeli 1984; \\                                              
                          & \ \ \ Non-Virgo example in S$^4$G catalogue: NGC 59)\\                                                                         
dS0                       & A ``dwarf S0" galaxy, a class of dwarfs which resemble dE galaxies except for \\                                               
                          & \ \ \ ``a change of slope in the radial gradient of the light distribution" \\                                                 
                          & \ \ \ showing ``either direct evidence of a disk, or ... evidence of two \\                                                    
                          & \ \ \ components" (Sandage \& Binggeli 1984) \\                                                                                
dE,N                      & A nucleated dwarf E galaxy (Sandage \& Binggeli 1984) \\                                                                       
dS0,N                     & A nucleated dwarf S0 galaxy (Sandage \& Binggeli 1984) \\                                                                      
dIm                       & A dwarf irregular galaxy, typically an Im galaxy having an absolute $B$-band \\                                                
                          & \ \ \ magnitude $M_B^o \gtrsim$ $-$17; low surface brightness and often resolved \\                                            
                          & \ \ \ in S$^4$G images \\                                                                                                      
Sph                       & A ``spheroidal" galaxy, a type of galaxy having the appearance of an E or S0  \\                                               
                          & \ \ \ galaxy, but the photometric characteristics of much later type, lower \\                                                 
                          & \ \ \ luminosity galaxies (like Sm, Im types); the dE and dS0 galaxies in the \\                                               
                          & \ \ \ Virgo Cluster are all of this basic type (Kormendy \& Bender 2012). \\                                                   
Sph,N                     & A nucleated spheroidal galaxy (Binggeli et al. 1985; Kormendy et al. 2009) \\                                                  
BCD                       & A star-forming, blue compact dwarf galaxy; at 3.6$\mu$m, can appear as a dE or \\                                              
                          & \ \ \ dS0 galaxy (example: NGC 1705) \\                                                                                        
\noalign{\vspace{3truemm}}                                                                                                                                 
\noalign{\centerline{\bf Family\phantom{000000}}}                                                                                                          
\noalign{\vspace{3truemm}}                                                                                                                                 
family                    & The characteristic of galaxy morphology that recognizes the apparent strength\\                                                
                          & \ \ \ of a bar or other type of nonaxisymmetric structure, such as an oval\\                                                   
SA                        & A nonbarred spiral or S0 galaxy (examples: NGC 488, 628, 1411, 4698; IC 1993, \\                                               
                          & \ \ \ 5267)\\                                                                                                                  
S$\underline{\rm A}$B     & A galaxy showing a trace of a bar, usually in the form of a broad oval or a \\                                                 
                          & \ \ \ very low contrast regular bar (examples: NGC 4203, 4899)\\                                                               
SAB                       & A barred galaxy of intermediate apparent bar strength (examples: NGC 4535, \\                                                  
                          & \ \ \ 5236, 7743)\\                                                                                                            
 \enddata
 \end{deluxetable*}
 %\clearpage
 \begin{deluxetable*}{ll}
 \tablewidth{0pc}
 \tablenum{1 (cont.)}
 \tablecaption{Explanation of CVRHS Symbols (cont.)\tablenotemark{a}
 }
 \tablehead{
 \colhead{Symbol} &
 \colhead{Description} 
 \\
 \colhead{1} &
 \colhead{2} 
 }
 \startdata
SA$\underline{\rm B}$     & A bar that is clear and well-defined but weaker-looking than a typical SB  \\                                                  
                          & \ \ \ galaxy bar (examples: NGC 4639, 4818, 5566, 5701) \\                                                                     
SB                        & A barred galaxy with a conspicuous bar, usually strong and obvious \\                                                          
                          & \ \ \ (examples: NGC 1300, 1365, 1452, 7513) \\                                                                                
SAB$_a$, SB$_a$           & A barred galaxy where the bar is defined by brightness enhancements (``ansae";  \\                                             
                          & \ \ \ Sandage 1961; Danby 1965) at its ends; these enhancements may be round \\                                                
                          & \ \ \ spots, short linear features, arcs, or star-forming clumps (Buta 2013) \\                                                
                          & \ \ \ [examples: NGC 2787, 5375, 7098 (see Figures 12 \& 13 for others)]\\                                                     
SAB$_x$, SB$_x$           & A galaxy showing a prominent X or box/peanut structure in its inner regions; \\                                                
                          & \ \ \ often seen in edge-on galaxies, the box/peanut/X shape is believed to \\                                                 
                          & \ \ \ be a manifestation of vertical resonant orbits in a bar potential, thus it \\                                            
                          & \ \ \ indicates the presence of a bar (Athanassoula 2005). An X may also be seen \\                                            
                          & \ \ \ in the clear bars of non-edge galaxies (Buta et al. 2007; Erwin \& Debattista \\                                         
                          & \ \ \ 2013) [examples: NGC 2654 (Figure 14; nearly edge-on); NGC 5377 \\                                                       
                          & \ \ \ (intermediate inclination)]\\                                                                                            
SAB$_{xa}$, SB$_{xa}$     & A galaxy showing both an X and a pair of ansae; the X is usually  \\                                                           
                          & \ \ \ three-dimensional while the ansae are flat. [example: NGC 4216 (Figures 12 \\                                            
                          & \ \ \ \& 13)]\\                                                                                                                
\noalign{\vspace{3truemm}}                                                                                                                                 
\noalign{\centerline{\bf Standard Inner Varieties\phantom{000000}}}                                                                                        
\noalign{\vspace{3truemm}}                                                                                                                                 
inner variety             & The characteristic of galaxy morphology that recognizes the presence or absence \\                                             
                          & \ \ \ of an inner ring\\                                                                                                       
(r)                       & An inner ring, a closed circular or oval feature enveloping either the ends of   \\                                            
                          & \ \ \ a bar if present or the central bulge (examples: NGC 1433, 3351, 3486, \\                                                
                          & \ \ \ 4245, 5566)\\                                                                                                            
($\underline{\rm r}$s)    & A well-defined inner ring, but made of tightly-wrapped spiral structure \\                                                     
                          & \ \ \ (examples: NGC 613, 1398, 3368, 3705) \\                                                                                 
(rs)                      & An inner ``pseudoring", a partial inner ring made of spiral arms (examples:\\                                                  
                          & \ \ \ NGC 779, 1232, 3346, 4548) \\                                                                                            
(r$\underline{\rm s}$)    & A weak inner pseudoring, usually fairly open and characterized by a pitch  \\                                                  
                          & \ \ \ angle only a little different from that of the outer spiral arms \\                                                      
                          & \ \ \ (examples: NGC 1365, 3513, 4501, 4548, 5383) \\                                                                          
(s)                       & A pure s-shaped spiral where the arms break from the ends of a bar or from \\                                                  
                          & \ \ \ the bulge without forming a pseudoring (examples: NGC 1300, 7721)\\                                                      
\noalign{\vspace{3truemm}}                                                                                                                                 
\noalign{\centerline{\bf Special Inner Varieties\phantom{000000}}}                                                                                         
\noalign{\vspace{3truemm}}                                                                                                                                 
(rr)                      & A galaxy having two inner rings (example: NGC 4698) \\                                                                         
(x$_1$r)                  & A ring-like feature that outlines a bar, possibly related to the x$_1$ family  \\                                              
                          & \ \ \ of bar orbits discussed by Regan \& Teuben (2004) (example: NGC 6012)\\                                                  
(r,s)                     & A variety where the inside of an inner ring includes a spiral pattern unrelated  \\                                            
                          & \ \ \ to the main outer spiral arms; prototype in this catalogue is NGC 5364 \\                                                
(rs,rs)                   & A galaxy having two inner pseudorings, usually of very different sizes  \\                                                     
                          & \ \ \ (examples: NGC 289, 4689) \\                                                                                             
 \enddata
 \end{deluxetable*}
 %\clearpage
 \begin{deluxetable*}{ll}
 \tablewidth{0pc}
 \tablenum{1 (cont.)}
 \tablecaption{Explanation of CVRHS Symbols (cont.)\tablenotemark{a}
 }
 \tablehead{
 \colhead{Symbol} &
 \colhead{Description} 
 \\
 \colhead{1} &
 \colhead{2} 
 }
 \startdata
(s,rs)                    & A galaxy having a strong s-shaped spiral in the presence of an inner pseudoring \\                                             
                          & \ \ \ (example: NGC 986) \\                                                                                                    
(l)                       & An inner lens, a type of feature, often seen in S0 galaxies, having a shallow  \\                                              
                          & \ \ \ brightness gradient interior to a sharp edge (Kormendy 1979) (examples: \\                                               
                          & \ \ \ NGC 1291, 1411, 4269, 5602) \\                                                                                           
(rl)                      & An inner ring-lens, recognized as a low contrast inner ring; types  \\                                                         
                          & \ \ \ ($\underline{\rm r}$l) and (r$\underline{\rm l}$) recognize different degrees of contrast enhancement \\                 
                          & \ \ \ (examples: NGC 2859, 4250) \\                                                                                            
(rs,rl)                   & A galaxy having an inner pseudoring and an inner ring-lens, the latter the  \\                                                 
                          & \ \ \ smaller (example: NGC 5055) \\                                                                                           
(r$^{\prime}$l)           & An inner pseudoring-lens, where the inner ring shows azimuthal contrast  \\                                                    
                          & \ \ \ differences like spiral arms (examples: NGC 210, 1415, 3147)\\                                                           
(ls)                      & An inner lens with a subtle embedded spiral pattern (example: NGC 3675) \\                                                     
(p)                       & A ``plume", usually seen as a secondary spiral arc positioned just off the  \\                                                 
                          & \ \ \ leading sides of a bright inner ring (Buta 1984); much rarer than inner rings \\                                         
                          & \ \ \ or lenses (example: NGC 1433)\\                                                                                          
(bl)                      & A ``barlens", a feature recognized by Laurikainen et al. (2011, 2013, 2014) \\                                                 
                          & \ \ \ and Athanassoula et al. (2014) as the inner part of an early-type bar \\                                                 
                          & \ \ \ [examples: NGC 1433, 2787, 3351 (see Figure 11)]\\                                                                       
\noalign{\vspace{3truemm}}                                                                                                                                 
\noalign{\centerline{\bf Nuclear Varieties\phantom{000000}}}                                                                                               
\noalign{\vspace{3truemm}}                                                                                                                                 
nuclear variety           & The nuclear variety classification refers to rings, lenses, bars, and other features\\                                         
                          & \ \ \ that are often found in the centers of barred galaxies, but which may also \\                                            
                          & \ \ \ be found in nonbarred galaxies \\                                                                                        
(nr)                      & A nuclear (or circumnuclear) ring, a small star-forming feature found in  \\                                                   
                          & \ \ \ the centers of barred galaxies but also sometimes seen in nonbarred galaxies \\                                          
                          & \ \ \ (example: NGC 1097; see Figure 20 for others) \\                                                                         
(nr$^{\prime}$)           & A nuclear pseudoring, where the ring appears formed by a wrapped spiral  \\                                                    
                          & \ \ \ pattern; in optical images, this character may sometimes be an artifact of \\                                            
                          & \ \ \ inner dust lanes in bars (examples: NGC 1068, 1090) \\                                                                   
(ns)                      & A nuclear spiral, a type of feature that may also be an artifact of dust in  \\                                                
                          & \ \ \ optical bands (example: NGC 1022)\\                                                                                      
(nl)                      & A nuclear lens; an excellent example in the S$^4$G catalogue is NGC 4250 \\                                                    
(nrl)                     & A nuclear ring-lens (examples: NGC 210, 1300, 1433, 5566)\\                                                                    
(nb)                      & A nuclear (or secondary) bar, a feature found in the centers of barred galaxies but  \\                                        
                          & \ \ \ which may also be found in SA galaxies (examples: NGC 1291, 1433, 4725)\\                                                
(nb$_a$)                  &  A nuclear ansae-type bar \\                                                                                                   
(nd)                      & A nuclear disk, usually seen in well-resolved edge-on disk galaxies \\                                                         
                          & \ \ \ (examples: NGC 24, 678, 1532, 3079, 3628, 4111, 5907) \\                                                                 
(np)                      & a ``nuclear pattern," a nuclear structure of uncertain nature \\                                                               
(tb)                      & a triaxial bulge, an elongated central component whose major axis is misaligned\\                                              
                          & \ \ \ with the galaxy major axis\\                                                                                             
(psb)                     & a pseudobulge, an elongated bulge component whose major axis is approximately \\                                               
                          & \ \ \ aligned with the disk major axis (example: NGC 4536)\\                                                                   
 \enddata
 \end{deluxetable*}
 %\clearpage
 \begin{deluxetable*}{ll}
 \tablewidth{0pc}
 \tablenum{1 (cont.)}
 \tablecaption{Explanation of CVRHS Symbols (cont.)\tablenotemark{a}
 }
 \tablehead{
 \colhead{Symbol} &
 \colhead{Description} 
 \\
 \colhead{1} &
 \colhead{2} 
 }
 \startdata
\noalign{\vspace{3truemm}}                                                                                                                                 
\noalign{\centerline{\bf Combined Inner and Nuclear Varieties\tablenotemark{b}}}                                                                           
\noalign{\vspace{3truemm}}                                                                                                                                 
(r,nr)                    & A galaxy having both an inner ring and a much smaller nuclear ring (examples:  \\                                              
                          & \ \ \ NGC 1326, 1512, 3351, 4274, 5850) \\                                                                                     
(l,nl)                    & A galaxy having both an inner lens and a nuclear lens (example: NGC 1411) \\                                                   
(rs,nr,nb)                & A galaxy having an inner pseudoring, a nuclear ring, and a nuclear bar (example: \\                                            
                          & \ \ \ NGC 4321) \\                                                                                                             
(r$\underline{\rm l}$,bl,nr) & A galaxy having an inner ring-lens (more lens than ring), a barlens, and a nuclear \\                                       
                          & \ \ \ ring (example: NGC 4314) \\                                                                                              
\noalign{\vspace{3truemm}}                                                                                                                                 
\noalign{\centerline{\bf Standard Outer Varieties\phantom{000000}}}                                                                                        
\noalign{\vspace{3truemm}}                                                                                                                                 
outer variety             & The characteristic of galaxy morphology which recognizes a large ring or ring-like \\                                          
                          & \ \ \ pattern in the outer regions of a galaxy\\                                                                               
(R)                       & An outer ring, a closed ring-shaped feature usually about twice the size of a  \\                                              
                          & \ \ \ bar in barred galaxies (examples: NGC 1291, 1350, 2859)\\                                                                
(R$^{\prime}$)            & An outer ``pseudoring", a near-outer ring made of spiral arms (examples:\\                                                     
                          & \ \ \ NGC 986, 1300, 1365, 5757, 7479) \\                                                                                      
\noalign{\vspace{3truemm}}                                                                                                                                 
\noalign{\centerline{\bf Special Outer Varieties\tablenotemark{b}\phantom{000000}}}                                                                        
\noalign{\vspace{3truemm}}                                                                                                                                 
(RR)                      & A galaxy having two outer rings, usually of very different sizes (examples: NGC  \\                                            
                          & \ \ \ 3898, 4457) \\                                                                                                           
(R$^{\prime}$R$^{\prime}$)& A galaxy having two outer pseudorings, usually of very different sizes (examples: \\                                           
                          & \ \ \ NGC 1425; UGC 8155; PGC 53093) \\                                                                                        
(R$^{\prime}$,R)          & A galaxy with an outer pseudoring and an outer ring, the latter being the  \\                                                  
                          & \ \ \ smaller (example: NGC 4984) \\                                                                                           
(L)                       & An outer lens, a type of feature, often seen in early-type galaxies, that is an  \\                                            
                          & \ \ \ outer analogue of an inner lens (Kormendy 1979) (examples: NGC 2787, 4262)\\                                             
(RL)                      & An outer ring-lens, recognized as a low contrast outer ring; types  \\                                                         
                          & \ \ \ ($\underline{\rm R}$L) and (R$\underline{\rm L}$) recognize different degrees of contrast enhancement \\                 
                          & \ \ \ (examples: NGC 5602, 5750)\\                                                                                             
(RL,R)                    & A galaxy with an outer ring-lens and an outer ring, the latter being the  \\                                                   
                          & \ \ \ smaller (example: NGC 3626)\\                                                                                            
(RL,R$^{\prime}$)         & A galaxy with an outer ring-lens and an outer pseudoring, the latter being the  \\                                             
                          & \ \ \ smaller (examples: NGC 1367, 4351, 4826)\\                                                                               
(R$^{\prime}$L)           & An outer pseudoring-lens, where the outer ring shows azimuthal contrast  \\                                                    
                          & \ \ \ differences like spiral arms (examples: NGC 1357, 2780)\\                                                                
\noalign{\vspace{3truemm}}                                                                                                                                 
\noalign{\centerline{\bf ``Outer Lindblad Resonance" Morphologies \phantom{000000}}}                                                                       
\noalign{\vspace{3truemm}}                                                                                                                                 
(R$_1$)                   & A closed outer ring showing a shape resembling a broad figure 8; a  \\                                                         
                          & \ \ \ subtly-dimpled oval ring, recognized as an OLR subclass (Buta \& Crocker 1991; \\                                        
                          & \ \ \ Buta 1995) (examples: NGC 1326, 4250, 5728; IC 1438, 4214)\\                                                             
 \enddata
 \end{deluxetable*}
 %\clearpage
 \begin{deluxetable*}{ll}
 \tablewidth{0pc}
 \tablenum{1 (cont.)}
 \tablecaption{Explanation of CVRHS Symbols (cont.)\tablenotemark{a}
 }
 \tablehead{
 \colhead{Symbol} &
 \colhead{Description} 
 \\
 \colhead{1} &
 \colhead{2} 
 }
 \startdata
(R$_1$L)                  & An R$_1$ outer pseudoring-lens (examples: NGC 2893, 5448)\\                                                                    
(R$_1^{\prime}$)          & An outer pseudoring made from arms that wind about 180$^o$ with respect to the  \\                                             
                          & \ \ \ bar ends; an OLR subclass (Buta \& Crocker 1991; Buta 1995) (examples:\\                                                 
                          & \ \ \ NGC 1566, 3504, 4192, 4314, 5377, 7051) \\                                                                               
(R$_2^{\prime}$)          & An outer pseudoring made from arms that wind about 270$^o$ with respect to the  \\                                             
                          & \ \ \ bar ends; an OLR subclass (Buta \& Crocker 1991; Buta 1995) (examples:\\                                                 
                          & \ \ \ NGC 2633, 7741, ESO 26-1) \\                                                                                             
(R$_1^{\prime}$L)         & An R$_1^{\prime}$ outer pseudoring-lens (example: NGC 4045)\\                                                                  
(R$_2^{\prime}$L)         & An R$_2^{\prime}$ outer pseudoring-lens (example: NGC 210)\\                                                                   
(R$_1$R$_2^{\prime}$)     & A combined outer ring-pseudoring pattern where the arms forming the  \\                                                        
                          & \ \ \ R$_2^{\prime}$ ring break from an R$_1$ ring; an OLR subclass \\                                                         
                          & \ \ \ (Buta \& Crocker 1991; Buta 1995) (example: NGC 5101)\\                                                                  
\noalign{\vspace{3truemm}}                                                                                                                                 
\noalign{\centerline{\bf Catastrophic Rings\phantom{000000}}}                                                                                              
\noalign{\vspace{3truemm}}                                                                                                                                 
catastrophic ring         & A ring-like feature formed in a catastrophic galaxy interaction event, such \\                                                 
                          & \ \ \ as a head-on collision or major disruption \\                                                                              
extraplanar disk          & A general term for any acquired disk-shaped feature inclined at an angle \\                                                    
                          & \ \ \ to a more massive disk system\\                                                                                          
RG                        & A ring galaxy, a term referring to the unusual rings thought to be \\                                                          
                          & \ \ \ produced in special galactic collisions (Theys \& Spiegel 1976; \\                                                       
                          & \ \ \ Madore et al. 2009) \\                                                                                                   
PRG                       & A polar ring galaxy or related type of object (Whitmore et al. 1990). In \\                                                    
                          & \ \ \ the strictest sense, an interacting system where a companion is \\                                                       
                          & \ \ \ disrupted along a polar or near polar orbit around another \\                                                            
                          & \ \ \ disk-shaped system (example: NGC 5122) \\                                                                                
IRG                       & An inclined ring galaxy, where a companion has been disrupted into a \\                                                        
                          & \ \ \ high angle but not necessarily polar orbit (example: NGC 660)\\                                                          
(R)E                      & A likely accretion ring formed from a companion disrupted into a large \\                                                      
                          & \ \ \ orbit around an elliptical galaxy (Schweizer et al. 1987)\\                                                              
\noalign{\vspace{3truemm}}                                                                                                                                 
\noalign{\centerline{\bf Edge-on Galaxies\tablenotemark{b}\phantom{000000}}}                                                                               
\noalign{\vspace{3truemm}}                                                                                                                                 
sp                        & A ``spindle", or highly-inclined (generally $i$ $>$ 65$^{\circ}$) disk-shaped \\                                               
                          & \ \ \ galaxy \\                                                                                                                
Scd sp                    & A highly-inclined Scd galaxy, too close to edge-on for the family or variety \\                                                
                          & \ \ \ to be reliably distinguished (example: NGC 100)\\                                                                        
SB($\underline{\rm r}$s)bc  sp  & A highly-inclined galaxy, but sufficiently far from edge-on that the family \\                                           
                          & \ \ \ and variety are distinguishable (example: NGC 253)\\                                                                     
spw                       & A highly-inclined disk-shaped galaxy where the disk shows evidence of warping  \\                                              
                          & \ \ \ in the nearly edge-on view [examples: NGC 522, 5084, 5403; UGC 10043 \\                                                  
                          & \ \ \ (Figure 18)]\\                                                                                                           
S0$^-$ sp / E(d)7         & An edge-on early S0 galaxy, having a disky thick disk of flattening E7 \\                                                      
                          & \ \ \ (example: NGC 1032); E(d) is Kormendy \& Bender (1996) notation \\                                                       
                          & \ \ \ and is used here only in a descriptive manner \\                                                                         
 \enddata
 \end{deluxetable*}
 %\clearpage
 \begin{deluxetable*}{ll}
 \tablewidth{0pc}
 \tablenum{1 (cont.)}
 \tablecaption{Explanation of CVRHS Symbols (cont.)\tablenotemark{a}
 }
 \tablehead{
 \colhead{Symbol} &
 \colhead{Description} 
 \\
 \colhead{1} &
 \colhead{2} 
 }
 \startdata
S0$^-$ sp / E(b)6         & An edge-on early S0 galaxy, having a boxy thick disk of flattening E6 \\                                                       
                          & \ \ \ (example: NGC 1332); E(b) is Kormendy \& Bender (1996) notation \\                                                       
                          & \ \ \ and is used here only in a descriptive manner \\                                                                         
S0$^-$ sp / E5-6          & An edge-on early S0 galaxy, having an E-like thick disk (neither boxy \\                                                       
                          & \ \ \  nor disky) of flattening E5-6 (example: NGC 3115) \\                                                                    
Sb sp / E5                & An edge-on Sb galaxy embedded in an E5 background (example: NGC \\                                                             
                          & \ \ \ 5078) \\                                                                                                                 
SA(rl,nrl,nb)0$^+$ / E(d)0 & An inclined disk with multiple features embedded within an E0 \\                                                              
                          & \ \ \ background (example: NGC 1553) \\                                                                                        
\noalign{\vspace{3truemm}}                                                                                                                                 
\noalign{\centerline{\bf Other Special Features or Categories\phantom{000000}}}                                                                            
\noalign{\vspace{3truemm}}                                                                                                                                 
Pec                       & A peculiar galaxy, a significantly disturbed interacting system that \\                                                        
                          & \ \ \ cannot be classified in any other way; typically an ongoing or \\                                                        
                          & \ \ \ advanced merger or other disturbed type of system (examples: NGC \\                                                      
                          & \ \ \ 520, 4038-9)\\                                                                                                           
pec                       & In a classification, this refers to a classifiable galaxy having some  \\                                                      
                          & \ \ \ peculiarity (such as some types of asymmetry) (examples: NGC 1087, \\                                                    
                          & \ \ \ 1097)\\                                                                                                                  
dark-spacer               & A galaxy which shows a distinctive area or areas of lower surface \\                                                           
                          & \ \ \ brighter within the bright part of a disk (Buta 2014). In some \\                                                        
                          & \ \ \ early-type galaxies, the appearance of an inner ring may be defined \\                                                   
                          & \ \ \ mainly by a subtle darker area around a bar (Laurikainen et al. 2013) \\                                                 
                          & \ \ \  (example: NGC 2787)\\                                                                                                   
counter-winding spiral    & A spiral galaxy having two sets of spiral structure which wind outward in \\                                                   
                          & \ \ \ opposite senses \\                                                                                                       
 \enddata
 \tablenotetext{a}{Examples are all based on Table 6    mean classifications}
 \tablenotetext{b}{For other combinations,              see Table 6}
 \end{deluxetable*}

The hallmark of VRHS classification is continuity of structure along
three morphological dimensions (in the form of a classification
volume): the {\it stage}, which refers to the E-S0-S-I position
along a modified Hubble sequence (the VRHS sequence); the {\it
family}, referring to the presence or absence of a bar; and the
{\it variety}, referring to the presence or absence of an inner
ring. In addition, there is a fourth dimension known as the {\it
outer ring classification}, referring to the presence of a large
ring in the outer disk. Because stage correlates with several basic
physical properties of galaxies (e.g., average surface brightness,
color, HI mass-to-blue light ratio), it is considered the fundamental
dimension of the system.

The positioning of galaxies in stage depends on specific morphological
characteristics: elliptical galaxies are defined by a smoothly
declining brightness gradient and little or no evidence for a disk
component; S0 galaxies are armless disk galaxies and form a sequence
(S0$^-$$\rightarrow$S0$^o$$\rightarrow$S0$^+$) of increasing structure
ranging from subtle inflections in the brightness distribution (type
S0$^-$) to prominent rings (type S0$^+$); spirals form a sequence
(S0/a-Sa-Sab-Sb-Sbc-Sc-Scd-Sd-Sdm-Sm) of decreasing bulge-to-total
luminosity ratio, increasingly open spiral arms, an increasing
degree of star formation, and increasing asymmetry; and Magellanic
irregular galaxies (Im) are the endpoint characterized by significant
asymmetries, often a high degree of scattered star formation, and
a significant range in luminosity.

The family classification ranges from SA for nonbarred galaxies to
SB for barred galaxies, with an intermediate category of SAB to
account for ``weakly-barred" galaxies, or galaxies intermediate in
apparent bar strength between SA and SB. Both relative bar length and 
bar contrast play a role in family classification. Although bars can
be recognized in edge-on spiral galaxies (Section 4.3), reliable
family classification is still something that can be done only for low
inclination galaxies. High inclination can considerably foreshorten
a bar, or confuse inner structure.

The variety classification ranges from (r) for a closed inner ring
to (s) for an open spiral, with an intermediate category (rs) to
account for partial inner rings having a spiral character (called
``inner pseudorings"). The outer ring classification ranges from
(R) for a closed outer ring to (R$^{\prime}$) for an outer pseudoring
made of variable pitch angle outer spiral arms. Because outer and
inner rings are similar aspects of galaxy morphology, we will
henceforth refer to the conventional variety as the ``inner variety"
and the outer ring classification as the ``outer variety" (Section
3.3).

The four parts of a VRHS spiral galaxy classification in order are:

\centerline{(outer variety)--family--(inner variety)--stage.} 
\centerline{\ \ \ \ \ \ \ \ (R$^{\prime}$)\ \ \ \ \ \ \ \ \ \ SB\ \ \ \
\ \ \ (r)\ \ \ \ \ \ \ \ \ ab  }

For example, the VRHS RC3 classification of NGC 1433 is
(R$^{\prime}$)SB(r)ab. If there is no outer ring or pseudoring, or
if the inner variety cannot be determined, these can be dropped
from a classification.

The distinction between inner and outer varieties is not
always clearcut. Inner and outer rings and pseudorings are easily
distinguished in barred galaxies because the bar usually fills an
inner ring or pseudoring in one dimension, while outer rings and
pseudorings are about twice the bar length in diameter. In the absence
of a bar, the distinction may be ambiguous unless more than
one ring is present. In some cases, it may not be possible to resolve
the ambiguity from visual inspection alone.

The VRHS provides information on inclination through the ``spindle"
(sp) notation. A spindle is a highly-inclined disk galaxy. For
example, the VRHS classification for NGC 4565 is Sb sp.  It is too
inclined to get a full classification with family and variety, but
stage is still distinguishable. The sp after the stage points to
its near edge-on orientation.

Peculiarities are recognized using ``Pec" as the classification, or
``pec" after a regular classification. For example, NGC 4038-9 is a
well-known merger system with distorted components and tidal tails.
The object does not fit into any VRHS ``cell" and so is classified
as ``Pec." If instead, ``pec" follows a classification, it implies
something unusual about the object, often uncharacteristic asymmetry
or odd shape.

de Vaucouleurs (1963) modified the VRHS to include {\it underline
notation} for stage, family, and (inner) variety, where in a combined
classification symbol such as Sab, Sbc, Scd, Sdm, SAB, and (rs), one
symbol is underlined to imply that it is ``closest to actual type." For
example, a classification like SA(s)a$\underline{\rm b}$ would indicate
a stage Sab galaxy that is more Sb than Sa, SB(s)$\underline{\rm c}$d
would be a stage Scd galaxy that is more Sc than Sd, etc. For family,
an S$\underline{\rm A}$B galaxy shows only a trace of a bar (often
merely an oval), while an SA$\underline{\rm B}$ galaxy is more barred
than nonbarred but not as strongly barred as an SB galaxy. Similarly,
for variety an ($\underline{\rm r}$s) galaxy shows a well-defined inner
ring only slightly broken by spiral structure, while an
r$\underline{\rm s}$ galaxy shows only a trace of an inner ring.

Although very useful [and applied in the Catalogue of Southern
Ringed Galaxies (CSRG, Buta 1995) and the dVA], for practical reasons
underline notation was not used in any of the reference catalogues
(RC1, de Vaucouleurs \& de Vaucouleurs 1964; RC2, de Vaucouleurs
et al. 1976; and RC3). de Vaucouleurs (1963) also used underline
notation sparingly: in his survey of 1500 bright galaxies, including
1263 spirals and S0s, underline notation for stages was used for
only 2\% of the galaxies, while underline notation for family and
variety was used for only 9-10\% of the galaxies. In VRHS
classifications, the bulk of classifications will be in the main
categories, while underline categories will generally be underrepresented.

\subsubsection{Comprehensive VRHS Classification}

What the CVRHS adds to the original VRHS system is recognition of
details whose significance to galaxy structure and evolution has only
recently been appreciated. For example, Kormendy (1979) showed that
{\it lenses}, disk morphological features characterized by a shallow
brightness gradient interior to a sharp edge, are prominent in barred
galaxies and could be intimately connected with the evolution of bars.
He argued that lenses were often misclassified as rings in RC2, and
noted that there was a lens analogue of each type of ring in the VRHS.
He suggested using the symbol (l) for inner lenses (analogue of inner
rings) and (L) for outer lenses (analogue of outer
rings).\footnote{Kormendy (2012) prefers the use of (lens) in place of
(l) for inner lenses, to avoid possible confusion with (1). Here we
continue to use (l) and (L) to be consistent with the dVA and the
CSRG.} The significance of lenses to galaxy morphology was further
established with the Near-Infrared S0 Survey (NIRS0S), a $K_s$-band
(2.2$\mu$m) survey of 206 early-type galaxies, including 160 S0-S0/a
galaxies (Laurikainen et al. 2011, 2013).

An important question is how lenses differ from bulges. In the
case of inner lenses, there is no ambiguity between these features
because the bar tends to fill the lens in one dimension (Kormendy
1979). However, there is another type of lens, called a ``barlens"
(Laurikainen et al. 2011) that can be mistaken for a classical bulge
(Athanassoula et al. 2014). These are discussed further in Section
4.3.1.

The recognition of lenses brings attention to other features known
as {\it ring-lenses}, where the apparently sharp edge of a lens is
slightly enhanced to appear as a low contrast ring. In CVRHS
classification, we use the notation (rl) for an inner ring-lens and
(RL) for an outer ring-lens, with pseudoring-lens equivalents of
(r$^{\prime}$l) and (R$^{\prime}$L), respectively. These are also
used with underline notation to emphasize the ring or lens aspect.

The CVRHS includes recognition of important nuclear features, such
as nuclear rings, pseudorings, lenses, ring-lenses, bars, and
disks (Buta \& Combes 1996). A nuclear ring (nr) is a small ring,
often defined by star-forming regions, typically found in the centers
of barred galaxies and on average $\approx$1 kpc in diameter
(Comer\'on et al. 2010). A nuclear lens (nl) is the lens analogue
of a nuclear ring. A nuclear bar (nb) is a small bar often found
within a nuclear ring or lens, or which may be present independent
of these features. A nuclear disk (nd) is typically a distinct,
highly-flattened feature seen most easily in edge-on S0 galaxies.
Nuclear ring-lenses (nrl), nuclear pseudorings (nr$^{\prime}$), and
nuclear spirals (ns) are also known. Similar to inner and outer
rings, in CVRHS classification the presence or absence of a nuclear
ring or related feature will be referred to as the ``nuclear variety."

The fact that there are three ring types (R,r,nr) and three types
of lenses (L,l,nl), with ring-lenses (RL, rl, nrl) as intermediate
categories, all with a similar relationship to bar extent, could
imply a close connection between rings and lenses in a dynamical
or even evolutionary sense. Kormendy (1979) originally argued that
inner lenses could be the result of secular dissolution of a primary
bar. Comer\'on (2013) has most recently examined the ring-lens
issue and concluded that once a star-forming ring exhausts or is
stripped of its gas, it may dissolve into a ring-lens in as little
as 200 Myr.

Other characteristics considered part of CVRHS morphology include
X-patterns, boxy/disky structures, outer Lindblad resonance (OLR)
ring morphologies, warps, spheroidal galaxies, and other features
described in more detail in the next Sections. CVRHS morphology
also considers alternative points of view, such as the parallel
sequence classification of van den Bergh (1976) where S0 galaxies
are stripped spirals on a sequence parallel to regular spirals.
Three recent studies have provided considerable support for this
idea: Laurikainen et al. (2011, 2013), Cappellari et al. (2011),
and Kormendy \& Bender (2012). In a cluster environment, parallel
sequence classification is a more accurate view of galaxy morphology
than the VRHS sequence, and provides a natural home for what Kormendy
\& Bender (2012) refer to as ``spheroidal galaxies." Nevertheless,
this does not negate the value of CVRHS morphology because CVRHS
classification is, for the most part, purely morphological and not
based on a ``whiff of theory" (Sandage 2005). [An exception is the
OLR outer ring/pseudoring morphological subclasses (Buta \& Crocker
1991), which were theoretically predicted by Schwarz (1981).]

Paper I described the application of the CVRHS to 200 S$^4$G galaxies,
where it was shown that, in spite of the significant differences
between modern mid-IR digital images and the optical photographic
blue-light plates that were the historical basis for the original VRHS
system, many galaxies show the same essential morphological features in
the mid-IR as in the $B$-band, allowing the effective application of
the VRHS system to S$^4$G images. There is no need to invent a new
classification system to accomodate mid-IR galaxy morphology; instead,
it should be sufficient to build on what we already have from studies
of blue-light images. This does not mean that the ``essential features"
are defined in the same way in the two wavebands. For example, the
degree of resolution into star-forming regions was one of Hubble's
original criteria for classifying spiral galaxies into Sa-Sb-Sc bins.
In the $B$-band, this resolution is determined by the distribution of
aggregates of young blue supergiants. In the mid-IR, however, these
stars are not prominent. Instead, we see the thermal emission from the local
dust heated by these stars. Paper I also noted that when CVRHS mid-IR
stages were compared to RC3 stages, galaxies of types Sc and later or
S0$^+$ and earlier were often classified the same as in RC3, while
intermediate stages such as S0/a to Sbc were classified about one stage
earlier.

Even if it is possible to use an existing classification system
effectively in the 3.6$\mu$m and 4.5$\mu$m IRAC bands, the
classification of galaxies based on S$^4$G mid-IR images can be
difficult because of competing factors. For example, the contrast of
the spiral structure in mid-IR light can be very low compared to the
brightness of the background disk light. This can make it difficult to
fit some galaxies into the CVRHS system, especially if a galaxy is
small or distant enough to be poorly resolved. Another issue is the
greater sensitivity of the IRAC bands to old stellar population bulges
[also known as classical bulges (Kormendy \& Kennicutt 2004,
Athanassoula 2005], while at the same time being less sensitive to
spiral structure.

The galaxies that show the most drastic differences between mid-IR and
$B$-band morphology are usually those having considerable internal dust
extinction, such as edge-on spirals, starburst galaxies, and major or
minor merger systems. Extinction in the mid-IR bands is less than 5\%
of that in the $B$-band, and generally allows fairly good penetration
into thick planar dust layers. S$^4$G images do allow us to improve our
interpretation of some edge-on galaxies, but even so classification is
still difficult for highly inclined galaxies.

\subsection{Application to the Full S$^4$G Sample}

The classification of the full S$^4$G sample was carried out by
R. Buta in three phases: Phase 1, the initial examination of the full
dataset as data were being collected; Phase 2, a re-examination of the
full dataset made more than a year after data collection ended and
without any reference to the Phase 1 results; and a partial (10\%)
Phase 3 made 6 months after Phase 2 (also without reference to the
previous phases) to better assess the internal consistency of the types
derived by {\it averaging} the full Phase 1 and Phase 2 catalogues
(Section 3.3).\footnote{We thank the referee for suggesting this
approach.}

Table 2 provides the classifications from Phases 1 and 2. (The
Phase 3 classifications are listed in the Appendix.) A visual
comparison shows generally good agreement between them, with
differences appearing to be mostly random. Of the 2412 galaxies, the
Phase 1 and 2 classifications are identical for 382 galaxies, or 16\%
of the sample. Restricting to the stage classification alone, the
agreement is better, with 1370 (57\%) of the sample classified
identically and 659 (27\%) differing by only $\pm$1 step (e.g., as in
Sab vs. Sb), 167 (7\%) differing by $\pm$2 steps (as in Sbc vs. Scd),
and 83 (3\%) differing by more than $\pm$2 steps. Some of the cases of
large disagreement are galaxies which appear to be of an early type,
yet have little or no bulge. These have classifications such as
``SB(s)0/a[d]," where the ``[d]" is meant to highlight the apparent
lack of a bulge [alluding to the van den Bergh (1976) parallel sequence
idea]. This inconsistency could be real, but could also partly be a
resolution effect in the sample.

 \begin{deluxetable*}{llll}
 \tablewidth{0pc}
 \tablenum{2}
 \tablecaption{Phase 1 and 2 CVRHS Classifications
 for 2412 S$^4$G Galaxies\tablenotemark{a,b}
 }
 \tablehead{
 \colhead{Galaxy} &
 \colhead{Phase 1 type} & 
 \colhead{Phase 2 type} & 
 \colhead{Fig. No.}  
 \\
 \colhead{1} &
 \colhead{2} &
 \colhead{3} &
 \colhead{4} 
 }
 \startdata
{\bf $\rightarrow$ RA: 0$^h$ $\leftarrow$} & & & \\
UGC 12893        & dSA(l)0$^o$ / Sph                                                  & dSA(l)0$^o$ / Sph                                                  &  II.1.0001   \\
PGC   143        & dIm                                                                & dI                                                                 &  II.1.0002   \\
ESO   12- 14     & IB(s)m                                                             & SB(s)m                                                             &  II.1.0003   \\
UGC    17        & IABm                                                               & IABm                                                               &  II.1.0004   \\
NGC  7814        & SA0/a spw                                                          & SA0/a spw                                                          &  II.1.0005   \\
NGC  7817        & SAB(rs,nd)bc                                                       & SA(rs,nd)b sp                                                      &  II.1.0006   \\
ESO  409- 15     & dIm                                                                & dIm                                                                &  II.1.0007   \\
ESO  293- 34     & S pec                                                              & S spw pec                                                          &  II.1.0008   \\
NGC     7        & Sd / Im sp                                                         & Sd sp                                                              &  II.1.0009   \\
\enddata
\tablenotetext{a}{The galaxies are listed in order of J2000 right
ascension; arrows indicate the beginning of a right ascension interval. 
See Table 1 for a summary of the meaning of CVRHS
notations, and the notes to Table 5 for galaxies which have a different
name in RC3. Figure numbers are either for paper I (I.1.nnn) or this
paper (II.1.nnnn). Galaxies with an asterisk after the name are on
S$^4$G frames but are not part of the formal sample. Some of these are
included in the atlas illustrations. If they are not, the name of the
galaxy image where they can be found (in the publicly available
database) is given.}
\tablenotetext{b}{Table 2 is published in its entirety in the
electronic edition of the Astrophysical Journal Supplement Series. A
portion is shown here for guidance regarding its form and content.}
\end{deluxetable*}

For 1709 galaxies where it was possible to reliably assign family
classifications, the Phase 1 and 2 classifications were identical
for 1160 (68\%) of the sample, and differed by one 
classification interval (e.g., SB vs SA$\underline{\rm B}$ for
264 (15\%), two intervals (e.g, SAB vs. SB) for 264 (15\%), 3
intervals (e.g., S$\underline{\rm A}$B vs.  SB for 6 (0.4\%), and
the full range (e.g., SB vs. SA) for 15 (0.9\%) of the sample.

The classification and recognition of inner and outer ring and lens
features can show variation, with some classifications in one phase
not being noted in the other phase. In 14 cases in the catalogue,
the same feature has been classified as an inner ring or pseudoring
in one phase, and as an outer ring or pseudoring in the other. As
noted in Section 3.1, such ambiguities can occur especially for
nonbarred or very weakly-barred galaxies (Buta 1995). We adopt
either the Phase 1 or Phase 2 classification for such cases, after
a re-inspection of the image.

\subsection{Comparison of Classifications}

In this Section, we use our independent classification phases to
examine the internal agreement of CVRHS stage, family, and variety
classifications of S$^4$G galaxies. The stages and families are also
compared with other sources.

For comparing classifications, and also for combining the Phase 1
and 2 catalogues, the letter classifications were coded with numbers
(Table 3). For stages, the standard numerical $T$ index, which ranges
from $T$ = $-$5 for E galaxies to $T$ = +10 for Im galaxies, was used
as in RC3.  For family classifications, following Baillard et al.
(2011) we defined an index $F$ which ranges from 0.0 for SA galaxies to
1.0 for SB galaxies. We then set $F$ = 0.25, 0.50, and 0.75 for
S$\underline{\rm A}$B, SAB, and SA$\underline{\rm B}$ galaxies,
respectively. Similar codings were used for inner
ring/pseudoring/spiral and outer ring/pseudoring classifications. In
Tables 3 and 4, these are referred to as the inner variety $IV$ and
outer variety $OV$, respectively. Although the parentheses in VRHS
classifications normally include only a single outer or inner feature
[e.g., as in (R)SB(rs)ab], CVRHS classifications can include multiple
inner and outer rings, lenses, or nuclear features. This complicates
combining Phase 1 and 2 classifications for some of the galaxies, which
had to be treated on an individual basis. For most of the galaxies, the
assignment of a $T$, $F$, $IV$, and $OV$ index was straightforward.
Nevertheless, we strongly emphasize that none of these numbers is a
{\it measured} quantity; they are merely convenient codings for a given
classification symbol. Table 4 shows how the Phase 1 and 2 $<T>$,
$<F>$, $<IV>$, and $<OV>$ values convert into final classifications for
most of the sample.

 \begin{deluxetable*}{lr}
 \tablewidth{0pc}
 \tablenum{3}
 \tablecaption{Numerical Codes}
 \tablehead{
 \colhead{Symbol} &
 \colhead{Numerical Index} 
 \\
 \colhead{1} &
 \colhead{2} 
 }
 \startdata
Stages            & $T$  \\
cE                &  $-$6 \\
E                 &  $-$5 \\
E$^+$             &  $-$4 \\
S0$^-$            &  $-$3 \\
S0$^o$            &  $-$2 \\
S0$^+$            &  $-$1 \\
S0/a              &     0 \\
Sa                &     1 \\
Sab               &     2 \\
Sb                &     3 \\
Sbc               &     4 \\
Sc                &     5 \\
Scd               &     6 \\
Sd                &     7 \\
Sdm               &     8 \\
Sm                &     9 \\
Im                &    10 \\
dE,dS0,Sph        &    11 \\
                  &        \\
Families          & $F$    \\
SA                &     0.00 \\
S$\underline{\rm A}$B  & 0.25 \\
SAB               &      0.50 \\
SA$\underline{\rm B}$ & 0.75 \\
SB                &     1.00 \\
                  &        \\
Inner and Outer Varieties    & $IV$,$OV$    \\
(s),(no outer feature)                &     0.00 \\
(r$\underline{\rm s}$)  & 0.25 \\
(rs),(R$^{\prime}$)               &      0.50 \\
($\underline{\rm r}$s) & 0.75 \\
(r),(R)                &     1.00 \\
(r$\underline{\rm l}$),(R$\underline{\rm L}$)  & 1.25 \\
(rl),(RL)               & 1.50 \\
($\underline{\rm r}$l),($\underline{\rm R}$L) & 1.75 \\
(l),(L)            & 2.00 \\
 \enddata
\end{deluxetable*}

 \begin{deluxetable*}{lc}
 \tablewidth{0pc}
 \tablenum{4}
 \tablecaption{Ranges for Combining Catalogues}
 \tablehead{
 \colhead{Symbol} &
 \colhead{Numerical Index Range} 
 \\
 \colhead{1} &
 \colhead{2} 
 }
 \startdata
(L)                            &           1.875 $\leq$ $<OV>$ $\leq$ 2.000      \\
(R$\underline{\rm L}$)        &           1.625 $\leq$ $<OV>$ $<$ 1.875      \\
(RL)                           &           1.375 $\leq$ $<OV>$ $<$ 1.625      \\
($\underline{\rm R}$L)       &           1.125 $\leq$ $<OV>$ $<$ 1.375     \\
(R)                            &           0.750 $\leq$ $<OV>$ $<$ 1.125      \\
(R$^{\prime}$)                 &           0.250 $\leq$ $<OV>$ $<$ 0.750      \\
no outer feature               &           0.000 $\leq$ $<OV>$ $<$ 0.250      \\
                               &                                                 \\
SA,IA                             &           0.00 $\leq$ $<F>$ $<$ 0.15 \\
S$\underline{\rm A}$B,I$\underline{\rm A}$B        &           0.15 $\leq$ $<F>$ $<$ 0.35 \\
SAB,IAB                           &           0.35 $\leq$ $<F>$ $\leq$ 0.65 \\
SA$\underline{\rm B}$,IA$\underline{\rm B}$        &           0.65 $<$ $<F>$ $\leq$ 0.85\\
SB,IB                             &           0.85 $<$ $<F>$ $\leq$ 1.00 \\
                                &                                       \\
(s)                             &           0.00 $\leq$ $<IV>$ $<$ 0.15 \\
(r$\underline{\rm s}$)        &           0.15 $\leq$ $<IV>$ $<$ 0.35 \\
(rs)                            &           0.35 $\leq$ $<IV>$ $\leq$ 0.65 \\
($\underline{\rm r}$s)        &           0.65 $<$ $<IV>$ $\leq$ 0.85 \\
(r)                             &           0.85 $<$ $<IV>$ $\leq$ 1.15 \\
                                &                                      \\
(l)                            &           1.85 $\leq$ $<IV>$ $\leq$ 2.00      \\
(r$\underline{\rm l}$)       &           1.65 $\leq$ $<IV>$ $<$ 1.85      \\
(rl)                           &           1.35 $\leq$ $<IV>$ $<$ 1.65      \\
($\underline{\rm r}$l)       &           1.15 $\leq$ $<IV>$ $<$ 1.35     \\
\enddata
\end{deluxetable*}

For some galaxies, the classification has two parts, as in
``S0$^-$ sp / E(d)7. This refers to an edge-on S0 showing a thin disk
embedded within a disky (pointy-ended) E-like thick disk. (An example
here is NGC 1032, shown in Figure 1.0193.) In such a case, only the
first part determines the $T$-index, which in this example is $-$3.
Note that normal ellipticals are very rarely more flattened than E4 and
genuine E galaxies are not necessarily found more flattened than E6
(van den Bergh 2009). The E(d) notation is from Kormendy \& Bender
(1996), and was originally proposed for genuine elliptical galaxies.
Two-part classifications recognizing a thick disk are only given for
highly-inclined galaxies.

\subsubsection{Stage and Family Phase 1,2 Comparisons}

\begin{figure}
\figurenum{2}
\plotone{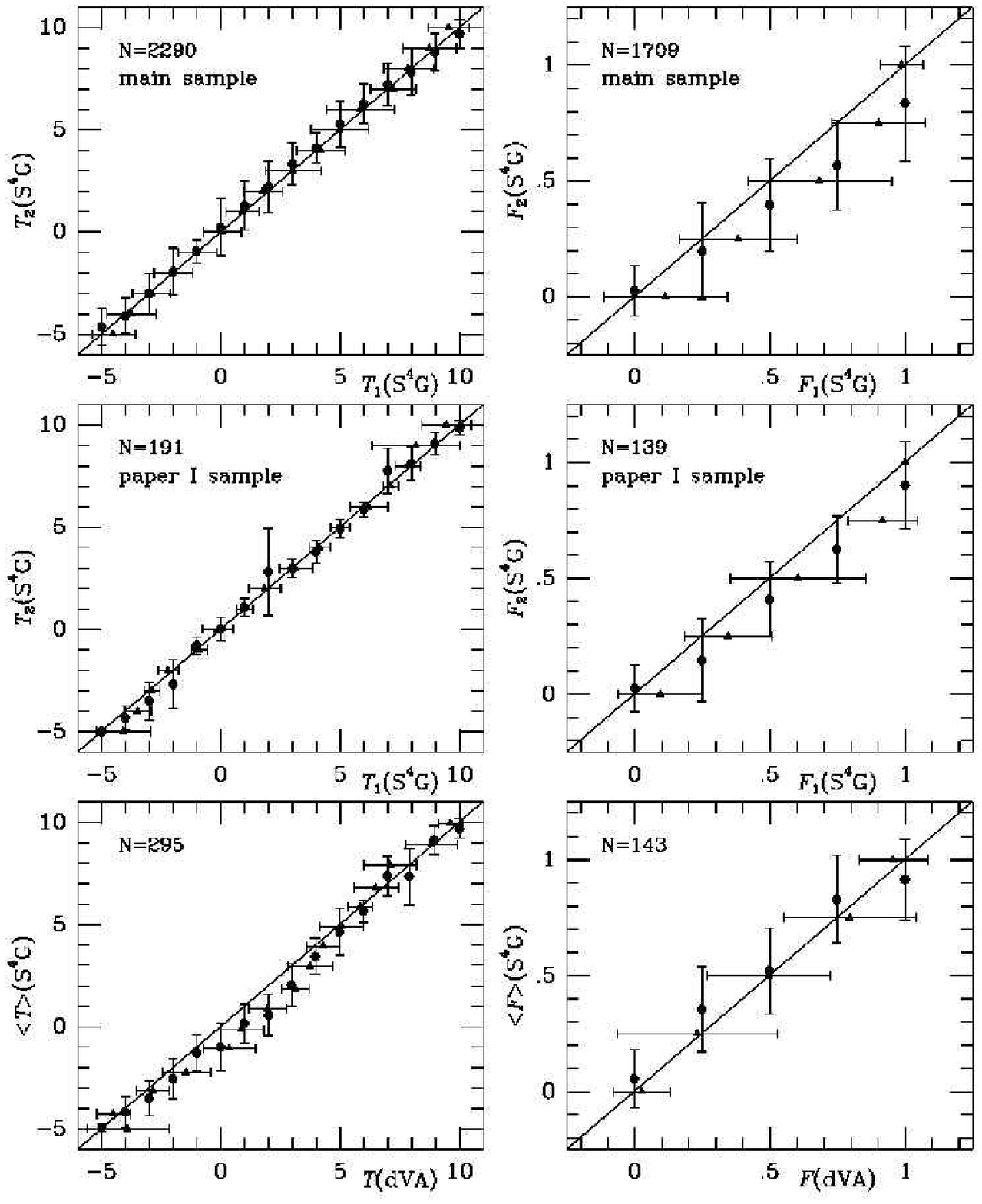}
\caption{{\it top frames} Comparisons between S$^4$G phase 1 and phase
2 stages and families for all S$^4$G galaxies for which both
characteristics could be determined; {\it middle frames}  Comparisons
between S$^4$G phase 1 and phase 2 stages and families for galaxies in
the paper I sample only; {\it bottom frames} Comparison between S$^4$G
mean stages and families with the $B$ and $g$-band classifications for
the same galaxies in the dVA. The error bars are 1$\sigma$ standard
deviations in all frames.}
\label{comp1}
\end{figure}

Figure~\ref{comp1} shows comparisons between the mean numerical type
and family indices for the Phase 1 and 2 classifications for the full
sample (top panels) and for the much smaller paper I subsample (middle
panels). Filled circles show the mean $T$ and $F$ values for Phase 2 at
each Phase 1 type, while the filled triangles show the mean $T$ and $F$
for Phase 1 at each Phase 2 type. The differences, $\Delta T$ =
$T_2-T_1$ and $\Delta F$ = $F_2-F_1$, can be used to judge how
consistent the Phase 1 and 2 classifications are with respect to the 
refined divisions of CVRHS morphology. 

The internal consistency of our Phase 1 and 2 stage and family
classifications can be quantified using the numerical codings in Table
3 to calculate $\sigma_{12}(T)$ = $\sigma (\Delta T)$ = ${\sqrt{{\Sigma
(T_2-T_1)^2} \over {N-1}}}$ and $\sigma_{12}(F)$ = $\sigma (\Delta F)$
= ${\sqrt{{\Sigma (F_2-F_1)^2} \over {N-1}}}$ for all galaxies in the
sample having stage and family classifications in both phases. From
these, we estimate the standard deviation of a single $T$
classification as $\sigma_1 (T)$ $\approx$ $\sigma_2 (T)$ = $\sigma
(T)$ = $\sigma (\Delta T)/\sqrt{2}$ and of a single $F$ classification
as $\sigma_1 (F)$ $\approx$ $\sigma_2 (F)$ = $\sigma (F)$ = $\sigma
(\Delta F)/\sqrt{2}$. In a similar manner, when the Phase 1 and 2
classifications are averaged, we estimate the standard deviation of the
mean types and families as $\sigma(<T>) = \sigma (\Delta T)/2$ and
$\sigma(<F>) = \sigma (\Delta F)/2$, respectively.

The results of this analysis are compiled in Table 5, which
gives these standard deviations in units of 1 stage interval for $T$
($\Delta T$ = 1.0) and 1 family interval for $F$ ($\Delta F$ = 0.25).
For the samples in the top frames of Figure~\ref{comp1}, the standard
deviations between the phases are $\sigma (\Delta T)$ = 1.04 stage
intervals and $\sigma (\Delta F)$ = 0.96 of a family interval. These
imply for a single estimate of $T$ an internal scatter of $\sigma(T)$ =
0.74 of a stage interval, and for a single estimate of $F$ an internal
scatter of $\sigma(F)$ = 0.68 of a family interval. In both cases, the
internal consistency of CVRHS classifications is slightly better than a
single interval of the classification system. These results are
supported by the partial Phase 3 analysis described in the Appendix.

 \begin{deluxetable*}{lrr}
 \tablewidth{40pc}
 \tablenum{5}
 \tablecaption{Internal error analysis of CVRHS classifications\tablenotemark{a}}
 \tablehead{
 \colhead{Dimension} &
 \colhead{Phase1,2} &
 \colhead{Phase$<12>$,3}
 \\
 \colhead{1} &
 \colhead{2} &
 \colhead{3} 
 }
 \startdata
Stage                  &            &               \\
Interval\tablenotemark{b}               &     1.00   &   1.00        \\
$\sigma (\Delta T)$ (intervals)   &     1.04   &   0.86        \\
$\sigma (T)$ (intervals)          &     0.74   &   0.69        \\
$\sigma (<T>)$ (intervals)        &     0.52   &   ....        \\
N                      &     2290   &   238         \\
                       &            &               \\
Family                 &            &               \\               
Interval\tablenotemark{b}               &     0.25   &   0.25        \\
$\sigma (\Delta F)$ (intervals)   &     0.96   &   0.92        \\
$\sigma (F)$ (intervals)          &     0.68   &   0.79        \\
$\sigma (<F>)$ (intervals)        &     0.48   &   ....        \\  
N                      &     1709   &   196         \\
                       &            &               \\
Inner variety          &            &               \\
Interval\tablenotemark{b}               &     0.25   &   0.25        \\
$\sigma (\Delta IV)$ (intervals)  &     1.16   &   1.32       \\
$\sigma (IV)$ (intervals)         &     0.82   &   1.19        \\
$\sigma (<IV>)$ (intervals)       &     0.58   &   ....        \\  
N                      &     1486   &   180         \\
                       &            &               \\
Outer variety          &            &               \\
Interval\tablenotemark{b}               &     1.00   &   1.00        \\
$\sigma (\Delta OV)$ (intervals)  &     1.80   &   1.57        \\
$\sigma (OV)$ (intervals)         &     1.27   &   1.29        \\
$\sigma (<OV>)$ (intervals)       &     0.90   &   ....        \\
N                      &     249   &   33         \\
 \enddata
\tablenotetext{a}{Col. (1): CVRHS classification dimension; (2) results
of Phase 1 and 2 comparison. The parameters listed for stage are
$\sigma (\Delta T)$ = $\sigma_{12} (T)$, where $\Delta T = T_2 - T_1$.
Assuming Phase 1 and 2 are independent, then $\sigma_1$ = $\sigma_2$ =
$\sigma (T)$ = $\sigma_{12}/\sqrt(2)$; $\sigma (<T>) = \sigma_{12}/2$,
where $<T>$ is the value given in column 2 of Table 6; the same kinds of uncertainties
are listed for the other classification dimensions; (3) results of
Phase 3 10\% experiment. In this column, $\Delta T = T_3 - <T>$ and
$\sigma (T) = \sqrt{\sigma (\Delta T)^2 - \sigma (<T>)^2}$. In both
columns 2 and 3, for stage and family, N is the number of galaxies in
the comparison, while for inner and outer variety, N is the number of
features.}
\tablenotetext{b}{Based on the numerical codings in Table 3, except for outer varieties, which
use numbers from 1 to 9 for 9 types of features from R$^{\prime}$ to (L).}
\end{deluxetable*}

\subsubsection{Inner and Outer Variety Phase 1,2 Comparisons}

\begin{figure}
\figurenum{3}
\plotone{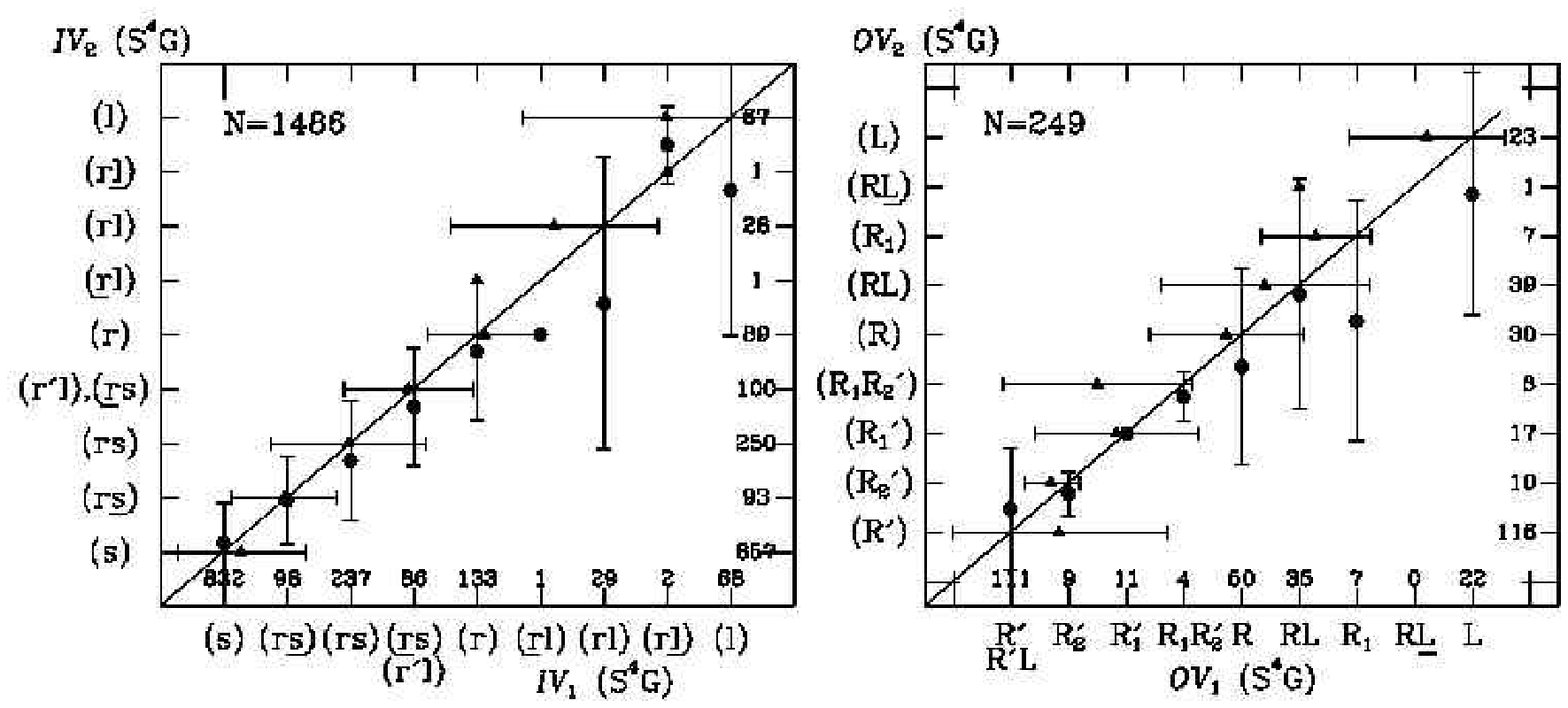}
\caption{Comparison of Phase 1 and 2 inner and outer variety
classifications
of S$^4$G galaxies. The number N in each panel is the number of features
and includes multiple features in some cases. The number in each
category is
also given to show that the dominant classifications are ``(s)" for
inner
variety and ``none" for outer variety.}
\label{ivov}
\end{figure}

Figure~\ref{ivov} shows comparisons between the Phase 1 and 2
inner and outer variety classifications. These comparisons are more
complicated than for stage and family because CVRHS classifications
have more categories (such as lenses and ring/pseudoring-lenses)
compared to the VRHS system, and multiple features are recognizeable in
some galaxies. Also, the classifications ``(r$^{\prime}$l)" and
``(R$^{\prime}$L)" do not fit well into the numerical codings in Table
3, and for the comparison, we have combined these categories with
``($\underline {\rm r}$s)" and ``(R$^{\prime}$)", respectively, based
on a mean stage analysis given in Figure~\ref{varieties} (Section 3.6).
Because of these differences compared to stage and family,
Figure~\ref{ivov} labels the axes using letter classifications rather
than numerical codes.

For inner features, the comparison shows an increased scatter among
lenses and ring-lenses, but nevertheless a good correlation between the
Phase 1 and 2 classifications is found. Within the framework of the
numerical codings in Table 3, and for a numerical inner variety interval of
0.25, the standard deviation between the Phase 1 and 2 inner variety
classifications (Table 5) is $\sigma (\Delta IV)$ = ${\sqrt{{\Sigma
(IV_2-IV_1)^2} \over {N-1}}}$ = 1.16 variety intervals (for $N$ = 1486
features in 1473 galaxies). A single estimate of variety thus has
$\sigma(IV)$ = 0.82 of a variety interval, comparable to what was derived
for stage and family.

The comparison of the Phase 1 and 2 outer variety classifications
(Figure~\ref{ivov}, right panel) uses a different numerical code from
Table 3 because these do not account for the OLR categories R$_1$,
R$_1^{\prime}$, R$_2^{\prime}$, and R$_1$R$_2^{\prime}$. Instead, the
comparison is made in terms of 9 bins ranging from outer pseudorings
(R$^{\prime}$) to outer lenses (L). The ordering is based on the mean
stage analysis in Figure~\ref{varieties}. In terms of a numerical code
ranging from 1 for category R$^{\prime}$ to 9 for outer lenses, and for
an outer variety interval of 1.0, the standard deviation of a single
estimate of outer variety is $\sigma(OV)$ = 1.27 variety intervals
(Table 5). The partial Phase 3 analysis in the Appendix gives a similar
result, but based on a much smaller subset of galaxies.

\subsubsection{Mean Classification Catalogue}

The comparisons in Figure~\ref{comp1} and Figure~\ref{ivov} show
good or at least reasonably good agreement between the Phase 1 and 2
classifications. For this reason, we do not favor one phase over the
other and present in Table 6 unweighted averages of these
classifications. In addition to the full average classification, the
Table gives the average stage and family numerical indices. The average
of the two phases was performed in the following manner:

 \begin{deluxetable*}{lrrll}
 \tablewidth{30pc}
 \tablenum{6}
 \tablecaption{Mean CVRHS Classifications
 for 2412 S$^4$G Galaxies\tablenotemark{a,b}
 }
 \tablehead{
 \colhead{Galaxy} &
 \colhead{$<T>$} & 
 \colhead{$<F>$} & 
 \colhead{$<$Type$>$} & 
 \colhead{Notes}  
 \\
 \colhead{1} &
 \colhead{2} &
 \colhead{3} &
 \colhead{4} &
 \colhead{5} 
 }
 \startdata
{\bf $\rightarrow$ RA: 0$^h$ $\leftarrow$} & & & & \\
UGC 12893        &   11.0 &   0.00 & dSA(l)0$^o$ / Sph & excellent case; face-on  \\
PGC   143        &   10.0 &  ..... & dIm & resolved dwarf; image defects  \\
ESO   12- 14     &    9.5 &   1.00 & S/IB(s)m              & .......... \\
UGC    17        &   10.0 &   0.50 & IABm              & .......... \\
NGC  7814        &    0.0 &   0.00 & SA0/a  spw         & excellent large edge-on   \\
NGC  7817        &    3.5 &   0.25 & S$\underline{\rm A}$B(rs,nd)$\underline{\rm b}$c  sp          & grand-design spiral; bright (nd);   \\
\phantom{NGC  }" & "\phantom{0} & "\phantom{0} & \phantom{00}" & bar is end-on and barely visible;  \\
\phantom{NGC  }" & "\phantom{0} & "\phantom{0} & \phantom{00}" & like N1365  \\
ESO  409- 15     &   10.0 &  ..... & dIAB:m & .......... \\
ESO  293- 34     &   .... &  ..... & S spw pec & warped disk  \\
NGC     7        &    7.0 &  ..... & Sd  sp          & large associations at ends of   \\
\phantom{NGC  }" & "\phantom{0} & "\phantom{0} & \phantom{00}" & major axis  \\
NGC    14        &   10.0 &   0.75 & (L)IA$\underline{\rm B}$(s:)m              & unusual for Im to have an outer   \\
\phantom{NGC  }" & "\phantom{0} & "\phantom{0} & \phantom{00}" & feature  \\
\enddata
\tablenotetext{a}{ The galaxies are listed in order of J2000 right
ascension; arrows indicate the beginning of a right ascension
interval. See Tables 3 and 4 for the numerical codes and ranges used
for combining the Phase 1 and 2 classifications. Arm classifications in
the notes are from Table 9. Galaxies with an asterisk after the name
are in S$^4$G image frames but are not part of the formal sample.}
\tablenotetext{b}{Table 6 is published in its entirety in the
electronic edition of the Astrophysical Journal Supplement Series. A
portion is shown here for guidance regarding its form and content.}

\end{deluxetable*}

\noindent
Example:

\noindent
NGC 4141:

\noindent
Phase 1: SB(r$\underline{\rm s}$)d  ($\underline{\rm s}$ means the s is
emphasized)

\noindent
Phase 2: SA$\underline{\rm B}$(s)dm ($\underline{\rm B}$ means the B 
is emphasized)

\noindent
Table 6 average: SB(s)$\underline{\rm d}$m ($\underline{\rm d}$ means 
the d is emphasized)

The situation above occurs for 71\% (1709) of the 2412 galaxies in the
catalogue. For 13\% (326), the following type of situation occurs:

\noindent
NGC 3044:

\noindent
Phase 1: SB(s)dm sp

\noindent
Phase 2: Sdm sp

\noindent
Table 6 average: SB:(s:)dm sp

\noindent
where the colons are used to indicate that the family and variety are
uncertain interpretations in this case. NGC 3044 is a nearly edge-on
galaxy (hence the ``sp"), and as we have already noted, it can still be
difficult to interpret the structure of an edge-on galaxy, even in a
dust-penetrated waveband.

As the above shows, the combination of the Phase 1 and 2
catalogues can lead to half-step stage classifications (e.g.,
S$\underline{\rm d}$m). It was noted in Section 3.1.1 that underline
stage notation like this is fully a part of VRHS galaxy classification,
but for single estimates of types was used sparingly by de Vaucouleurs
(1963). It was also used sparingly in Phase 1 classifications, and not
at all in Phase 2 classifications. Table 5 shows that the mean stage
classifications in Table 6 have $\sigma (<T>)$ = 0.52 of a stage
interval, essentially equal to one half-step. This implies that half
step mean stages have only marginal significance, a conclusion
supported by the partial Phase 3 analysis (Appendix). Nevertheless, we
retain the underline notation in the mean stages in Table 6 in order to
preserve as much information from the two phases as possible.

For the mean families in Table 6, the Table 5 analysis gives
$\sigma (<F>)$ = 0.48 of a family interval, or half an underline
interval. In this case, the de Vaucouleurs (1963) underline family
classifications SA, S$\underline{\rm A}$B, SAB, SA$\underline{\rm B}$,
and SB have more significance than underline stages. The standard
deviation of the mean inner varieties is $\sigma (<IV>)$ = 0.58 of an
inner variety interval while the standard deviation of mean outer
varieties is $\sigma (<OV>)$ = 0.90 of an outer variety interval,
also giving marginal significance to both classifications.

The mean classifications in Table 6 can be viewed as our ``final" types
because these have the benefit of two independent inspections which
should average out most mistakes and misinterpretations. Nevertheless,
the Phase 1 and 2 classifications listed separately in Table 2 still
have value and are useful to highlight uncertain cases and to see how
consistently a particular object has been interpreted.

\subsubsection{Other Comparisons}

The lower panels in Figure~\ref{comp1} compare the Table 6 $<T>$ and
$<F>$ parameters with the $B$- and $g$-band classifications in the dVA.
In Paper I, we showed that mid-IR types generally agree well with
$B$-band types, with a tendency for galaxies of $B$-band types S0/a to
Sbc being classified slightly earlier in mid-IR type. This is the
``earlier effect" in IR galaxy classification; it results from the
decreased prominence of star-forming regions and the increased
prominence of the bulge in IR images of $B$-band intermediate-type
galaxies (Eskridge et al. 2002). The effect can be seen
directly in the lower left panel of Figure~\ref{comp1}, where the mean
points tend to lie just below the line in this type range.  There is
also a fairly good correlation between 3.6$\mu$m families and dVA
families.

\begin{figure}
\figurenum{4}
\plotone{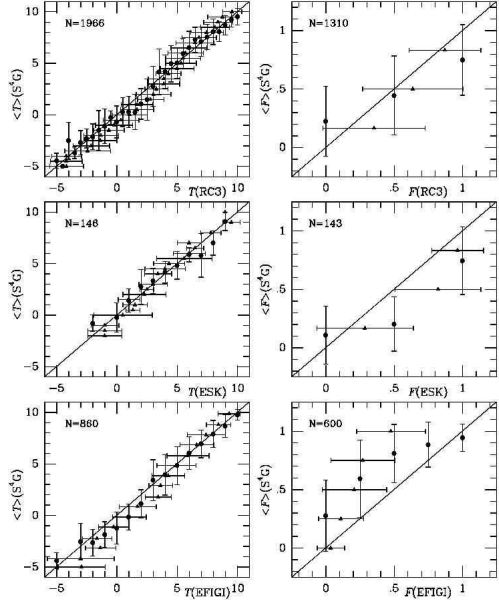}
\caption{{\it top frames} Comparisons between S$^4$G mean stages and
families and the $B$-band classifications for the same galaxies from
RC3; {\it middle frames} The same graphs for the $H$-band
classifications of Eskridge et al. (2002). {\it lower frames} The same
graphs for the optical SDSS classifications of Baillard et al. (2011,
the ``EFIGI" survey). The error bars are 1$\sigma$ standard
deviations in all of the frames.}
\label{comp2}
\end{figure}

Figure~\ref{comp2} ({\it top panels}) compares the mean S$^4$G stage and
family with the same classifications in RC3. The agreement on stages is
similar to that for the dVA, showing the bending of points below the
line in the type range S0/a to Sbc.  (This graph uses only galaxies
having $|\Delta T|$ $\leq$ 4.0 stage intervals, in order to best show
the systematic differences.) The comparison between S$^4$G families and
RC3 families has less resolution than does that for the dVA, but a
reasonable correlation is still found.

The Ohio State University Bright Spiral Galaxy Survey (OSUBSGS;
Eskridge et al. 2002) was an optical/near-IR imaging survey of 205
nearby galaxies designed for a direct comparison between optical
and near-IR galaxy classifications. The main optical filter used
in the survey was the $B$-band and the main near-IR filter used was
the $H$-band. $H$-band (1.65$\mu$m) images are similar to 3.6$\mu$m
images in that the effects of extinction are greatly reduced and
the light mostly traces the stellar mass. However, the Eskridge et
al. (2002) $H$-band images lack the depth of the 3.6$\mu$m IRAC
images and are also less sensitive to star forming regions. The
Eskridge et al. (2002) study demonstrated not only the ``earlier"
effect in near-IR galaxy classification, but also challenged previous
suggestions that there was little correlation between optical and
IR galaxy morphology (e.g., Block \& Puerari 1999).

Figure~\ref{comp2}, middle frames shows a good correlation between
the Eskridge et al. $H$-band types and the Table 6 average $T$
values. The correlation of family classifications for the Eskridge
et al. sample shows, in contrast, that galaxies classified as SA
and SAB by Eskridge et al. are mostly classified as S$\underline{\rm
A}$B in Table 6.

The lower panels compare S$^4$G mid-IR stages and families with the
optical classifications from the ``EFIGI" survey, where a team of 10
astronomers used SDSS images to classify 4458 nearby galaxies (Baillard
et al. 2011). The lower left graph shows the comparison of stages, and
like the comparisons with the dVA and RC3, the ``earlier effect" is
evident at intermediate stages. The comparison with EFIGI bar
classifications shows an unusual pattern that is mainly due to
methodology: the Baillard et al. family classifications are based on
relative bar length, while S$^4$G family classifications are based on
apparent bar strength or contrast (as well as length). While there is
some correlation, it appears S$^4$G bar classifications are stronger on
average than EFIGI bar classifications.

Our general conclusion from all of the comparisons described in this
section is that CVRHS classifications have both good internal and
external consistency. This consistency could, however, still mask
resolution and especially inclination biases in the classifications,
even if there were perfect agreement between phases or between
different sources.  Campbell et al. (2014) describe ``imaging
classification bias" in the context of the GalaxyZoo citizen
morphological catalogue, which they argue has excessive numbers of
distant early-type galaxies that are likely to be misclassified
spirals. As the authors correctly note, the ability to make reliable
distinctions between galaxy types demands ``the most stringent imaging
requirements." Although S$^4$G images are of much greater depth than
groundbased near-IR images, the limited resolution coupled with high
inclination could introduce some ``imaging classification bias." This
is explored further in the next sections.

\subsection{The Distribution of Mid-IR Galaxy Types}

For the purpose of examining the distribution of mid-IR galaxy types,
families, and varieties, we restrict the analysis to the actual
(``formal") sample of 2352 galaxies selected for the S$^4$G (Section
2). The reason for this is that some of the extra galaxies that
were on frames of formal sample galaxies are background objects
beyond the survey's distance limit of 40 Mpc. 

\begin{figure}
%\vspace{6.0truein}
\figurenum{5}
%\vspace{-3.0truein}
\plotone{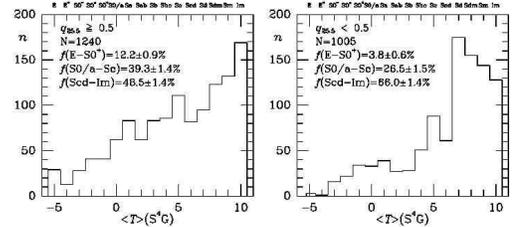}
\caption{The distribution of S$^4$G galaxies versus mean mid-IR type
index (Table 6, column 2) for 2245 formal sample galaxies divided
according to the 3.6$\mu$m isophotal axis ratio, $q_{25.5}$, measured
at the (AB) magnitude surface brightness level of 25.5 mag
arcsec$^{-2}$. The ``low-$i$" sample is at left and the ``high-$i$"
sample is at right. The fractions of galaxies out of the total numbers
N in different type ranges are indicated. The distributions are plotted
in unit stage intervals, with half-step averages rounded to the nearest
full type.}
\label{dist-types}
\end{figure}

Figure~\ref{dist-types} shows the distribution of these mean stages for
the low inclination (low-$i$) and high inclination (high-$i$) galaxies
in the S$^4$G catalogue. The distinction is made using the 3.6$\mu$m
isophotal minor-to-major axis ratio, $q_{25.5}$, fitted at the
(AB-magnitude) surface brightness level of 25.5 mag arcsec$^{-2}$
(Munoz-Mateos et al. 2014). The low-$i$ sample is restricted to
$q_{25.5}$ $\geq$0.5, corresponding to an inclination of
$\approx$60$^{\circ}$. The high-$i$ sample is restricted to $q_{25.5}$
$<$ 0.5. Spindle (``sp") galaxies in the catalogue tend to have
$q_{25.5}$ $<$ 0.4, and are mostly excluded from the low-$i$
sample. For both histograms in Figure~\ref{dist-types}, the $T$ values
have been rounded to the nearest whole type.

The low-$i$ plot shows that nearly half of the formal S$^4$G sample
consists of extreme late-type spiral and irregular galaxies of mid-IR
stages Scd-Im. These constitute 48.5$\pm$1.4\% of the 1240 low-$i$
formal sample galaxies having a mean stage.\footnote{For a percentage
$p$ out of a number $N$ of sample objects, the error is calculated as
$\Delta p (\%) = \sqrt{(100.0-p)p/N}$ (e.g., Laurikainen et al. 2013).}
The right panel in Figure~\ref{dist-types} shows, however, that this
type range constitutes 66.0$\pm$1.4\% of the high-$i$ sample. This
difference is likely to be partly due to the two-part manner in which
we have classified some highly-inclined galaxies (Section 4.4.2). Also,
for highly-inclined galaxies, mid-IR morphology often involves low
contrast features and central concentrations viewed against a very
bright background. These can lead to later-type classifications, and it
is possible that some Sc galaxies are misclassified as Sd when the
inclination is high. The problem can be exacerbated further if the
image of a high inclination galaxy has relatively poor resolution.

The emphasis of the formal S$^4$G sample on extreme late-type galaxies
not only reflects its nature as a volume-limited sample, but also is
due to the fact that the galaxies were selected on the basis of an HI
radial velocity. The sample is rich in dwarf irregulars and bulgeless
spirals, because these tend to be HI-rich systems (e.g., Buta et al.
1994).

\subsection{The Distribution of Mid-IR Families}

\begin{figure}
\vspace{4.0truein}
\figurenum{6}
\vspace{-4.0truein}
\plotone{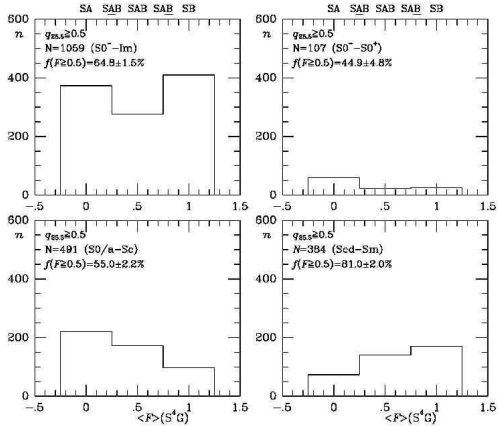}
\caption{{\it top left} The distribution of S$^4$G galaxies versus mean
mid-IR family index (Table 6, column 3) for 1059 ``low-$i$" sample
galaxies where a family classification was recorded in both Phase 1 and
Phase 2.  The other frames show distributions for subsets of this
sample divided according to mid-IR stage. The fractions, $f$, of SAB,
SA$\underline{\rm B}$, and SB galaxies out of the total numbers N
listed are indicated. The distributions are plotted after combining SA
and S$\underline{\rm A}$B into the SA bin and SB and SA$\underline{\rm
B}$ into SB.}
\label{allfams}
\end{figure}

The distribution of mean Phase 1 and 2 family classifications ($<F>$)
is shown in Figure~\ref{allfams}. The graphs are based only on formal
sample galaxies where a family classification was recorded in both
Phase 1 {\it and} Phase 2. Although bars are detectable in
highly-inclined galaxies (Section 4.3), for the purpose of examining
the relative frequency of bars it is still best to restrict the
analysis to the low-$i$ subsample. Also, the graphs are plotted after
combining SA and S$\underline{\rm A}$B into SA, and SA$\underline{\rm
B}$ and SB into SB, since the underline categories are always
under-represented compared to the main categories.

The barred family classification fraction, $f(F \geq 0.5)$, is defined as

$$f (F \geq 0.5) = {{\rm N(SAB)}+{\rm N(SA\underline B}) + {\rm N(SB)}\over {\rm N}}$$

\noindent
where N is the total number of galaxies in the sample that can be
subdivided in this manner [i.e., N = N(SA) + N(S$\underline{\rm A}$B) +
N(SAB) + N(SA$\underline{\rm B}$) + N(SB)] and the counts are based on the
mean letter-classifications in Table 6, not the $<F>$ index. This assumes that SAB
galaxies have obvious bars, even if these bars might be less striking
than those seen in a classical SB galaxy. We anticipate that a sample
like the S$^4$G might give a reliable assessment of this fraction,
because no bars would be overlooked due to star formation or excessive
internal extinction, two problems which complicate $B$-band visual bar
fractions (Eskridge et al. 2000).

Figure~\ref{allfams} shows estimates of $f (F \geq 0.5)$ for the 1059
low-$i$ formal S$^4$G galaxies having Phase 1 and 2 family
classifications, and for subsets of low-$i$ galaxies separated by
stage. Over the type range S0$^-$ to Im, $f(F \geq 0.5)$ is about 2/3,
similar to previous studies such as de Vaucouleurs (1963), Eskridge et
al. (2000), and Marinova \& Jogee (2007), although the latter two
studies are based on the Ohio State University Bright Spiral Galaxy
Survey (Eskridge et al. 2002) which did not include very many galaxies
having RC3 stages later than Scd.

When divided by type, the histograms are distinctly different: for
Scd-Sm galaxies, the SB bin has the highest number of galaxies, while
for S0/a-Sc galaxies, the SA bin has the highest number. As a result,
the barred family classification fractions are significantly different:
$f (F \geq 0.5)$ = 81.0$\pm$2.0\% for Scd-Sm galaxies and
55.0$\pm$2.2\% for S0/a-Sc galaxies.

\begin{figure}
\figurenum{7}
%\vspace{-2.0truein}
\plotone{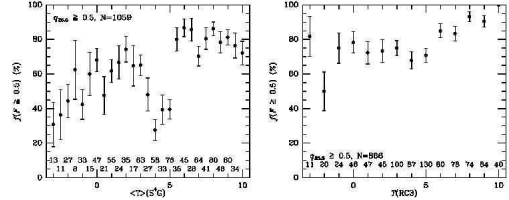}
\caption{({\it left}): Graph of the mid-IR barred family classification
fraction as a function of mean mid-IR stage (Table 6, column 2) for
1059 galaxies in the low-$i$ S$^4$G subsample.  ({\it right}): Graph of
the $B$-band barred classification fraction (SAB+SB) as a function of
stage for 866 of the same galaxies as in the left graph, based on RC3
classifications. Error bars are derived as $\Delta p (\%) =
\sqrt{(100.0-p)p/n}$, where $p$ is the fraction and $n$ is the number
of galaxies at each stage (indicated on each graph).}
\label{barfrac}
\end{figure}

Further insight into these histograms is provided in
Figure~\ref{barfrac}, left, which shows $f (F\geq 0.5)$ as a function
of individual stages from S0/a to Im, again restricted to the low-$i$
sample. Much of the difference between the early and late samples is
due to what appears to be a substantial drop in $f$ at stages
S$\underline{\rm b}$c to Sc. As a check on this, Figure~\ref{barfrac},
right, shows the same graph for 866 of the same galaxies using RC3
classifications. The large drop in $f$ for mid-IR classification is
much less evident when $B$-band RC3 classifications are used. The drop
in the mid-IR fractions occurs in the type domain where the ``earlier
effect" begins to be noticeable. In spite of the drop, both graphs
support a higher barred family classification fraction for galaxies of
types Scd and later.

Table 7 lists the dependence of $f$ in the early-type, late-type, and
full spiral subsamples as a function of $q_{min}$, the minimum mid-IR
isophotal axis ratio for the sample. (The ``low-$i$" sample we have
been using corresponds to $q_{min}$ = 0.5). The table shows that
the significant difference between the early and late-type samples
diminishes but is not completely eliminated except for the most
face-on subsample ($q_{min}$ = 0.9).

 \begin{deluxetable*}{ccrcrcr}
 \tablewidth{0pc}
 \tablenum{7}
 \tablecaption{S$^4$G Bar Classification Fraction\tablenotemark{a}}
 \tablehead{
 \colhead{$q_{min}$} &
 \colhead{$f_e (F\geq 0.5)$} &
 \colhead{N$_e$} &
 \colhead{$f_l (F\geq 0.5)$} &
 \colhead{N$_l$} &
 \colhead{$f (F\geq 0.5)$} &
 \colhead{N} 
 \\
 \colhead{} &
 \colhead{(S0/a-Sc)} &
 \colhead{} &
 \colhead{Scd-Sm} &
 \colhead{} &
 \colhead{S0/a-Sm} &
 \colhead{} 
 \\
 \colhead{1} &
 \colhead{2} &
 \colhead{3} &
 \colhead{4} &
 \colhead{5} &
 \colhead{6} &
 \colhead{7}
 }
 \startdata
  0.1 &  55.8$\pm$ 1.9 &   694 &  84.9$\pm$ 1.3 &   720 &  70.6$\pm$ 1.2 &  1414 \\
  0.2 &  55.7$\pm$ 1.9 &   691 &  84.2$\pm$ 1.4 &   689 &  69.9$\pm$ 1.2 &  1380 \\
  0.3 &  55.7$\pm$ 1.9 &   659 &  82.7$\pm$ 1.6 &   584 &  68.4$\pm$ 1.3 &  1243 \\
  0.4 &  55.2$\pm$ 2.1 &   585 &  82.2$\pm$ 1.8 &   466 &  67.2$\pm$ 1.4 &  1051 \\
  0.5 &  55.0$\pm$ 2.2 &   491 &  81.0$\pm$ 2.0 &   384 &  66.4$\pm$ 1.6 &   875 \\
  0.6 &  57.7$\pm$ 2.5 &   378 &  78.0$\pm$ 2.4 &   291 &  66.5$\pm$ 1.8 &   669 \\
  0.7 &  58.5$\pm$ 3.0 &   265 &  73.7$\pm$ 3.2 &   194 &  64.9$\pm$ 2.2 &   459 \\
  0.8 &  54.8$\pm$ 4.0 &   157 &  71.3$\pm$ 4.2 &   115 &  61.8$\pm$ 2.9 &   272 \\
  0.9 &  67.9$\pm$ 6.4 &    53 &  76.5$\pm$ 7.3 &    34 &  71.3$\pm$ 4.9 &    87 \\
 \enddata
\tablenotetext{a}{Col. (1) minimum isophotal axis ratio at
$\mu_{AB}$(3.6$\mu$m) = 25.5 mag arcsec$^{-2}$; (2) fraction (percent)
of classified bars in galaxies having isophotal axis ratio $q_{25.5}$
$\ge$ $q_{min}$ in the earlier type range S0/a - Sc; (3) no. of galaxies in the
S0/a - Sc subsample; (4,5) same as (2,3) for the later type sample
(Scd - Sm); (6,7) same as (2,3) for the combined samples.}
\end{deluxetable*}

These results are actually fairly consistent with previous
studies. For example, Barazza et al. (2008) used ellipse fits to infer
the presence of bars in a large sample of nearby SDSS galaxies over the
absolute magnitude range $-$18.5 $\leq$ $M_B$ $\leq$ $-$22.0, and
concluded that 87\% of bulgeless disk galaxies are barred as opposed to
44\% of bulge+disk galaxies. Galaxies of CVRHS types Scd-Sm not only do
not have classical bulges, they also generally do not have
pseudobulges. Barazza et al. also found that the optical bar fraction
rises with decreasing total galaxy mass, a result which is consistent
with the mass-dependent bimodality of the bar fraction noted by Nair \&
Abraham (2010). Scd-Sm galaxies tend to be less luminous on average
than S0/a-Sc galaxies (e.g., the dVA), and therefore are likely less
massive than S0/a-Sc galaxies.

The upper right panel of Figure~\ref{allfams} shows the distribution
of family classifications for 107 S0 galaxies in the low-$i$ subsample.
The most common family classification for S$^4$G S0 galaxies is SA, and
the barred family fraction is 45$\pm$5\%. This smaller fraction of
bars is also consistent with previous studies (Laurikainen et al. 2009;
Aguerri et al. 2009; Buta et al. 2010b), but a more reliable assessment
must await the S$^4$G extension.

Note that the ``barred family classification fraction" is not
necessarily the same as the cosmologically significant ``bar fraction"
(e.g., Sheth et al. 2008; 2014a,b). This is because the types of bars
seen in nearby late-type galaxies may not necessarily be the ones we
see at high redshift. The distinction is discussed further in Section
4.3.3.

\subsection{Inner and Outer Varieties}

In this Section we examine some of the systematics of variety
classifications in the S$^4$G catalog. Inner varieties range from
the pure (s) shape to inner pseudorings (rs) to inner rings (r)
and finally to inner lenses (l). Outer varieties include no feature for
an unclosed spiral, outer pseudorings (R$^{\prime}$), outer rings (R),
outer ring-lenses (RL), outer lenses (L), and OLR subclass features
R$_1$, R$_1^{\prime}$, R$_2^{\prime}$, and R$_1$R$_2^{\prime}$.

\begin{figure}
\figurenum{8}
%\vspace{-4.0truein}
\plotone{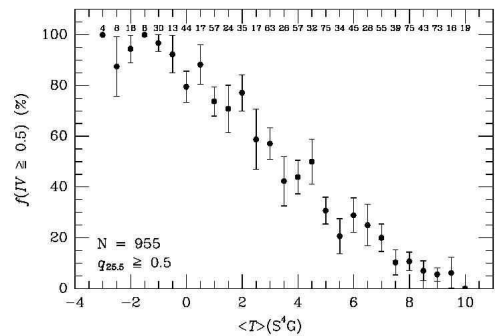}
\caption{Graph of the inner ring/lens classification fraction (rs,
$\underline{\rm r}$s, r, rl, and l) as a function of mean mid-IR stage
(Table 6, column 2) for 955 galaxies from the low-$i$ S$^4$G subsample.
The number of galaxies at each stage is indicated. Error bars
calculated in the same manner as for Figure~\ref{barfrac}.}
\label{ring-lens-frac}
\end{figure}

Figure~\ref{ring-lens-frac} shows the inner ring-lens fraction as
a function of mid-IR stage. The fraction strongly declines with
advancing stage from early to late.  Although S0s can and do appear
in (s) variety form, this is far less abundant than ring and lens
varieties.  Comparison of Figure~\ref{ring-lens-frac} with
Figure~\ref{barfrac} shows that where the barred classification
fraction is highest, the fraction of inner rings and lenses is
lowest. This appears to imply that gas-rich galaxies are less capable
of forming lenses or the kinds of rings that are typically seen in
early-type barred and nonbarred galaxies.

Similar to Table 7, Table 8 shows the inner ring/pseudoring fraction
for S0/a-Sc, Scd-Sm, and S0/a-Sm galaxies as a function of the value of
$q_{min}$ used to define the subsamples. These features are found in
$\approx$50\% of S0/a-Sc galaxies and $\approx$13\% of Scd-Sm galaxies.
Across the whole spiral sequence, about 1/3 of the galaxies has an inner
ring/pseudoring.

 \begin{deluxetable*}{ccrcrcr}
 \tablewidth{0pc}
 \tablenum{8}
 \tablecaption{S$^4$G Inner Ring/Pseudoring Fraction\tablenotemark{a}}
 \tablehead{
 \colhead{$q_{min}$} &
 \colhead{$v_e$ (rs,$\underline{\rm r}$s,r)} &
 \colhead{N$_e$} &
 \colhead{$v_l$ (rs,$\underline{\rm r}$s,r)} &
 \colhead{N$_l$} &
 \colhead{$v$ (rs,$\underline{\rm r}$s,r)} &
 \colhead{N} 
 \\
 \colhead{} &
 \colhead{(S0/a-Sc)} &
 \colhead{} &
 \colhead{Scd-Sm} &
 \colhead{} &
 \colhead{S0/a-Sm} &
 \colhead{} 
 \\
 \colhead{1} &
 \colhead{2} &
 \colhead{3} &
 \colhead{4} &
 \colhead{5} &
 \colhead{6} &
 \colhead{7}
 }
 \startdata
  0.1 &  48.6$\pm$ 2.0 &   621 &   9.6$\pm$ 1.2 &   634 &  28.9$\pm$ 1.3 &  1255 \\
  0.2 &  48.8$\pm$ 2.0 &   619 &   9.9$\pm$ 1.2 &   614 &  29.4$\pm$ 1.3 &  1233 \\
  0.3 &  49.6$\pm$ 2.1 &   591 &  11.2$\pm$ 1.4 &   528 &  31.5$\pm$ 1.4 &  1119 \\
  0.4 &  51.0$\pm$ 2.2 &   524 &  13.0$\pm$ 1.6 &   431 &  33.8$\pm$ 1.5 &   955 \\
  0.5 &  51.1$\pm$ 2.4 &   444 &  13.0$\pm$ 1.8 &   354 &  34.2$\pm$ 1.7 &   798 \\
  0.6 &  51.3$\pm$ 2.7 &   343 &  13.1$\pm$ 2.0 &   274 &  34.4$\pm$ 1.9 &   617 \\
  0.7 &  52.7$\pm$ 3.2 &   243 &  14.3$\pm$ 2.6 &   182 &  36.2$\pm$ 2.3 &   425 \\
  0.8 &  50.4$\pm$ 4.2 &   141 &  15.0$\pm$ 3.4 &   107 &  35.1$\pm$ 3.0 &   248 \\
  0.9 &  52.1$\pm$ 7.2 &    48 &  24.2$\pm$ 7.5 &    33 &  40.7$\pm$ 5.5 &    81 \\
 \enddata
\tablenotetext{a}{Col. (1) minimum isophotal axis ratio at
$\mu_{AB}$(3.6$\mu$m) = 25.5 mag arcsec$^{-2}$; (2) fraction (percent)
of classified inner rings and pseudorings in galaxies having isophotal
axis ratio $q_{25.5}$ $\ge$ $q_{min}$ in the earlier type range S0/a -
Sc; (3) no. of galaxies in the S0/a - Sc subsample; (4,5) same as (2,3)
for the later type sample (Scd - Sm); (6,7) same as (2,3) for the
combined samples.}
\end{deluxetable*}

\begin{figure}
\figurenum{9}
\plotone{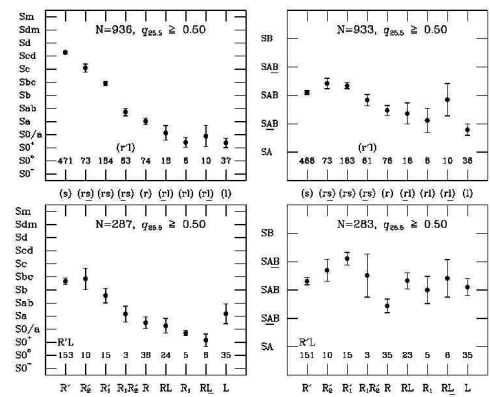}
\caption{Graphs of mean mid-IR stage and family versus inner and outer
variety classifications. In these graphs, $r^{\prime}$l is combined
with $\underline{\rm r}$s and R$^{\prime}$L is combined with
R$^{\prime}$. The number $n$ of galaxies at each classification is
indicated. Error bars are mean errors derived as $\sigma_1$/$\sqrt{n}$,
where $\sigma_1$ is the standard deviation.}
\label{varieties}
\end{figure}

Figure~\ref{varieties} ({\it top left}) shows how both mean stage and
mean family correlate with inner variety classifications. A strong
correlation between mean stage and variety is seen in the sense that
(s)-shaped spirals are on average mid-IR stage Scd, while (r)-variety
spirals average at mid-IR stage Sa. Ring-lenses average at stages
S0$^+$ to S0/a. The upper right panel of Figure~\ref{varieties} shows
the average visual bar strength for (s), (r$\underline{\rm s}$), and
(rs) varieties is a little higher than for ($\underline{\rm r}$s), (r),
(rl), and (l) varieties.

Similar results are found for the outer variety features
(Figure~\ref{varieties}, {\it lower left}). Galaxies having outer
pseudorings average between stages Sb and Sbc, while closed outer rings
are found on average in S0/a and Sa galaxies.  Outer ring-lenses (RL),
OLR subclass rings R$_1$, and outer lenses (L) are found on average in
S0$^+$ to S0/a galaxies. The lower right frame of
Figure~\ref{varieties} shows that average visual bar strength for outer
pseudorings is only slightly higher than for outer rings and
ring-lenses. In the NIRS0S sample of bright early-type disk galaxies
(Laurikainen et al. 2011), outer lenses are common in the non-barred
S0s. Some of these are missing from S$^4$G because of the sample
selection, which picked up only the fairly gas rich galaxies.

\section{Morphological Highlights of the Catalogue}

In this Section, we examine some of the highlights of mid-IR galaxy
morphology. The ``stellar structures" we discuss are mostly aspects of
normal galaxies rather than strongly-interacting or merging galaxies.
Stellar rings, characteristics of early and late-type bars,
dust-penetrated views of classical edge-on galaxies, unusual spirals,
the possible connection between late-type and spheroidal galaxies, and
other special cases of interest are what we focus on here. These
complement the highlights described in Paper I. All of the
galaxies we illustrate in this section were selected for special
attention from the large database because they are considered to be
good examples of the morpholological characteristics that we have
chosen to highlight.

\subsection{Mid-IR CVRHS Sequence}

\begin{figure}
\figurenum{10}
\plotone{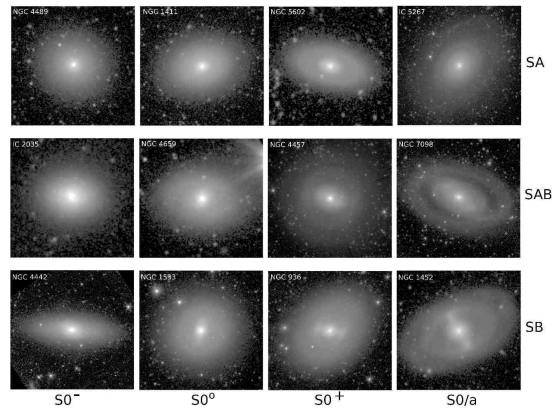}
\caption{Montage of early-type galaxies covering some of the range
in mid-IR morphologies of S0 and S0/a galaxies. Unless otherwise
noted, all of the images displayed in this and succeeding figures
are in the IRAC 3.6$\mu$m filter, and are in units of mag arcsec$^{-2}$
within the range of surface brightnesses given in the caption for
each galaxy in Figure 1. Figure 1 numbers are given in Table 2.
}
\label{S0s-names}
\end{figure}

\begin{figure}
\figurenum{11}
%\plotone{mid-ir-tuning-fork-names-w4449.ps}
\begin{center}
\includegraphics[height=6.0in,angle=0]{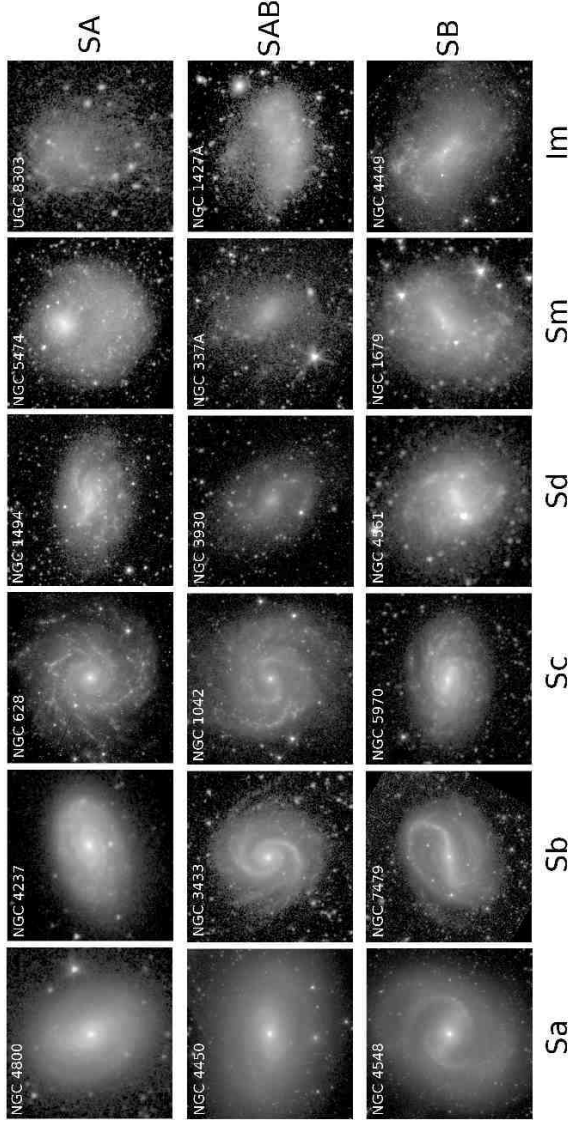}
\end{center}
\caption{Montage covering some of the range in mid-IR morphologies
of spiral and irregular galaxies.
}
\label{spirals}
\end{figure}

The first major conclusion one can draw from examining the mid-IR
morphologies of a large sample of galaxies is that the stellar mass
structure of galaxies is not hidden too deeply by dust in optical
imaging. The montages in Figures~\ref{S0s-names} and ~\ref{spirals}
show the same kinds of morphologies as are seen in blue light,
ordered in a 3-pronged ``tuning fork" by VRHS family (SA, SAB, SB)
along the standard VRHS sequence of stages.

S0 galaxies are subdivided into stages S0$^-$, S0$^o$, and S0$^+$
in a sequence of increasing structure (but not necessarily decreasing
bulge-to-total luminosity ratio). In S0$^-$ galaxies, bars and
lenses are usually fairly subtle, and the disk is virtually
featureless. S0$^o$ galaxies have more obvious lenses, ring-lenses,
and bars. The most structured S0s are stage S0$^+$, which is the
type where rings (without spirals) are most prominent (Section 4.2).
In S0/a galaxies, the rings may break into tight spiral structure.

The spiral and irregular sequences also show well-defined and
representative types. The decrease in average bulge-to-total
luminosity ratio is evident. In general, the same three criteria
Hubble used for spirals: the relative prominence of the bulge, the
degree of resolution of the arms, and the degree of openness of the
arms, are still applicable in the mid-IR, as already noted in paper
I. 

Several of the galaxies in Figure~\ref{spirals}, if displayed in blue
light, would appear in a different cell. For example, NGC 7479 is
classified as stage Sc in the $B$-band but as stage Sb in the mid-IR,
while NGC 4548 is classified as stage Sb in the $B$-band and Sa in the
mid-IR. This is the ``earlier" effect noted in Section 3.3.

\subsection{The Mid-IR Morphology of Galactic Rings and Lenses}

Rings, pseudorings, and lenses are important features of disk-shaped
galaxies. Rings are often sites of active star formation and come
in several different types: nuclear, inner, and outer, that occur
on very different linear scales. Rings and lenses have distinctive
shapes and orientations that tie them to aspects of internal dynamics
(Buta \& Combes 1996). Generally highly flattened, rings and lenses
are most commonly seen in early-type disk galaxies and are likely
some of the clearest products of secular evolution that we can
observe in galaxies (Knapen 2010).

Recognizing all of the variations of the stellar mass morphology
of galactic rings and lenses can add considerable complexity between
the brackets of CVRHS 3.6$\mu$m galaxy classifications (Table 1).
In this Section, we examine this complexity, beginning with CVRHS
stage S0$^+$, the first stage where all of these phenomena become
prominent, and then look at multi-feature systems.  Stage S0$^+$
is a part of the CVRHS sequence where galaxies that are likely
highly-evolved are mixed with others that have been
environmentally-modified. By ``highly-evolved", we mean that
well-defined stellar rings and lenses are end-products of long-term
internal evolution of what were originally zones of intense star
formation. By ``environmentally modified," we mean that these same
galaxies were stripped of their gas which eventually greatly lowered
or even turned off the star formation in their resonant features,
allowing secular evolution of the stellar mass distribution to
eventually reduce the contrast of the features. Thus, ``highly-evolved"
and ``environmentally-modified" are not mutually exclusive categories.

\subsubsection{S0$^+$ Galaxies: Diversity at a Single Stage}

\begin{figure}
\figurenum{12}
\plotone{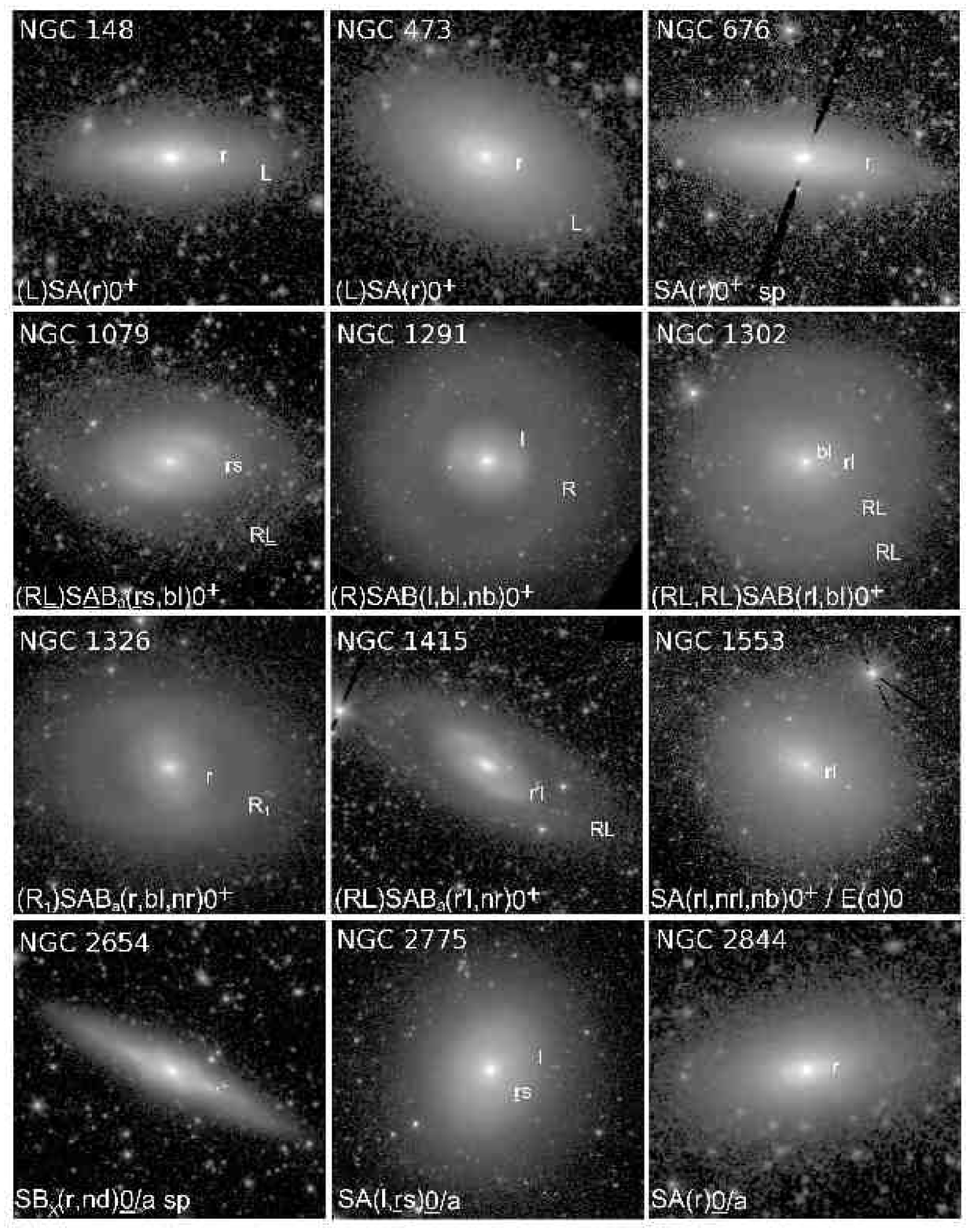}
\caption{}
\label{S0plus}
\end{figure}

In optical imaging, late S0s can be dusty, and some galaxies classified
in the $B$-band as stages S0/a to Sab (e.g., NGC 1079, 1291, 2775,
4594) can appear to be stage S0$^+$ in the mid-IR.  Forty-one cases of
mid-IR type S0$^+$ are collected in the montages of
Figure~\ref{S0plus}. Most interesting is that 20 of these objects are
of family SA (see Table 6 classification in each frame), and in seven
of these, a bright knotty ring is the main morphological feature. If
rings are best understood as resonant products of secular evolution in
barred galaxies (Buta \& Combes 1996), then these nonbarred cases with
bright stellar or star-forming rings are a rather important subset to
know about. Invariant manifold theory (Romero-G\'omez et al. 2006,
2007; Athanassoula et al. 2009a,b) also describes the structure and
secular evolution of rings, but would also have trouble explaining
rings in nonbarred galaxies. The large number of nonbarred S0$^+$
galaxies in the sample differs from the NIRS0S sample, where SB0$^+$ and
SAB0$^+$ galaxies are more abundant (Laurikainen et al. 2013). This
could largely be due to the S$^4$G sample bias towards gas-rich S0s.

\begin{figure}
\figurenum{12 (cont.)}
\plotone{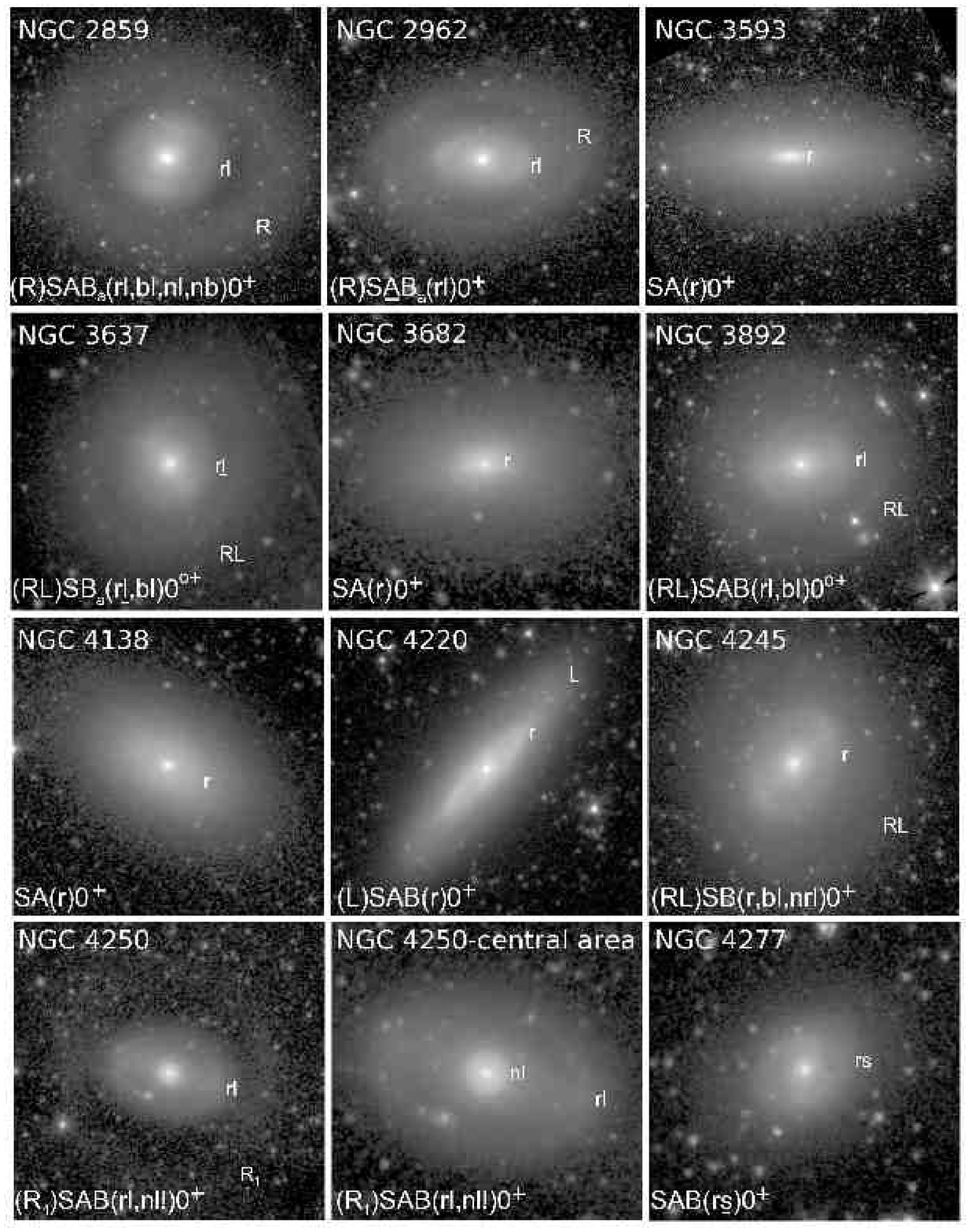}
\end{figure}
\begin{figure}
\figurenum{12 (cont.)}
\plotone{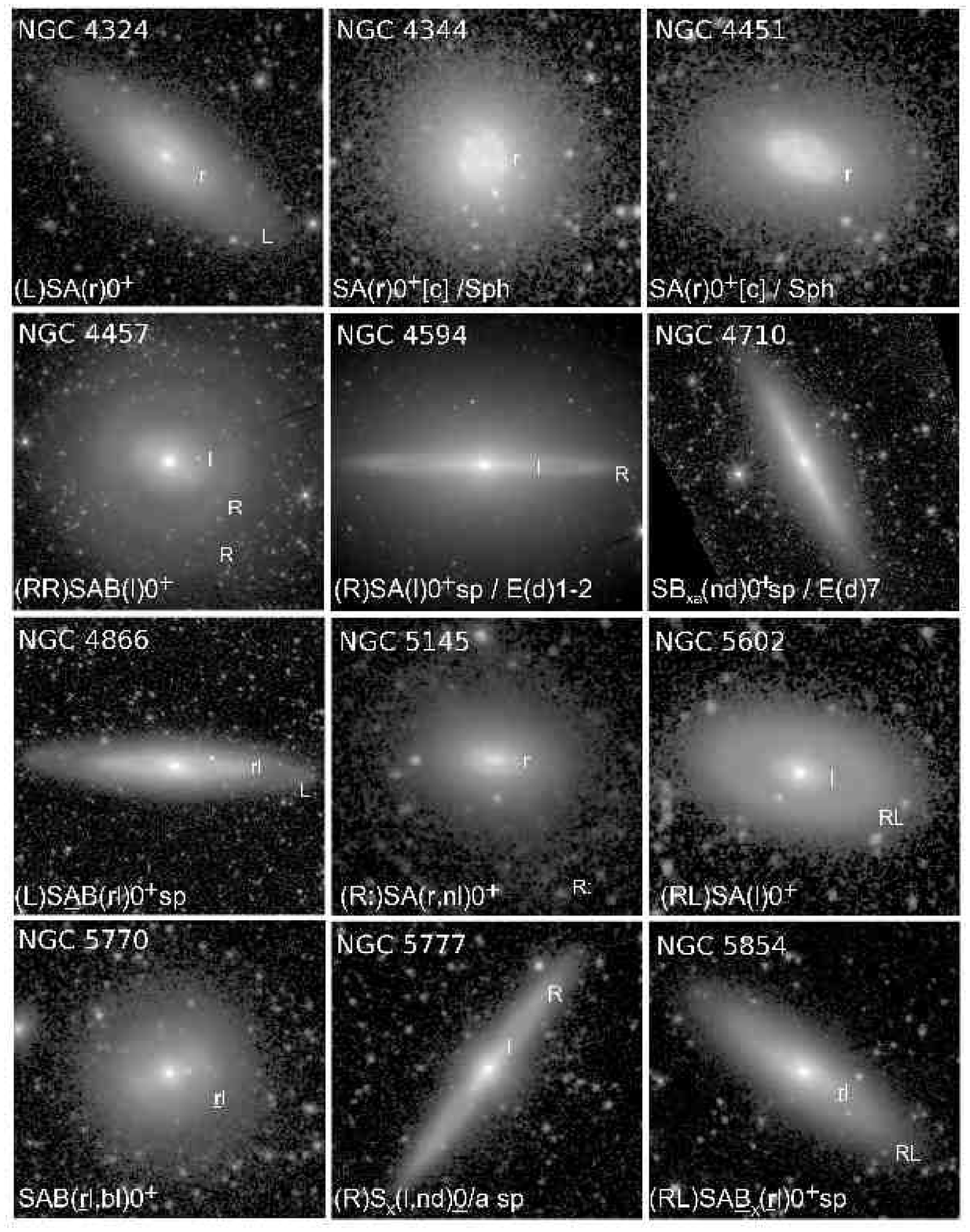}
\end{figure}
\begin{figure}
\figurenum{12 (cont.)}
\plotone{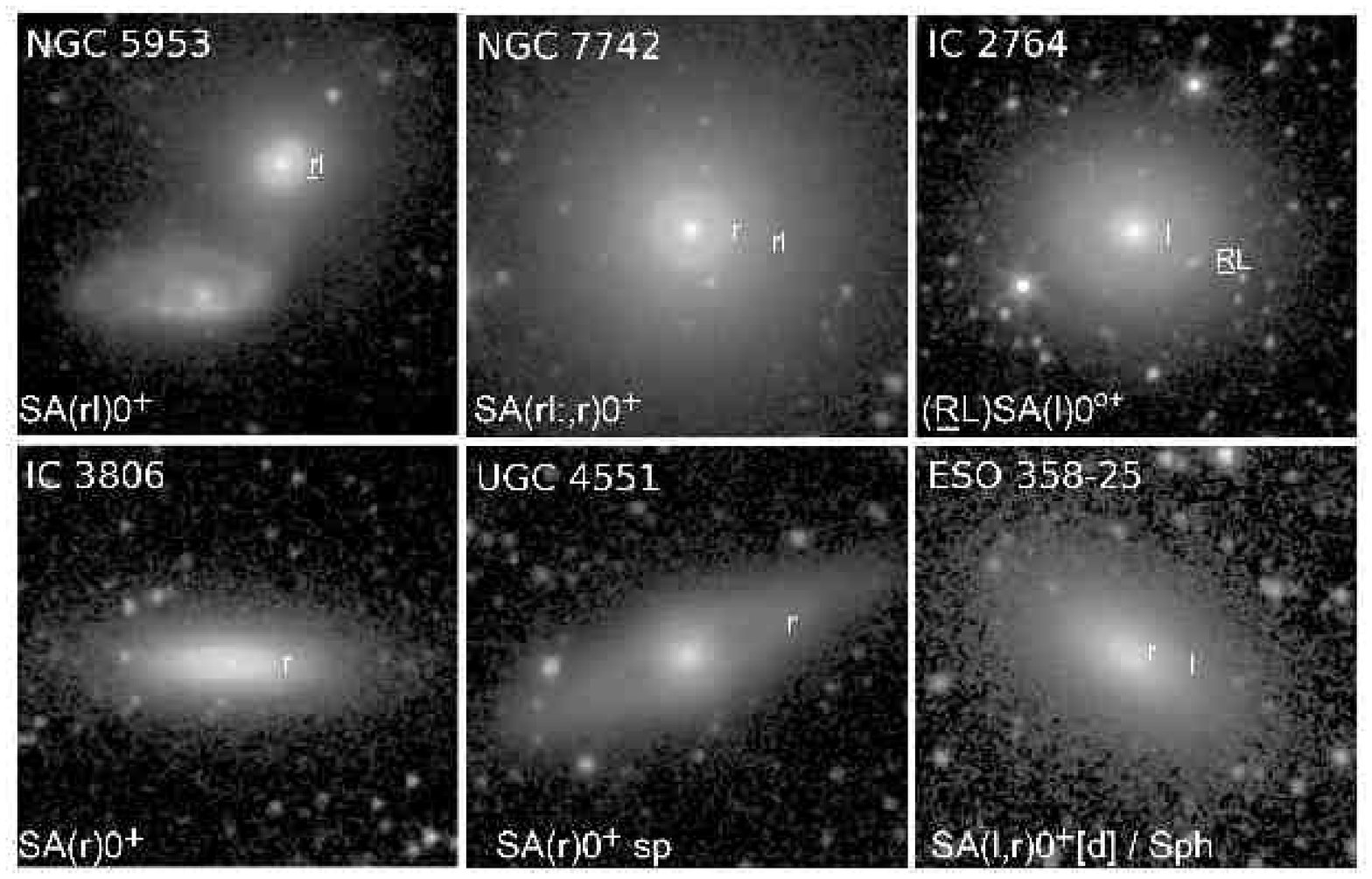}
\caption{Montages showing the diversity of the mid-IR morphologies of
S0$^+$ galaxies. In addition to the galaxy name, each frame includes
the final
mean classification from Table 6. The most important inner and outer
variety
features are labeled on the images.
}
\end{figure}

The following galaxies in Figure~\ref{S0plus} deserve special attention
because they are good examples of the characteristics they display:

\noindent{\it NGC 1553} - This galaxy, classified as type
SA(rl,nr$^{\prime}$l)0$^+$, is the well-known lens (here a ring-lens)
galaxy studied by Kormendy (1984). The lens is significantly elongated
in projection, as are the isophotes just outside this zone. In the
outer regions, however, the isophotes are nearly circular. The lens
appears to be part of an inclined, embedded disk (see also Section
4.4.2). The galaxy also has a nuclear bar that was detected in a
near-IR image (Laurikainen et al. 2011).

\noindent{\it NGC 4250} - This galaxy, classified in Table 6 as type
(R$_1$)SAB(rl,nl!)0$^+$, is noteworthy for its remarkably high surface
brightness nuclear lens. The feature has an angular diameter of
16$^{\prime\prime}$ and a linear diameter of about 2.4 kpc for a
distance of 29.4 Mpc (NED\footnote{NASA/IPAC Extragalactic Database,
http://nedwww.ipac.caltech.edu}). This is larger than the typical
nuclear ring, but is within the range of sizes of these features
(Comer\'on et al. 2010). The presence of a very faint, only slightly
elongated outer ring indicates that NGC 4250 may be inclined as little
as 30$^o$. Thus, the near circular shape of the nuclear lens is
intrinsic, while the inner ring-lens is highly elongated, both of which
are typical characteristics (Buta 1995, 2012, 2013).

\noindent{\it NGC 4344 and NGC 4451} - Two low-luminosity members of
the Virgo Cluster with small cores and bright partial inner rings of
star formation. The small cores suggest that the rings are embedded
within Kormendy spheroidals (Kormendy \& Bender 2012).  The
classification for both galaxies in Table 6 is SA(r)0$^+$[c] / Sph,
where the [c] indicates a small central concentration for such
early-type galaxies. As noted by Ilyina \& Sil'chenko (2011), such
star-forming rings (prominent in UV images) are unusual for
lenticulars, especially nonbarred lenticulars such as NGC 4344 and
4451.  Since bar dynamics are unlikely to account for these rings, an
interaction in a cluster environment (e.g., accretion of a small
gas-rich companion) may explain them. Ilyina et al. (2014) have also
argued that accretion of gas from a large, gas-rich companion can
account for similar rings in other S0 galaxies. Sph galaxies are discussed
further in Section 4.5.

\noindent{\it NGC 4594} - The 3.6$\mu$m image reveals a well-defined
outer ring and a smooth, bright inner disk. The absence of a clear
X feature suggests the galaxy is nonbarred. Classified as type
SA(s)a sp in RC3 and Sa$^+$/Sb$^-$ in the Carnegie Atlas of Galaxies
(Sandage \& Bedke 1994), the S$^4$G image reveals no clear spiral
structure in this well-studied galaxy.

\noindent{\it NGC 7742} - This galaxy, which is known to exhibit
counter-rotation (de Zeeuw et al. 2002), has a single bright inner ring
and a much fainter, larger ring that is not likely to be an outer ring.
The latter could be a weak pseudoring formed by faint spiral arms. The
two features are the reason for the Phase 1 variety ``(rr)" in Table 2.
A double inner variety was also recognized in a NIRS0S $K_s$-band image
by Laurikainen et al. (2011).

\noindent{\it ESO 358-25} - This galaxy is distinctive in the S$^4$G
sample because it appears to include a small, very bright ring in the
central area, but no clear trace of a nucleus or central concentration.
The galaxy is included in the ACS Fornax Cluster Survey (Jord\'an et al.
2007) as object FCC 152, and a high resolution color image posted on
the ACSFCS website shows a small blue ring of knots, but again little
evidence of a central nucleus. ESO 358$-$25 is an intermediate to low
luminosity member of the Fornax Cluster similar to other galaxies in
the Virgo Cluster. The classification in Table 6 is SA(l,r)0$^+$[d] / Sph,
similar to NGC 4344 and 4451, but with even less central concentration
than those Virgo Cluster galaxies.

Other galaxies in Figure~\ref{S0plus} show what we mean by lenses (l),
ring-lenses (rl), and pseudoring lenses (r$^{\prime}$l), or the outer
counterparts of these features. For example, NGC 1291 shows a very
bright inner lens at about half the diameter of a subtle outer ring
embedded in a more circular background. The feature has a shallow
brightness gradient interior to a sharp edge, but no rim enhancement.
The galaxy is virtually face-on, so the slightly elongated shape of the
(l) is intrinsic.

NGC 4457 shows a similar slightly elongated inner lens in a face-on disk. 
But most interesting are the conspicuous inner lenses in the nonbarred
galaxies NGC 5602 and IC 2764. Both of these also have a prominent
outer ring-lens (RL).

NGC 2859 is also very similar to NGC 1291 except the rim of the
prominent (l) is slightly enhanced, and the classification is (rl).
Other classified inner ring-lenses in Figure~\ref{S0plus} include NGC
2962, 3637, 3892, 4250, and 5770. The classified inner rings (r) in NGC
1326 and 4245 tend to have more of an enhancement, but even these
features are not especially strong as inner rings. NGC 1415 is
classified as having an (r$^{\prime}$l) (where ``r$^{\prime}$" is an
alternate notation for inner pseudoring) because of a subtle spiral
structure at the rim of the lens.

\subsubsection{Double Inner and Outer Varieties}

\begin{figure}
\figurenum{13}
\begin{center}
\includegraphics[height=5.0in,angle=0]{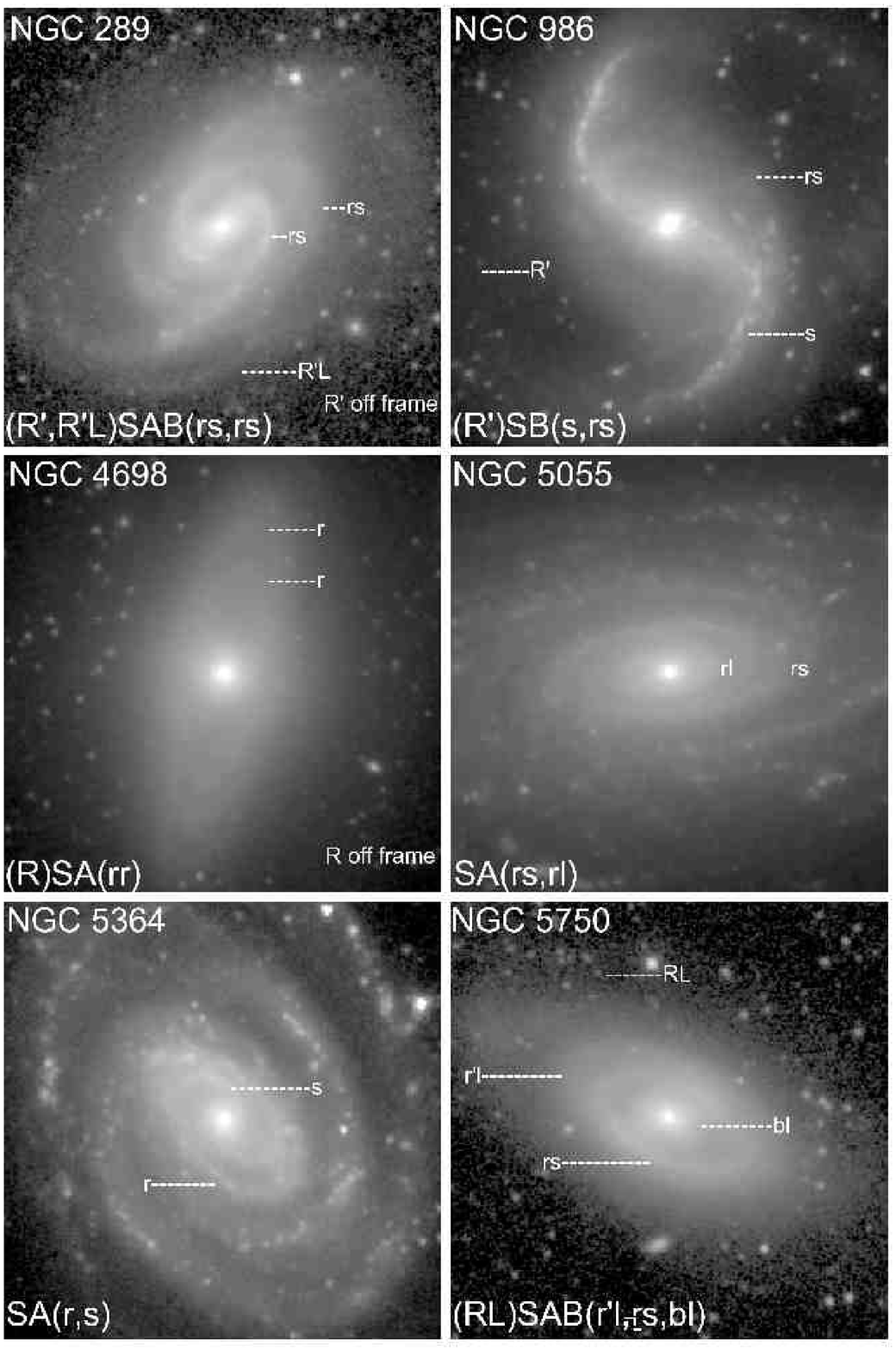}
\end{center}
\caption{Six galaxies with two recognized inner features in the CVRHS
classifications in Table 6. These features are labeled on each frame
to show what the symbols are referring to. In two cases (NGC289 and
4698), an additional outer feature is outside the field of the image.
}
\label{double-IV}
\end{figure}

\begin{figure}
\figurenum{14}
\begin{center}
\includegraphics[height=5.0in,angle=0]{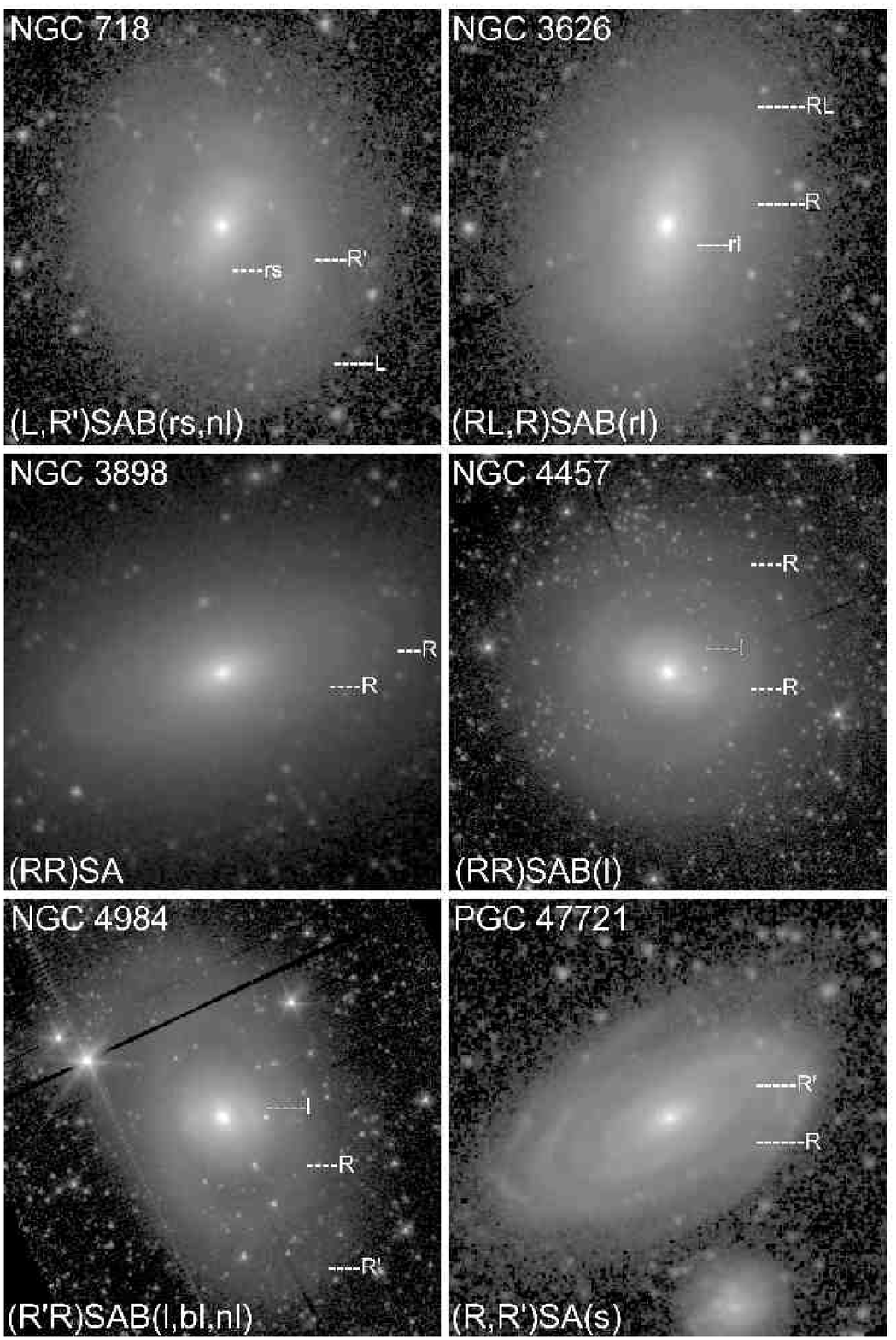}
\end{center}
\caption{Six galaxies with two recognized outer features in the CVRHS
classifications in Table 6. These features are labeled on each frame
to show what the symbols are referring to. In some cases, an inner
feature is also labeled.
}
\label{double-OV}
\end{figure}

Most of the galaxies in Figure~\ref{S0plus} are single inner and
outer variety systems, i.e., one outer feature (R, R$^{\prime}$,
RL, L) and/or one inner feature (r, rs, r$^{\prime}$l, rl, l) between
CVRHS classification bracketts. Several multiple inner and outer
variety systems are shown in Figures~\ref{double-IV} and ~\ref{double-OV}.
The recognized features are labeled with the classifications shown.

Comments on the six galaxies in Figure~\ref{double-IV} are as follows:

\noindent
NGC 289 - an extensive, smooth spiral showing multi-scale individual,
pseudoring-like patterns. The inner (rs) envelops a weak bar, while
the second (rs) is morphologically distinct and about the twice the
diameter of the first (rs). In the outer regions, an outer
pseudoring-lens, R$^{\prime}$L, is seen as a subtle enhancement.
Not seen in the illustration is a fourth, much fainter feature (an
R$^{\prime}$) approximately twice the diameter of the R$^{\prime}$L.

\noindent
NGC 986 - A barred spiral with a strong s-shaped pattern mixed with
a clear inner pseudoring pattern. Both the (s) and the (rs) are
nevertheless normal-looking features. The main arms of NGC 986 also
form an outer pseudoring.

\noindent
NGC 4698 - The Table 6 classification recognizes two inner rings and
a single, much larger outer ring that is outside the illustrated frame.
The two inner rings are closely spaced and very subtle, and could be parts
of a tightly-wound spiral pattern. The inner rings are embedded in a
much rounder component.

\noindent
NGC 5055 - The two inner features consist of a well-defined (rl)
and an (rs) about twice as large. The latter is fairly open but is
still a distinct pattern.

\noindent
NGC 5364 - This galaxy has an extremely well-defined spiral pattern
outside of a bright inner ring. What is interesting is the unrelated
spiral pattern lying {\it inside} the inner ring. The two patterns
together form the unusual variety (r,s) [as opposed to (s,rs) for
NGC 986].

\noindent
NGC 5750 - This early-type system shows a mostly closed inner ring
enveloping a foreshortened bar. Close outside this ring is a second
feature we classify as an (r$^{\prime}$l), giving the galaxy a double
variety character. A partial outer ring-lens is seen beyond these two
features.

Comments on the six galaxies in Figure~\ref{double-OV} are as follows:

\noindent
NGC 718 - This galaxy has a well-defined outer pseudoring, bar, and
inner pseudoring, but this whole pattern set is embedded in a relatively
uniform background which we recognize as an outer lens (L) that extends
to more than twice the diameter of the R$^{\prime}$.

\noindent
NGC 3626 - The (rl) and (R) are normal features, but like NGC 718, these
are embedded in a more extensive background including what we classify
as an (RL).

\noindent
NGC 3898 - an early-type nonbarred galaxy having two closely space large
rings that we interpret as a pair of outer rings, (RR). 

\noindent
NGC 4457 - The deep 3.6$\mu$m image reveals not only the previously
known bright outer ring in this well-known Virgo Cluster galaxy
(Sandage 1961), but also what appears to be a second outer ring at
roughly twice the radius of the first feature. Both features are
very nearly circular. The second ring is also detectable in an SDSS
color image. An unsharp-masked near-IR image previously revealed
the weak bar in the galaxy (Laurikainen et al. 2011).

\noindent
NGC 4984 - The inner regions consist of a normal lens and faint bar with
ansae. Surrounding this zone is an outer ring, but most unusual is the
second outer feature, an R$^{\prime}$, at approximately twice the
diameter of the first outer ring. This accounts for the outer variety
(R$^{\prime}$R).

\noindent
PGC 47721 - An M81-like galaxy which has what appears to be a mostly
closed ring within its main outer spiral pattern. Within this feature
is a very large pseudoring. The two features account for the
classification of (R,R$^{\prime}$), although in the absence of a
bar, these could also be interpreted as very large inner rings.

\subsection{The Mid-IR Morphology of Bars}

Bars are a logical focus for mid-IR morphological examination. It
is well-known that some bars which appear weak or even absent in
blue light imaging can be more prominent in IR bands (e.g., Eskridge
et al.  2000). Heavy internal extinction (due, for example, to
strong leading dust lanes or other factors) can account for some
of these, while stellar populations can account for others in the
sense that IR bands are more sensitive to the older stellar populations
seen in many bars than is the $B$-band. In paper I, we argued that
the ``stronger bar effect" is apparent in S$^4$G images, but that
the rankings of bars (by actual strength) are not greatly changed.
What we mean by this is that, while weaker bars can appear more
conspicuous in the IR as compared to blue light, such that a $B$-band
SAB-type bar can be classified as IR type SB, a $B$-band SB type
bar has no stronger category to be placed in, yet it also looks
stronger in the IR. The rankings of bars (strong versus weak) thus
is not changed greatly from blue light to the mid-IR.  This does
not undermine the case for IR imaging: it is still best to judge
the importance of any bar in an IR waveband.

Bars can also be reliably recognized in both face-on and edge-on
galaxies, as well as across the entire CVRHS sequence from stage S0$^-$
to Im. No other aspect of galaxy morphology has this characteristic,
which is considerably enhanced in the mid-IR. In this Section, we look
at the mid-IR morphology of bars with two goals: (1) to highlight the
diversity of bar morphologies, especially among early-type galaxies,
and (2) to clarify the meaning of family classifications between early-
and late-type galaxies that impacts how we interpret the barred family
fraction.

\subsubsection{Ansae Bars and Barlenses}

One of the interesting aspects of the galaxies in Figure~\ref{S0plus}
is how weak-looking most of the apparent bars are, even if previously
classified as SB. For example, NGC 1291 is classified as (R)SB(s)0/a in
RC3 and as SBa by Sandage \& Tammann (1981), and yet, viewed against
the bright inner lens in a logarithmic, background-subtracted image,
the bar classification that seems appropriate for NGC 1291 is SAB at
best (see also the dVA). The same is true for NGC 2859 and others shown
in the Figure. Only NGC 3637 and 4245 have bars classified as SB in
Table 6.

\begin{figure}
\figurenum{15}
\plotone{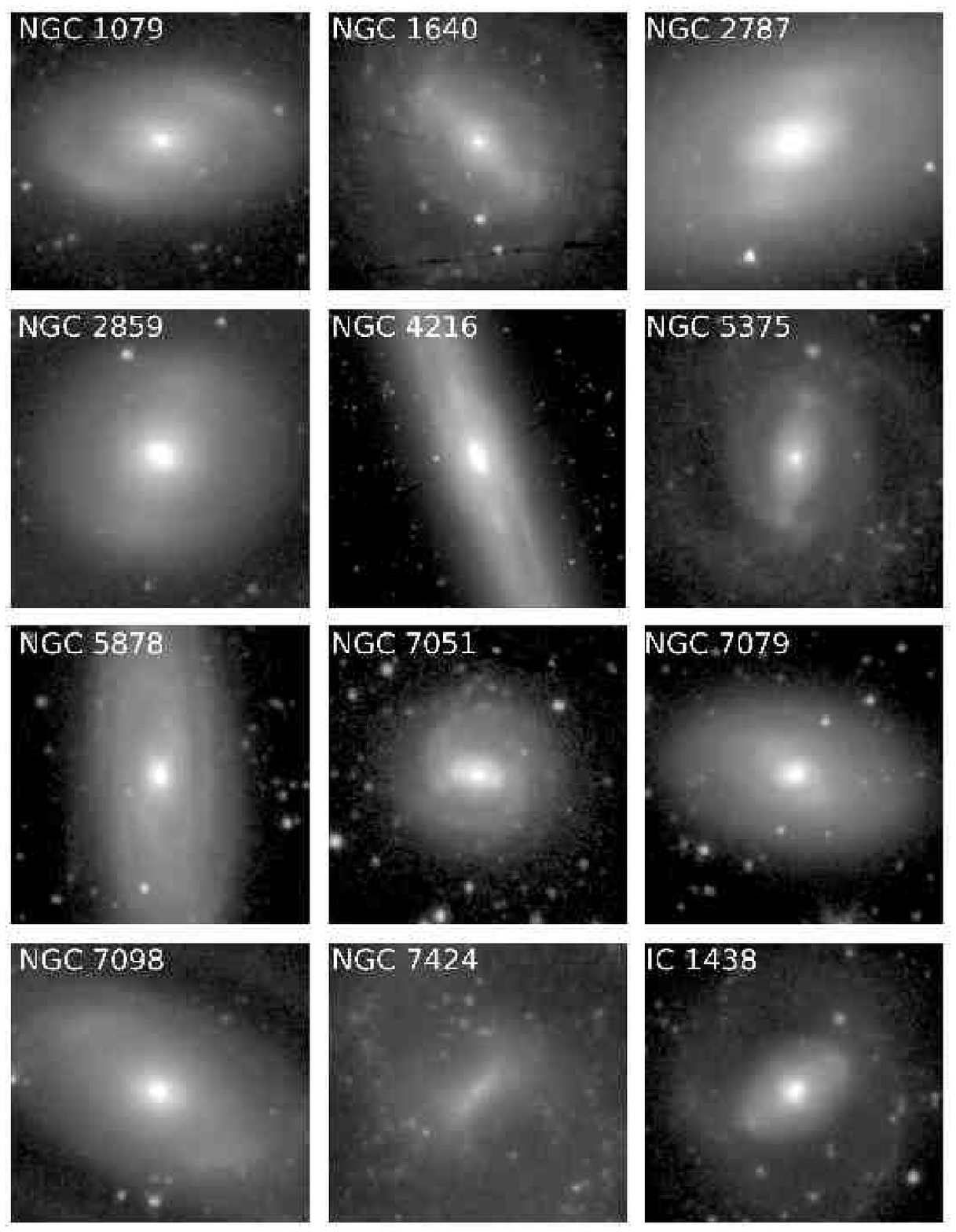}
\caption{Montage showing the mid-IR morphology of ansae bars in 12
S$^4$G galaxies.
}
\label{ansae-bars}
\end{figure}

\begin{figure}
\figurenum{16}
\plotone{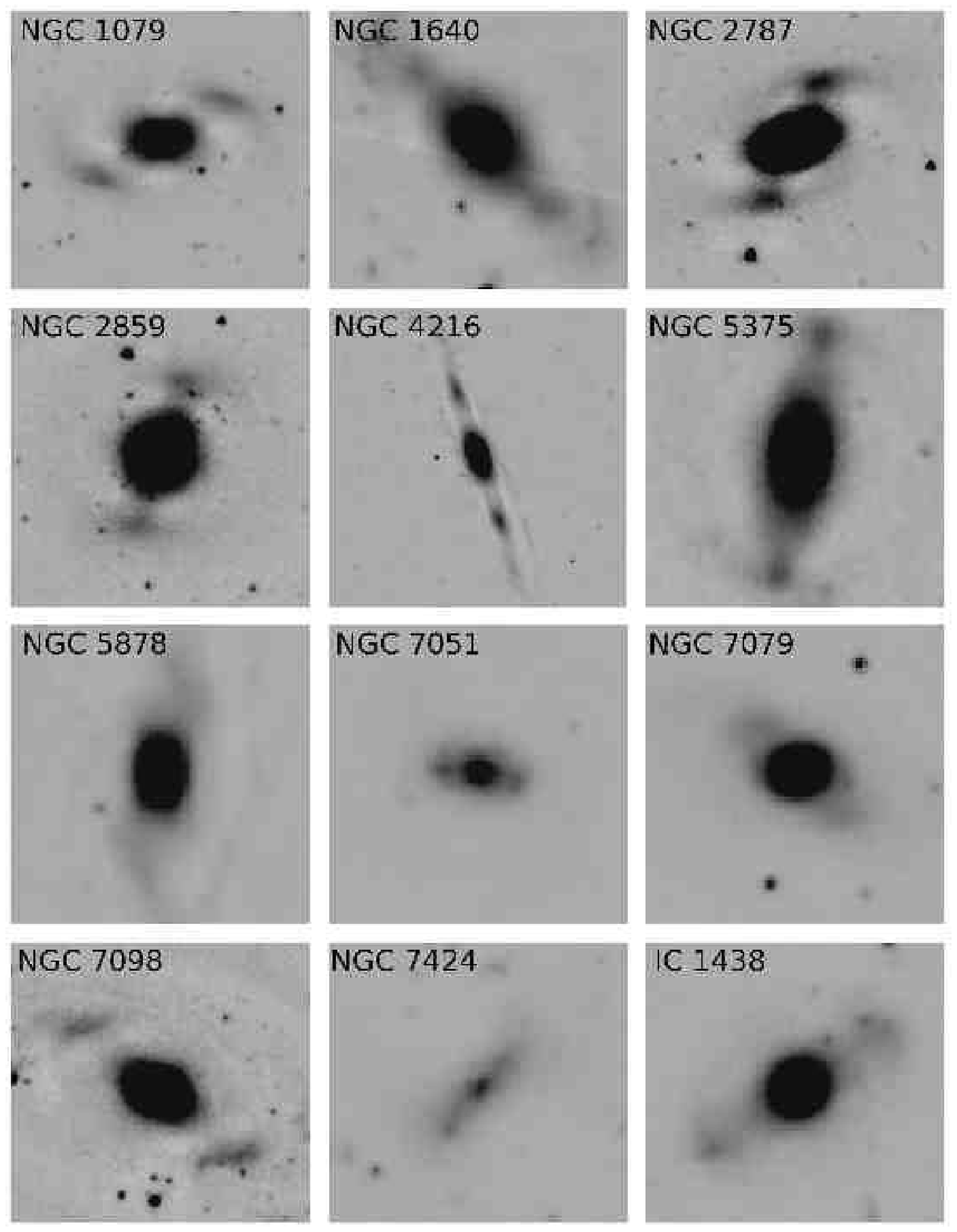}
\caption{The same galaxies as in Figure~\ref{ansae-bars}, displayed
using unsharp-masking (and on different scales in some cases) to make
the ansae more easily visible. The 3.6$\mu$m images are in intensity
units.}
\label{unsharp-masked}
\end{figure}

Many of the apparent bars in the galaxies in Figure~\ref{S0plus}
are of the ansae (``handles") type, where the ends of the bar are
defined by subtle brightness enhancements.  Martinez-Valpuesta et
al. (2007) made a statistical study of these features, and concluded
that ansae are mainly a phenomenon of early-type galaxies and are
very rare at stages later than Sb.  Figure~\ref{ansae-bars} shows
a selection of 12 of the best-defined ansae bars in the S$^4$G
sample. Because these features are subtle, Figure~\ref{unsharp-masked}
shows unsharp-masked images of the same 12 galaxies. These show how
ansae come in different morphologies, ranging from curved or nearly
linear arcs to rounder spots. Although no detailed quantitative
measurements have yet been made, SDSS color images and dVA color index
maps have indicated that ansae may be purely stellar in nature
or include star formation. In Tables 2 and 6, ansae are recognized
using the symbols SB$_a$ and SAB$_a$.

Although the detection of ansae does not necessarily require observing
galaxies in the mid-IR, mid-IR imaging can still shed some light
on the properties of the features, such as their vertical structure.
NGC 4216 is the most highly inclined example shown in
Figure~\ref{ansae-bars}, and in the optical shows strong planar
dust. In the mid-IR, the galaxy shows an inner boxy structure and
two intense brightness enhancements that, in Figure~\ref{unsharp-masked},
appear at the ends of the major axis of a thin inner ring. These
enhancements are likely to be ansae in the inner ring. Although we
cannot rule out that these features are merely due to line-of-sight
effects through the major axis points of an inner ring, other nearly
edge-on galaxies with highly-inclined rings (e.g., NGC 4594 in
Figure~\ref{S0plus}) do not show conspicuous ansae.

The boxy inner zone of NGC 4216 shows a very faint X-pattern in the
unsharp-masked image, indicating that the galaxy does indeed have
a bar with vertical resonant structure. However, the ansae are
clearly much flatter features. Thus, mid-IR imaging shows that a
bar in a $B$-band intermediate-type barred spiral consists of a 3D
inner section and much flatter ends, consistent with the general
structure of bars noted by Athanassoula (2005).

In Figure~\ref{ansae-bars}, NGC 7424, a face-on SBcd galaxy, is
much later than any ansae barred galaxies identified by Martinez-Valpuesta
et al. (2007). Because bars in extreme late-type spirals can be
linear chains of star forming regions, it is not clear that the
apparent ansae in NGC 7424 are in any way dynamically related to
those seen in the other galaxies in Figure~\ref{ansae-bars}. Late-type
galaxy bars as compared to early-type ones are more carefully
examined in Section 4.3.3.

Figure~\ref{ansae-bars} shows an additional feature of barred galaxies
only recently recognized. This is the ``barlens," symbolized by (bl).
A barlens refers to the inner part of a typical early-type galaxy bar.
As a class of structures, barlenses were first recognized by
Laurikainen et al. (2010), who used high-quality near-IR images
and multi-component decompositions to establish the distinct nature of
the features. Barlenses can be mistaken for a bulge because they
generally have less elongated isophotes than the bar ends. They appear
to be part of the bar.

Laurikainen et al. (2013) proposed that a barlens represents an
evolution of the bar, and that inner lenses in nonbarred galaxies could
be the former inner part of an evolved, disintegrated bar (see their
Figure 11). The curved ansae in NGC 1079 appear to be the bar ends
dispersing into the inner ring-lens area. Athanassoula et al. (2014)
and Laurikainen et al. (2014) use both numerical simulations and
observations to show that barlenses are indeed likely to be the more
face-on view of the 3D inner part of a bar, that when seen edge-on
shows the familiar X or box/peanut shape. It is likely that NGC 4216,
for example, in Figure~\ref{ansae-bars} would show a barlens in the
face-on view, with the barlens corresponding to the broad inner zone.

Five galaxies (NGC 1079, 2787, 2859, 5375, 7079) with a barlens are
included in Figure~\ref{ansae-bars}, the best example being seen in NGC
2787. More than 70 examples are included in the whole catalogue. A
barlens and a regular inner lens differ in scale: in a barred galaxy
with an (l), the bar typically fills the (l) in one dimension (Kormendy
1979), while the (bl) is smaller than the bar, as measured from the
NIRS0S atlas (Laurikainen et al. 2011) and illustrated in Athanassoula
et al.  (2014). This is in agreement with the theoretical predictions
by Athanassoula (2005) about the relative extents of the thick and the
thin part of the bar. A (bl) will also generally be larger than and
distinct from a nuclear lens (nl).\footnote{This is well-illustrated by
Buta et al. (2001), who used Fourier analysis to remove the bar of NGC
1433 from a deprojected near-IR $H$-band image, leaving behind ``a
large round area surrounding the secondary lens and bar," a preliminary
isolation of a barlens (see their Figure 5).} Laurikainen et al. (2013)
showed that ansae occur even in 52\% of galaxies having barlenses,
compared to 24\% when no barlens is present. However, multiple lenses
are rare in such galaxies.

In spite of what ansae may imply about bar structure and evolution, the
features are still poorly understood. No simulations have yet accounted
for the diversity of the morphologies of ansae (round spots, arcs,
linear shapes), or for their spread in color (young versus old stellar
population ansae; e.g., Martinez-Valpuesta et al. 2007; Buta 2013).

\subsubsection{X-structures}

\begin{figure}
\figurenum{17}
\plotone{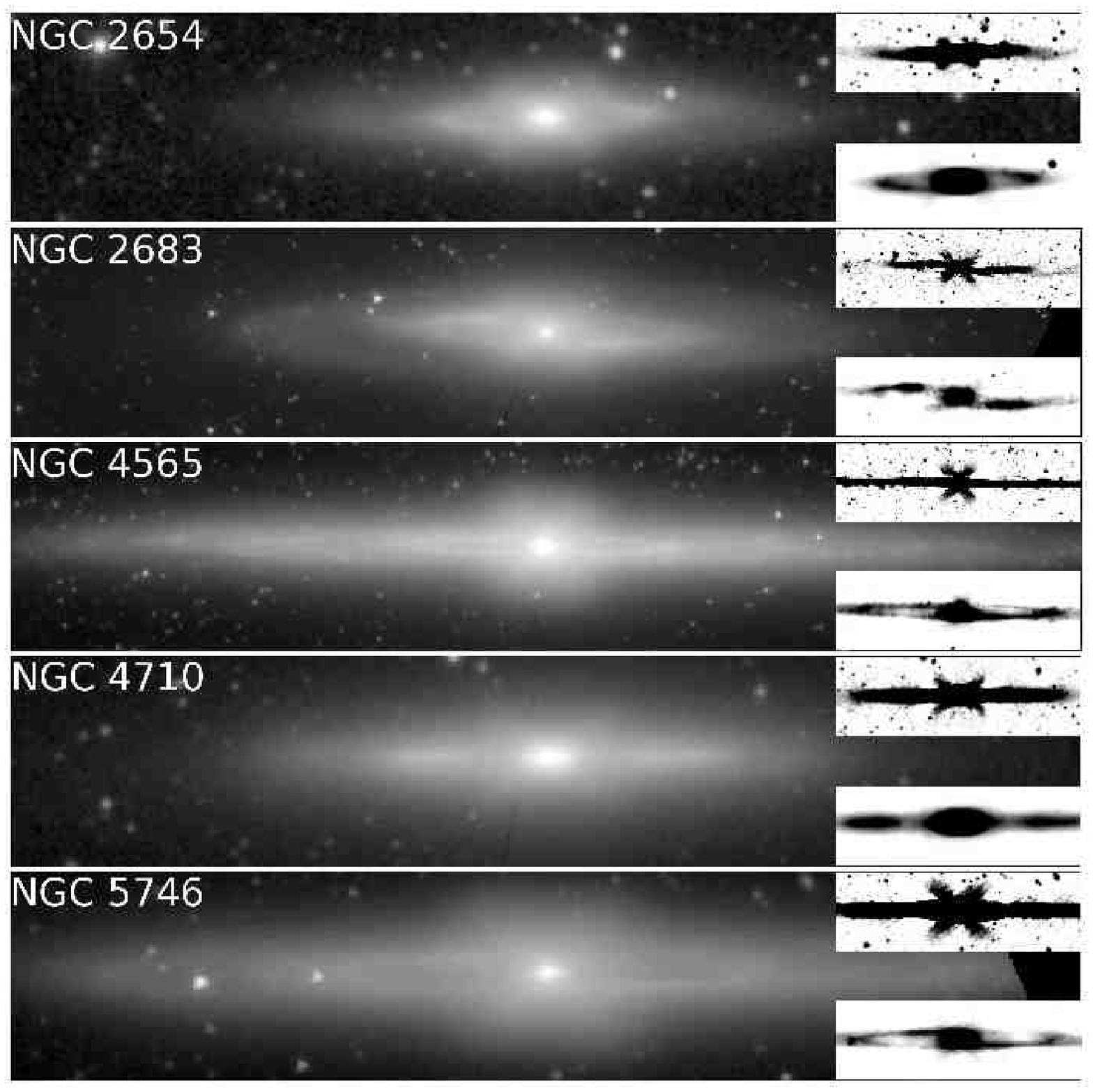}
\caption{Montage showing the mid-IR morphology of five S$^4$G galaxies
showing strong inner X-patterns. The units of these 3.6$\mu$m images
are mag
arcsec$^{-2}$ over the same ranges for the galaxies given in the
captions to Figure 1. The upper right inserts show unsharped-masked
versions
of the intensity images after subtraction of a heavily
median smoothed image. These make the X-patterns more visible. The
lower right inserts show the same unsharp-masked images of the inner
disk regions to highlight inner rings and bar ansae. The scales of
these
inserts are twice that of the upper right inserts, except for NGC 5746
where the scale of the lower right insert is 1.25 times that of the
upper right insert.}
\label{X-gals-with-rings}
\end{figure}

X-structures are strongly evident in some S$^4$G galaxies, five of
which are shown in Figure~\ref{X-gals-with-rings}. Like the ansae in
Figure~\ref{unsharp-masked}, these unusual patterns can be made more
obvious using unsharp-masking (upper right inserts in
Figure~\ref{X-gals-with-rings}). Detailed observations and numerical
simulations (e.g., Bureau \& Freeman 1999; Athanassoula \& Bureau 1999;
Bureau et al. 2004, 2006) have definitively shown that X-structures are
related to the edge-on view of a bar. The X is believed to show the
vertical resonant structure of the bar. Often, the X pattern is a
subtle aspect of what is commonly referred to as a ``box/peanut bulge"
(Luetticke, Detmar, \& Pohlen 2000a,b). Here we follow the de
Vaucouleurs nomenclature and use the generic term X-structure for any
box/peanut/X (B/P/X) structure or bulge. In Table 6, these are
recognized using the symbols SB$_x$ and SAB$_x$ (Table 1). For edge-on
galaxies, the SB in SB$_x$ is inferred if the X is especially strong,
but if the X is fairly weak, then SAB$_x$ is used instead.

For all of the galaxies in Figure~\ref{X-gals-with-rings}, internal
dust complicates the view of the X in blue light but did not prevent
the box/peanut shape from being detected. All were recognized as peanut
or boxy bulges by Luetticke, Detmar, \& Pohlen (2000a) based on
inspection of Digitized Sky Survey images. Most interesting is how in
four of these galaxies, the X-pattern is situated within a conspicuous
inner ring or pseudoring whose projected shape indicates that the
galaxies are not exactly edge-on. These rings were not necessarily
obvious in blue light. The lower right inserts in
Figure~\ref{X-gals-with-rings} show unsharp-masked images of the rings
that reveal the presence of ansae in at least two cases (NGC 2654 and
2683). NGC 4710 appears exactly edge-on and only ansae are visible on
each side of the center. Because inner rings and pseudorings commonly
envelop bars in more face-on galaxies, the detection of strong X
patterns situated within bright inner rings helps to clinch the
argument that X-patterns are related to bars. The detected inner
ring in NGC 4565 is also described by Kormendy (2012) and de Looze et al.
(2012).

The five galaxies in Figure~\ref{X-gals-with-rings} also show that the
width of the X pattern relative to its vertical height can vary.  In
NGC 4565, for example, the nearly round pseudobulge shows a tighter X
feature than is seen in NGC 2654, 2683, and 5746, suggesting that our
perspective on the bar of NGC 4565 must be more end-on than for these
galaxies.

Table 6 includes more than 60 visually recognized box/peanut/X-galaxies
in the S$^4$G sample.  While X-patterns are often most easily
recognized in nearly edge-on galaxies, they are also seen in much less
inclined galaxies (e.g., the dVA; Erwin \& Debattista 2013).
An example in the S$^4$G sample is NGC 5377 (paper I). In such cases,
the classifications SB$_x$ and SAB$_x$ are not inferences of a bar
as they would be for edge-on galaxies.

Note that Erwin \& Debattista (2013) did not show or statistically
examine X-shaped bars, only box/peanut features, or bars having
spurs. They examined a few moderately inclined galaxies with X-shapes
from the NIRS0S atlas assuming that physically they represent the same
phenomenon. The inclination distribution of the galaxies with X-shaped
bars for the combined S4G+NIRS0S sample is given in Laurikainen et al.
(2014; see their Figure 2). This study showed that X-shapes appear only
in bright galaxies, and not in the latest Hubble types. The
identifications of the X-shapes in Laurikainen et al. (2014) was based
on unsharp masks made for the complete S4G and NIRS0S samples.

\subsubsection{Early- versus Late-type Bars}

Elmegreen \& Elmegreen (1985) examined the photometric properties of
bars over a range of types, and found a dichotomy: early-type galaxy bars
are long, strong, and have a relatively flat luminosity profile, while
late-type bars are short, relatively weak, and have an exponential
luminosity profile. This dichotomy is evident also in S$^4$G images.
By ``strong bar," we will generally mean a bar having a maximum
relative bar torque parameter $Q_b \geq$ 0.2 (Buta et al. 2006).

CVRHS family classifications provide visual information on apparent bar
strength in galaxies, but little insight on the actual diversity of bar
morphologies. The recognition of ansae bars, X-structures, and
barlenses [through SB$_a$, SB$_x$, etc., and (bl), respectively]
provides extra information about a bar in a given galaxy, but fails to
account for the significant differences between the bars seen in early
and late-type galaxies. In Section 3.5 it was shown that the distribution
of family classifications in the S$^4$G catalogue depends on mid-IR
stage, with SA classifications being most abundant in the type range
S0/a to Sc, and SB classifications being most abundant in the Scd-Sm
type range. The barred family classification fractions were found to
be 55\% in the range S0/a to Sc and 81\% in the range Scd to Sm.

\begin{figure}
\figurenum{18}
%\plotone{etbar-types-labels.ps}
\begin{center}
\includegraphics[height=5.0in,angle=0]{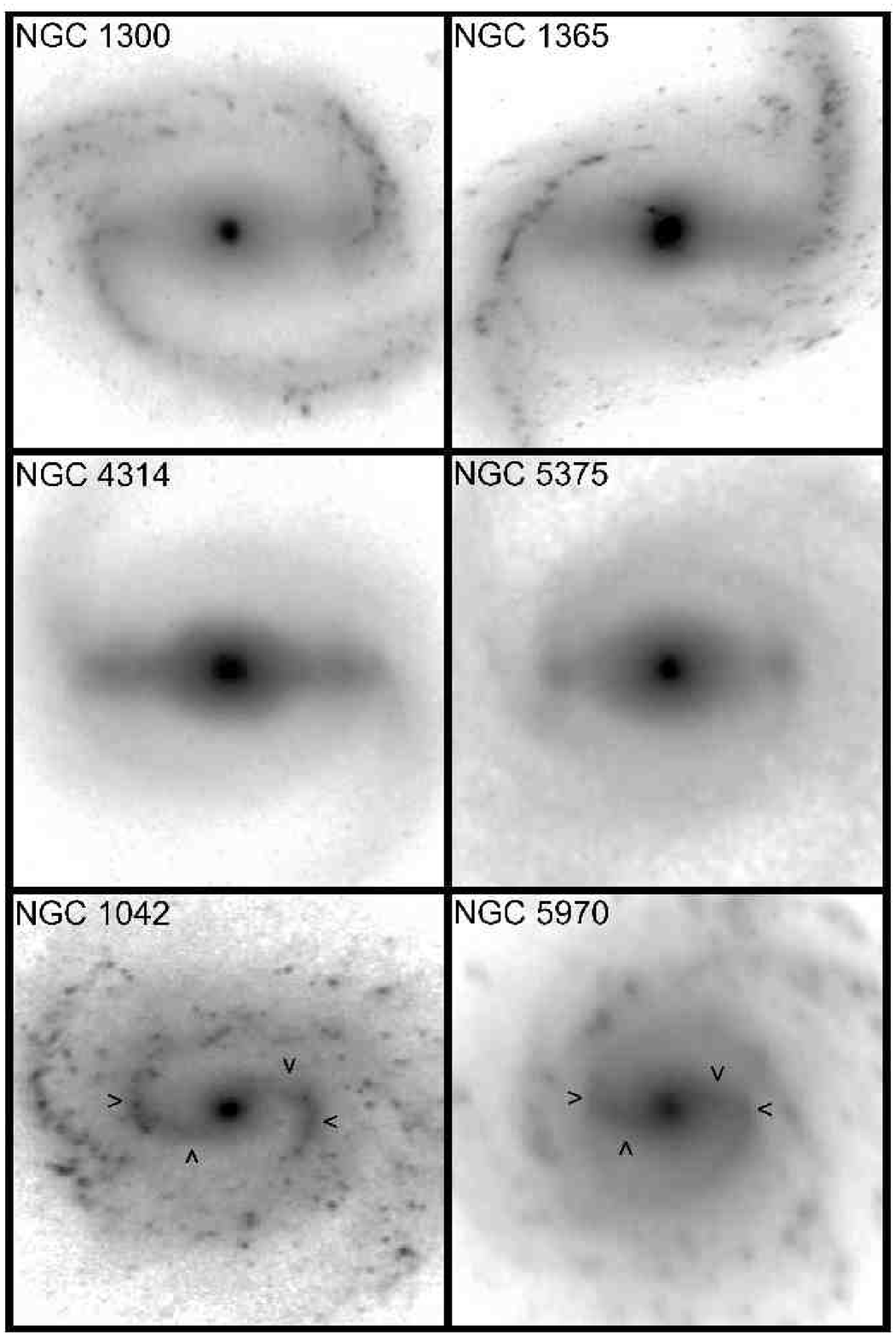}
\end{center}
\caption{Deprojected 3.6$\mu$m images of 6 galaxies in the mid-IR type
range S0/a to Sc, rotated such that the bar is oriented horizontally.
The bars in the lower two examples (NGC 1042 and 5970) show a spiral
character that is outlined with the circumflexes. Units of the images
are mag arcsec$^{-2}$.}
\label{et-bars}
\end{figure}

\begin{figure}
\figurenum{19}
%\plotone{ltbar-types-labels.ps}
\begin{center}
\includegraphics[height=5.0in,angle=0]{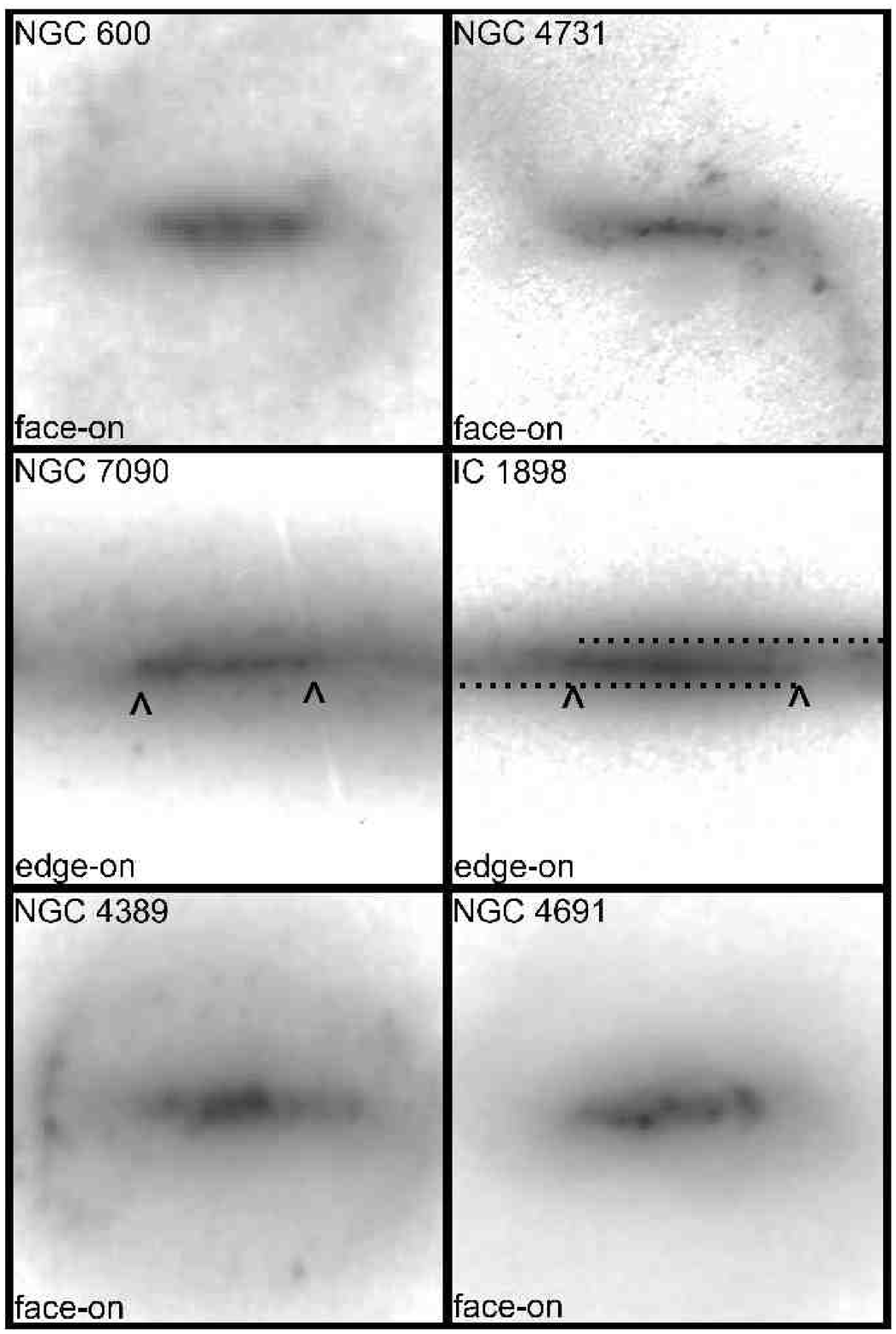}
\end{center}
\caption{({\it top row}): Deprojected 3.6$\mu$m images of 2 galaxies of
mid-IR types Sc$\underline{\rm d}$ (NGC 600) and Sd (NGC 4731), rotated
such that the bar is oriented horizontally.  ({\it middle row}): Two
edge-on galaxies showing a short, linear feature (indicated by the
circumflexes) that is likely to be an edge-on view of the same kind of
bar seen in NGC 600 and 4731.  The dotted lines in the IC 1898 image
highlight two spiral arms that break from the ends of the linear
feature.  The disk major axes for both galaxies are rotated to be
horizontal. ({\it bottom row}): Deprojected images of two lower
inclination early-type galaxies showing the same kind of bar seen in
NGC 600 and 4731, rotated such that the bar is oriented horizontally.
Units are mag arcsec$^{-2}$.
}
\label{lt-bars}
\end{figure}

At stage Scd, bars begin to show a distinct knotty morphology.
Ansae, barlenses, and X-patterns seem to be no longer relevant. To
see this, Figure~\ref{et-bars} shows the bar regions of six relatively
low-inclination S$^4$G galaxies having mid-IR stages in the range
Sa to Sc, while Figure~\ref{lt-bars} shows the bar regions of six
galaxies in the type range Sc$\underline{\rm d}$-S$\underline{\rm
d}$m. In each case, the IRAC 3.6$\mu$m image has been deprojected
and rotated to make the bar horizontal. Foreground and background
objects have also been removed. 

In the Figure~\ref{et-bars} montage, NGC 4314 has a prominent barlens,
and indeed the feature was used as a prototypical example of a (bl) in
the NIRS0S atlas (Laurikainen et al. 2011). Laurikainen et al. (2014)
also show further evidence in favor of the barlens interpretation for
this galaxy (see their Fig. 1). NGC 5375 has two strong circular ansae
in addition to a barlens (see also Figures~\ref{ansae-bars} and
~\ref{unsharp-masked}), NGC 1300 and 1365 have very strong spirals
breaking from the ends of their bars, while NGC 1042 and 5970 each have
a bar with a clear spiral character (highlighted by the circumflexes)
that, at least in the case of NGC 1042, is not likely to be due to a
projection effect between the 3D inner part of the bar and the flatter
bar ends (e.g., Erwin \& Debattista 2013). The bars in all of these
galaxies are also relatively smooth and, except for the sharp,
right-angle-turning ansae of the bar in NGC 1042 (indicated by the $><$
symbols), are made largely of old stars.

The bars seen in Figure~\ref{et-bars} can be contrasted with those
shown in Figure~\ref{lt-bars}.  The top panels of Figure~\ref{lt-bars}
show the deprojected 3.6$\mu$m images of NGC 600 and NGC 4731, types
SB(r$\underline{\rm s}$)c$\underline{\rm d}$ and SB(s)d, respectively.
In both cases, the bar is a line of star-forming knots (see also Martin
1996). That such bars are likely to be highly flattened and confined to
the thin disk is suggested by NGC 7090 and IC 1898
(Figure~\ref{lt-bars}, middle), two nearly edge-on galaxies classified
as type SB(s)$\underline{\rm d}$m sp, where a distinct linear feature
(whose extent is indicated by circumflexes) is seen in the inner
regions. In both cases, subtle spiral structure (highlighted by
horizontal dotted lines for IC 1898) breaks from near the ends of the
linear feature, suggesting that the feature is a bar and not merely an
inner disk.

The fact that these late-type galaxy bars show no inner 3D component
and are defined to a signifcant extent by star formation indicates that
the dichotomy in $f(F\geq 0.5)$ seen in Figure~\ref{allfams} reflects
something more fundamental than a mere change in the bar fraction with
galaxy types. The bars in Figure~\ref{et-bars} are very different
structures from those shown in Figure~\ref{lt-bars}. While numerical
simulations of the bar instability have had considerable success in
explaining the structure of the types of bars seen in galaxies like NGC
1300 and 4314 (e.g., Athanassoula et al. 2014), the same is not true
for the bars of NGC 600 and 4731 which have no 3D component, no X-pattern,
and in general no ansae or obvious spiral character.

The lower two frames in Figure~\ref{lt-bars} show two Virgo Cluster
galaxies whose stage is ``early" but whose bar is clearly ``late." NGC
4389 and NGC 4691 are classified in Table 6 as SB(rs)a[d] and
(R$^{\prime}$L,R$^{\prime}$)SB(s)0/a[dm], respectively. The ``B" in
each case is a line of star-forming knots just as in NGC 600 and 4731.
The disks, in contrast, are mostly devoid of star formation as would be
found in S0/a or Sa galaxies.

The types of bars shown in Figure~\ref{et-bars} are rarely, if ever,
seen in galaxies of types Scd and later. Because these types of
galaxies tend to have a lower average luminosity than S0/a to Sc
types (e.g., dVA, Figure 1.16), we may deduce that the fraction of
Figure~\ref{et-bars} bars drops precipitously for lower mass disk
galaxies. We have excluded from the discussion the bars of Magellanic
irregulars that have classifications such as IABm, IB(s)m,
etc. The reason is that the bars of such galaxies are often
ill-defined, and the large spread in luminosity of Im galaxies means
that many members of the class do not have the sophistication of
structure needed to support the existence of a bar.

\subsubsection{Nuclear Rings, Lenses, and Secondary Bars}

Nuclear rings, lenses, and secondary bars are described in this
Section as features of the morphology of primary bars. All are
dynamically important structures found in the centers of
early-to-intermediate type barred galaxies (e.g., Buta \& Crocker
1993). Nuclear rings are especially heterogeneous in their
morphological, metric, and star-forming characteristics.  In some
galaxies, a nuclear ring is the site of a starburst, and may be the
only place in a galaxy where active star formation is occurring.
In optical imaging, a nuclear lens may be interpreted as a quiescent
nuclear ring in a non-star-forming phase, although not all nuclear
lenses are necessarily dead nuclear rings.

Comer\'on et al. (2010) have provided the most extensive recent
study of the morphology and metric characteristics of nuclear rings,
and include not only star-forming and stellar features in the class,
but also dust nuclear rings. The main requirement to be defined as
a nuclear ring is proximity of the feature to the nucleus of a
galaxy. Also, Comer\'on et al. (2010) required that the width of
the ring is no more than half of its radius. As for inner and outer
rings, there can be ambiguity in interpreting nuclear rings, even
in some barred galaxies. These ambiguities are discussed in detail
by Comer\'on et al. (2010).

\begin{figure}
\figurenum{20}
%\plotone{nr-nl-new2.ps}
\begin{center}
\includegraphics[height=5.0in,angle=0]{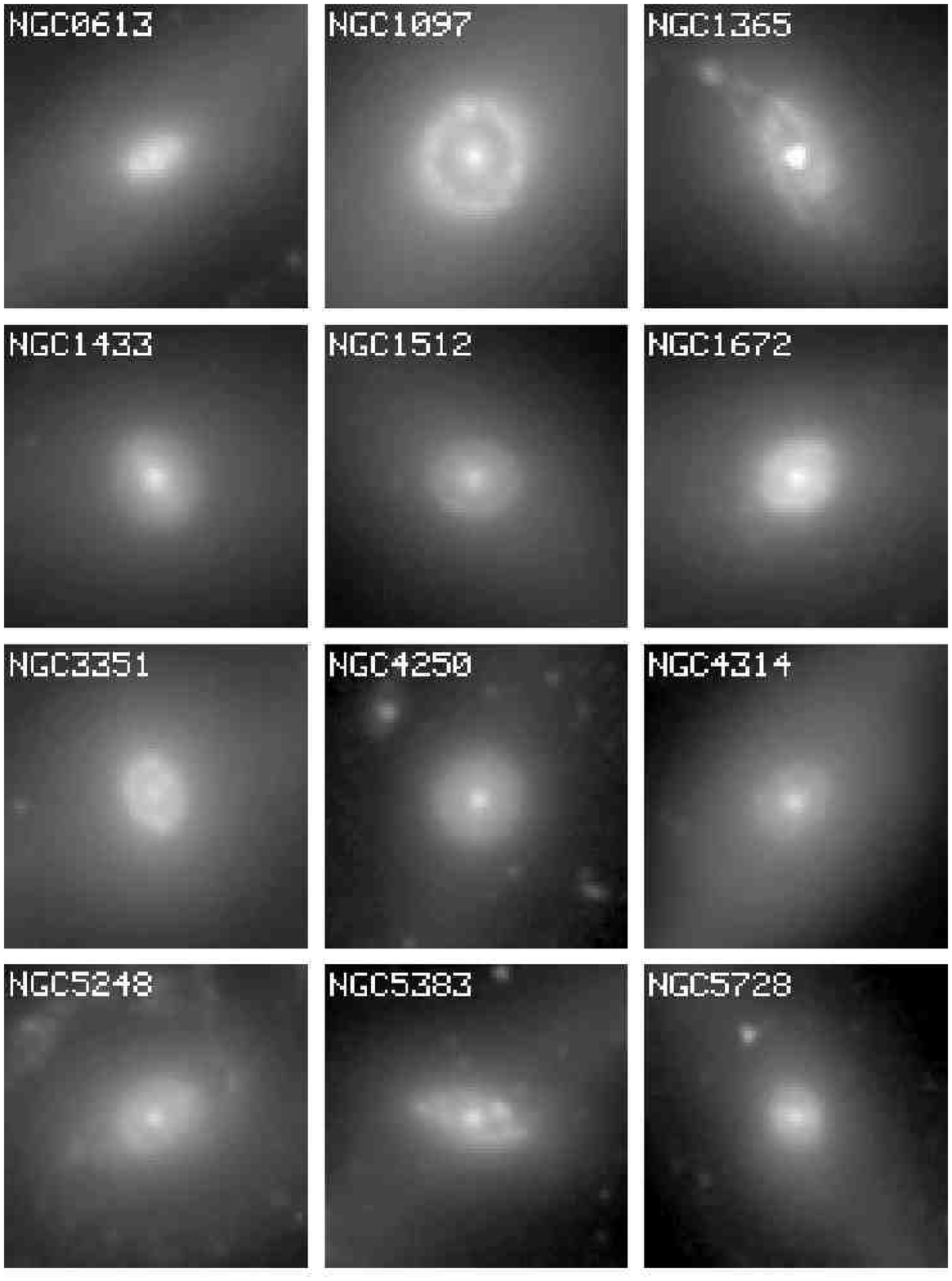}
\end{center}
\caption{Montage showing the mid-IR morphology of the nuclear rings and
lenses in 12 S$^4$G galaxies.
}
\label{nrings}
\end{figure}

Because the centers of intermediate-stage barred spirals can be
affected by complex dust patterns, mid-IR imaging is especially useful
for seeing the actual structure of nuclear rings, lenses, and secondary
bars. Figure ~\ref{nrings} shows the mid-IR morphology of the nuclear
rings or lenses of 12 S$^4$G galaxies. In spite of the emphasis on an
older stellar population, the mid-IR morphology of nuclear rings can
still display a knotty structure due to intense star formation. 

In the CVRHS, nuclear rings are recognized with the symbol (nr). In
some cases, an optical nuclear ring appears only as a subtle
enhancement at the edge of a lens; these are recognized using the
symbol (nrl). Nuclear lenses are recognized with the symbol (nl).
Table 6 includes 38 galaxies classified as having an (nr), an (nrl),
or a nuclear pseudoring nr$^{\prime}$, and 33 galaxies as having an
(nl). The metric characteristics of these features are provided in
Comer\'on et al. (2010, 2014). Note that poor resolution can, in some
cases, make a nuclear ring look more like a nuclear lens.

Most of the nuclear ring/lens features shown in Figure~\ref{nrings} are
detectable as rings in blue light. Exceptions are the features seen in
NGC 1365 and 5383, which appear as nuclear spirals carved by dust in
blue light and more like rings in the mid-IR. Even the ring in NGC 1097
is much more spiral-like in blue light (dVA) compared to its very
regular, almost circular shape in the mid-IR.

Secondary bars, also commonly known as nuclear bars, are well-known
features of early-type barred galaxies (e.g., Laine et al. 2002; Erwin 2004).
These small features often have a linear scale similar to
nuclear rings. For example, NGC 5728 in Figure~\ref{nrings} shows a
nuclear bar (oriented roughly horizontally) within its small nuclear
ring (seen also in a NIRS0S $K_s$-image; Laurikainen et al.  2011). In
the CVRHS, nuclear bars are recognized with the notation (nb). NGC 1291
(Figure~\ref{S0plus}) also shows a nuclear bar, but no nuclear ring
surrounds this bar. The nuclear bar of NGC 1291 is actually a more
prominent-looking feature than the galaxy's primary bar. Table 6
recognizes a definite or possible nuclear bar in 15 galaxies. There are
undoubtedly many more present that would be detectable with better
resolution.

\subsubsection{x$_1$ rings}

S$^4$G images provide an extinction-free view of a rare type of
ring called an ``x$_1$ ring." These are highly-elongated rings which
lie within and along a bar. The term x$_1$ refers to the main family
of central orbits that support a bar (Contopoulos \& Grosbol 1989).
Regan \& Teuben (2004) used numerical orbit calculations to show
that inner rings in SB galaxies could be tied to looping 4:1 resonant
orbits, and that if a bar is especially strong, shocks in the loops
can collect gas into a highly-elongated x$_1$ orbit within the bar.
Star formation in this orbit would then form an ``x$_1$ ring."

\begin{figure}
\figurenum{21}
\plotone{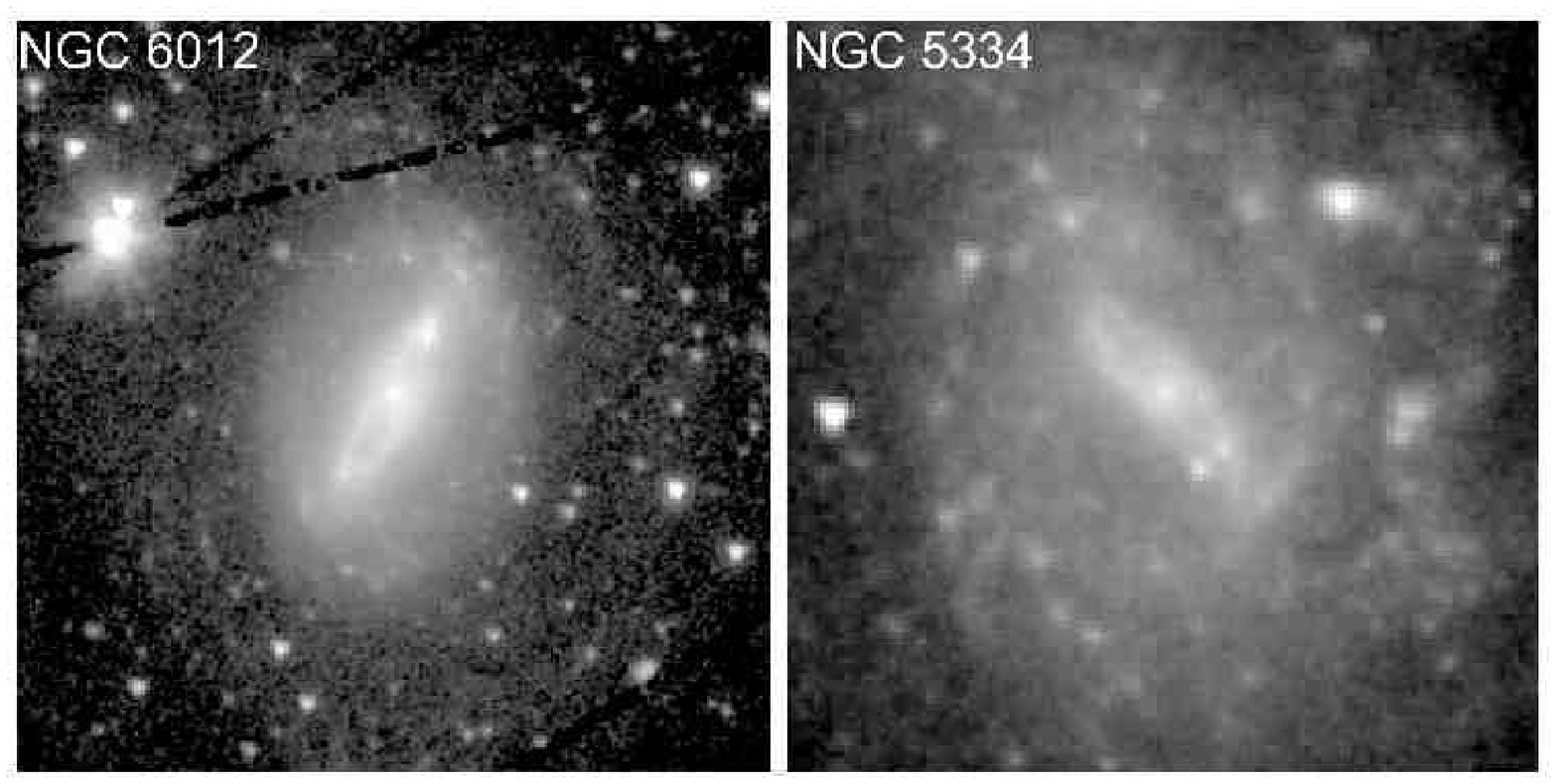}
\caption{Two S$^4$G galaxies showing x$_1$ rings, which are the highly
elongate
d
features in the bar.
}
\label{x1r-gals2}
\end{figure}

Only four possible examples of this type of feature are recognized in
the S$^4$G catalogue (NGC 4569, 5334, 6012, and UGC 7848). The best
example, NGC 6012, is shown in the left panel of Figure~\ref{x1r-gals2}
(see also Regan \& Teuben 2004). The highly-elongated, slightly
miscentered feature extends to a little over half of the bar radius and
is brighter at the ends. The galaxy also has a regular inner ring-lens
($\underline{\rm r}$l) and a very faint outer pseudoring.

NGC 5334 is an interesting late-type spiral that, in blue light, shows
a bar that appears to have centered dust lanes (dVA). However,
Figure~\ref{x1r-gals2} shows that the split character of the bar is
not due to dust, but to a highly-elongated ring-like morphology.  This
is similar to NGC 6012 and we interpret the feature as another x$_1$ ring.

\subsection{The Mid-IR Morphology of Galactic Disks}

The considerable depth and dust penetration of 3.6$\mu$m S$^4$G images
provides excellent views of galaxy disks, especially thin disks which
may be obscured by planar dust in optical images. In this Section, we
examine the mid-IR morphology of highly inclined and edge-on disks,
focussing on warps, embedded disks in 3D systems, thick disks, and
other special cases.

\subsubsection{Warped and Flared Disks}

Warped disks are easily recognizable in many S$^4$G galaxies (see
also Saha et al. 2009). Warping can appear as an edge-on disk twisted
into an integral sign shape. The inner thin disk may be perfectly
flat, while the outer parts of the disk bend in opposite senses
from one end of the major axis to the other. Kinematically, a warp
can be described in terms of circular orbits having a radius-dependent
inclination and line of nodes position angle (e.g., Briggs 1990).

Since warps are most easily recognized in edge-on galaxies that
would be classified as spindles (``sp"), the CVRHS recognizes warps
using the notation ``spw". Disk warping may be caused by a bending
instability (Binney \& Tremaine 2008), among other possible
explanations (e. g., Radburn-Smith et al. 2014). An up-to-date
review of the theory of warps is presented by Sellwood (2013).

\begin{figure}
\figurenum{22}
\plotone{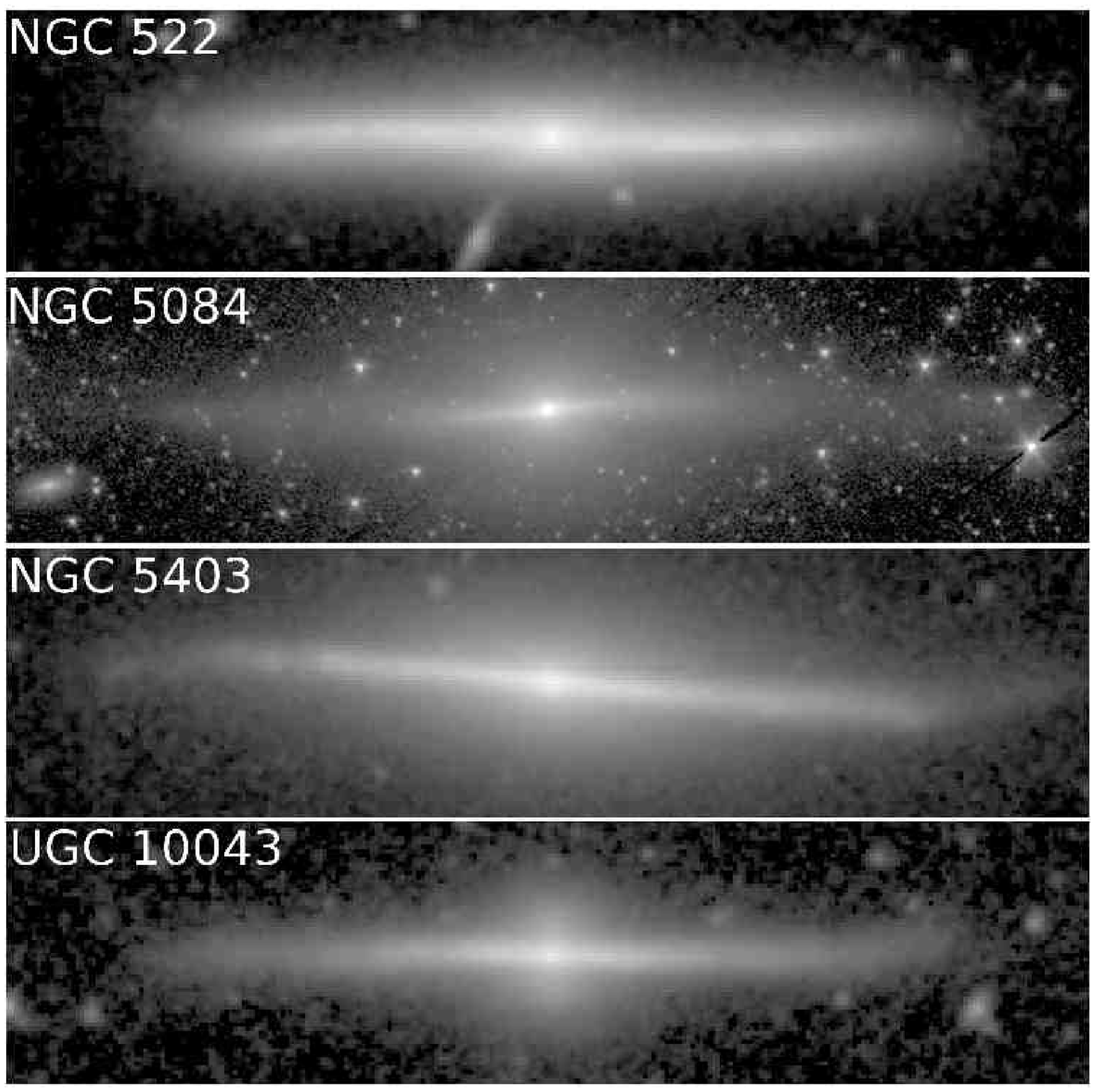}
\caption{Montage showing the mid-IR morphology of four S$^4$G galaxies
showing
strong outer
disk warping.
}
\label{warped-gals}
\end{figure}

Four S$^4$G galaxies having obvious warps are shown in
Figure~\ref{warped-gals}. These are selected merely as good examples
that are well-resolved in S$^4$G images. NGC 522 is an edge-on
galaxy whose mid-IR thick disk structure has been studied by Comer\'on
et al.  (2011a). The galaxy is classified as type Sbc in RC3, but
in the mid-IR it appears almost smooth enough to be classified as
type S0. A subtle knottiness along the thin disk component favors
a later type. The image shows strong warping and a subtle X in the
central area that may signify the presence of a bar.

NGC 5084 shows a thin disk that flares within the prominent bulge
region, and then bends farther out. No planar dust is present in the
thin disk as seen in an SDSS color image, and thus optical and mid-IR
images provide similar views of the galaxy's structure.

NGC 5403 shows a bright thin disk embedded within a flattened
spheroidal component. The thin disk bends and flares at large
radii. An SDSS color image shows that the galaxy has a strong
planar dust lane and is likely almost exactly edge-on. 

UGC 10043 is an exceptional example with strong warping and
a central bulge with an interesting peculiarity: inside the 
roundish region, the isophotes are elongated perpendicular to
the inner disk. The appearance suggests that the inner part of
the bulge has a {\it prolate} shape. This galaxy was recently
studied in detail by Matthews \& de Grijs (2004), who concluded
the galaxy may have experienced an accretion or merger event
that can account for some of its unusual characteristics, such
as the warped disk.

\subsubsection{Embedded Disks in Three-dimensional Early-Type Galaxies}

The S$^4$G sample includes numerous examples where a clear
highly-flattened disk-shaped system is embedded in a more
three-dimensional early-type system. Such systems have been known
for a long time, and highlight how some disks appear to be ``living
within an elliptical" [or how some bulges are ``ellipticals living
within a disk" (e.g., Kormendy \& Kennicutt 2004)]. The mid-IR
provides a dust-penetrated view of these interesting systems that
highlights how thin disks can fill, overextend, or underextend the
3D systems they are embedded within.

\begin{figure}
\figurenum{23}
\begin{center}
\includegraphics[height=5.0in,angle=0]{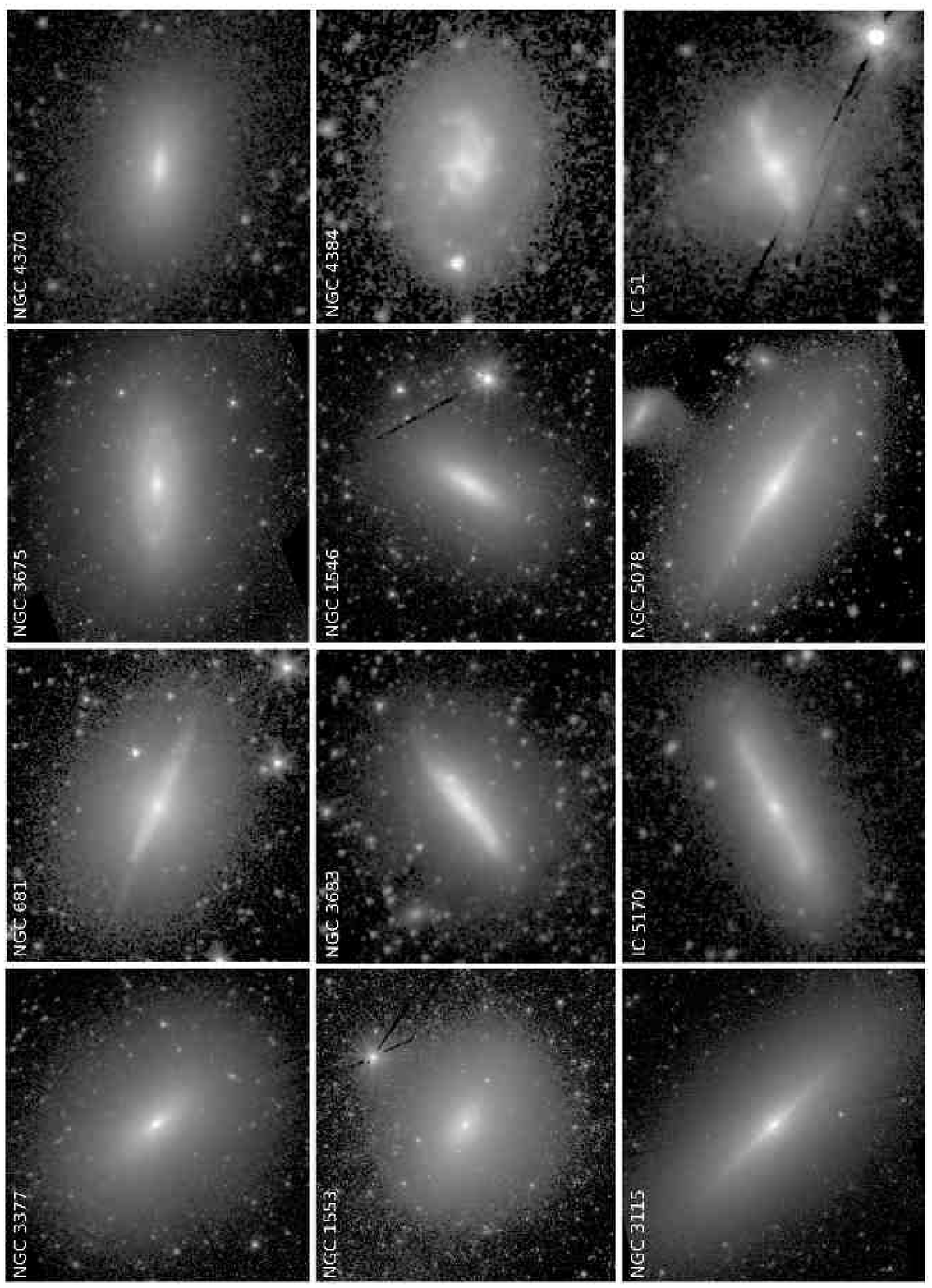}
\end{center}
\caption{Montage showing galaxies having embedded disks in 3D
early-type
systems.
}
\label{embeddeds}
\end{figure}

Figure~\ref{embeddeds} shows a sampling of examples of these kinds of
cases beginning with NGC 3377, a normal ``disky elliptical" galaxy
[type E(d)4, Kormendy \& Bender 1996]. The disk in NGC 3377 is very
subtle, featureless, purely stellar in nature, and possibly not exactly
edge-on. The presence of the disk was quantitatively established by
Jedrzejewski (1987), who showed that the $\cos 4\theta$ relative Fourier
deviation from perfect ellipses is positive across a wide range of
radii in this galaxy.

NGC 681 is very similar to NGC 3377, except that the disk appears to be
a nearly edge-on late-type spiral. The published classifications of NGC
681: SAB(s)ab sp (RC3) and Sab (Sandage \& Tammann 1981) seem
inadequate to describe this galaxy. It is not clear that NGC 681 is
merely an early-type spiral galaxy with a large bulge. The disk part of
NGC 681 is a clear late-type spiral with a high degree of knottiness
and star formation. The logarithmic, background-subtracted display
shows that the embedded disk neither greatly overfills nor underfills
the 3D component, but fades to a similar extent.

In Table 6, embedded disk systems such as NGC 681 are classified
generally in a two-part manner. The 3D early-type component is
usually denoted with ``E" because of appearance only, not because
these components necessarily have $r^{1\over 4}$ luminosity profiles.
For example, NGC 681 is classified in Table 6 as SA:(s:)$\underline{\rm
b}$c sp / E(d)3, where the E(d)3 refers to the 3D component as a
``disky elliptical" (using Kormendy \& Bender 1996 notation). The
adopted classification in Table 6 recognizes both the similarity
to NGC 3377 and the greater significance of the embedded disk.

Very similar to NGC 681 is NGC 3683, whose disk is less edge-on and has
a somewhat smaller bulge. It also is embedded in an E3 background. The
Phase 1 and 2 classifications in Table 2 disagree on the presence of a
bar in this galaxy; the mean classification of SAB(s)b$\underline{\rm
c}$ sp / E3 is an average of SA and SB. NGC 5078 is another example,
but the embedded disk is more edge-on; its average type is S(r:)b sp / E5.
The lack of a ``(d)" in these two cases [as in E(d)5] only means that
the outer isophotes of the E5 part are neither pointy nor boxy as
judged visually in a color display, although the inner isophotes would be
disky.

Related to these is the remarkable case of NGC 3675, where the embedded
disk is far from edge-on and its structure is clearly visible. The
Phase 1 classification recognizes two outer pseudorings, but only the
inner one is well-defined and was also recognized in Phase 2. The final
classification is (R$^{\prime}$,R$^{\prime}$)SA(ls)b / E4, recognizing
an inner lens with a faint spiral pattern. The main outer pseudoring is
about twice the diameter of the inner lens and bright spiral structure
breaks from it to larger radii, forming another outer pseudoring. The
diskiness of the E4 isophotes is not evident because the disk
inclination is so low.

Like NGC 681, NGC 3683 and 5078 are cases of comparable extent embedded disks,
where the disk appears to nearly fill the detectable outer isophotes of
the 3D component. Also shown in Figure~\ref{embeddeds} are cases of
what we will call ``limited extent disks," where an edge-on disk
appears to fade well within the bright bounds of the 3D component. An
excellent example is NGC 4370, where the embedded disk is so small it
was classified as a nuclear disk in Phase 2.  Phase 1 also recognized
boxy outer isophotes of the 3D component. The adopted average
classification is S0$^{-/o}$ sp / E(b,nd)4. An interesting aspect of NGC
4370 is the slight misalignment between the inner disk and the outer,
boxy isophotes.  A galaxy similar to NGC 4370 is described by Graham et
al. (2012).

NGC 3115 is the well-known example originally classified by Hubble
(1936) as type E7. It was reclassified as S0$^-$ sp in RC3 and as
S0$_1$(7)/a by Sandage \& Bedke (1994). Although the morphology of NGC
3115 in the mid-IR hardly differs from its morphology in blue light, it
is a good example to compare with the other embedded disks illustrated
in Figure~\ref{embeddeds}. The bright thin disk appears to fade well
before it fills the major axis of the 3D component. In this case the
outer isophotes of the 3D component are not obviously disky, and the
classification adopted in Table 6 is S0$^-$ sp / E5-6.

NGC 1546 is an example where a non-edge-on embedded disk includes a
lightly patchy pair of closely spaced rings. The outer isophotes of the
3D component are slightly boxy. This was recognized in both the phase
1 and 2 classifications, and the average classification is
(R$^{\prime}$)SA(r)a / E(b)3-4. We have already noted a similar embedded
disk ring/lens in NGC 1553.

IC 5170 is a case where what appears to be an abruptly terminating
edge-on disk is embedded within a boxy 3D component. The mean Phase 1
and 2 type is S0/$\underline{\rm a}$ spw / E(b)5. NGC 4634 
(Figure~\ref{thick-disks}) shows a similar sharply ending limited extent
thin disk embedded in a larger, boxy zone.

IC 51 is a very unusual case where the embedded disk is not settled
within the projected major axis of the rounder background component. In
this case, the background component could be a more face-on disk, and
IC 51 would be an example of a polar ring galaxy where only the
disrupted disk is seen edge-on. The Table 6 classification is SA(s)b
sp / SA0$^o$ (PRG?). IC 51 is also listed as a ``good candidate" for a
polar ring galaxy by Whitmore et al. (1990).

The final example in Figure~\ref{embeddeds} is NGC 4384. The inner part
of the galaxy is a clear SB(rs)dm type with virtually no bulge. This
appears embedded in a smooth, relatively uniform background interpreted
in both Phase 1 and Phase 2 as an outer lens (L). 

\subsubsection{Thick Disks}

Thick disks are an important morphological feature in many S$^4$G
galaxies, especially among late-type systems with little bulge
contribution. A thick disk is defined to be a highly-flattened
component with a vertical scale height a few times larger than that
of the thin disk (Comer\'on et al. 2011a). Because of the considerable
depth of S$^4$G images and the almost negligible impact of extinction,
thick disks are very prominent and are especially made visible with
the logarithmic display that we use. S$^4$G images allow us to see
the relation between thick and thin disks only minimally affected
by extinction of the thin component.

\begin{figure}
\figurenum{24}
\plotone{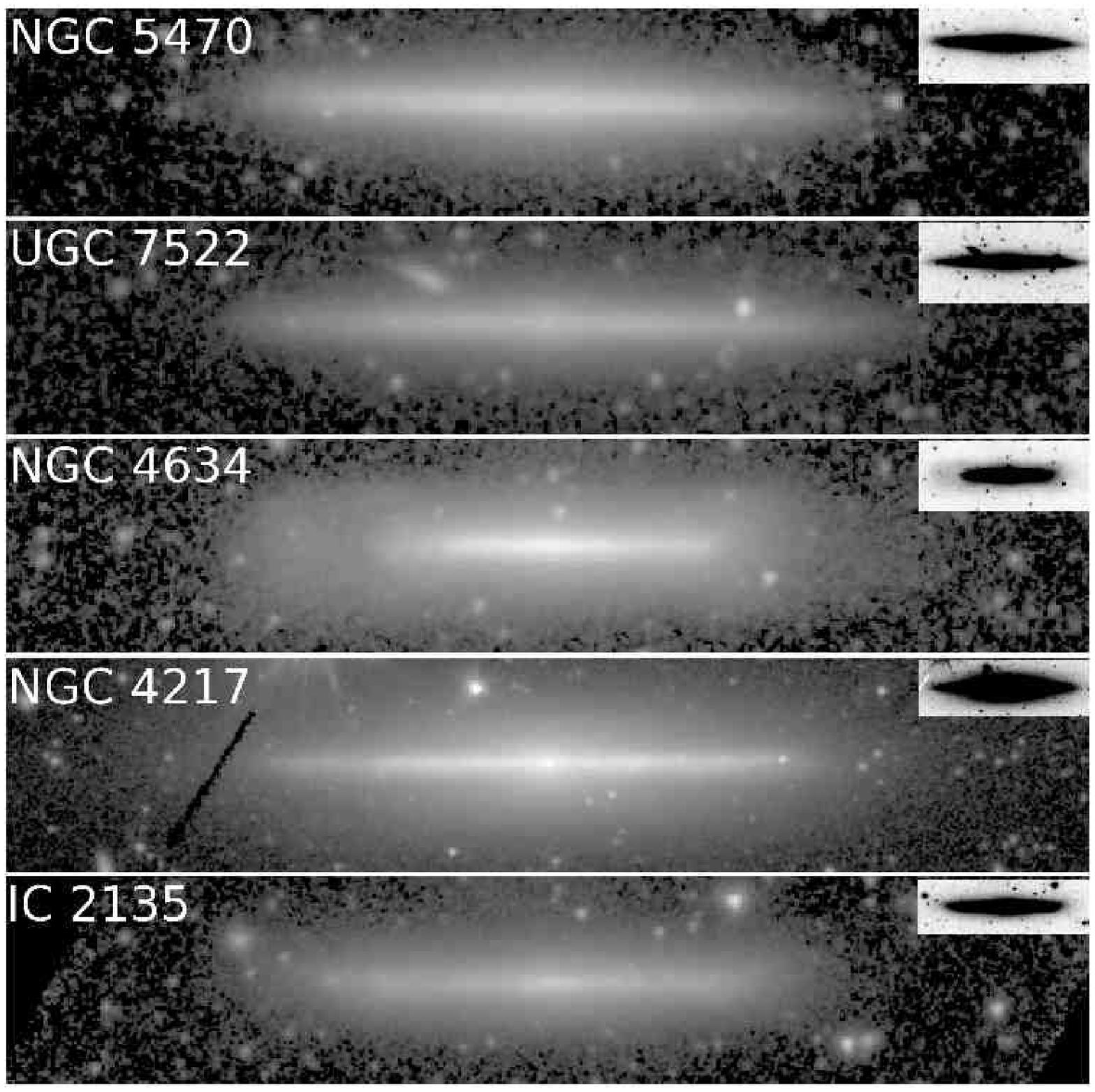}
\caption{Montage showing five S$^4$G galaxies having typical thick
disks.
}
\label{thick-disks}
\end{figure}

Figure~\ref{thick-disks} shows five S$^4$G galaxies having
well-defined but typical thick disks. The features have a range of
morphologies. For example, the thick disks of NGC 5470, NGC 4217, and
UGC 7522 are pointed ovals at the faintest light levels detected, which
may not be unusual except for the fact that some thick disks are either
more elliptical than pointy at the ends [as in IC 2135 (bottom frame,
Figure~\ref{thick-disks})] or are even boxy [as in NGC 4634 (middle
frame, Figure~\ref{thick-disks})].  Each galaxy in
Figure~\ref{thick-disks} has a two part classification as described in
Section 3.3. The thick disk is treated as a highly elongated disky,
boxy, or plane elliptical feature using Kormendy \& Bender (1996)
notation.

Notes on several of the galaxies in Figure~\ref{thick-disks}:

NGC 4217 is classified as an edge-on Sb spiral in RC3, but the
3.6$\mu$m image shows only a small central concentration, more
characteristic of Sc than Sb.  The star-forming disk is very thin, and
is embedded in a thick disk. The galaxy appears to be an Sc embedded in
an E6-7 thick disk [Table 6 classification: Sc sp/ E(d)6-7], and is
exactly edge-on. The inserts in Figure~\ref{thick-disks} for NGC 4217
show how the shape of the fainter isophotes changes with decreasing
surface brightness.

NGC 5470 is an almost exactly edge-on disk galaxy also classified as
type Sb in RC3. In the mid-IR, however, NGC 5470 is a completely
bulgeless, pure disk galaxy that shows an S0-like appearance.  The
Table 6 classification, S0$^o$[d] sp / E(d)8, alludes to the van den
Bergh (1976) parallel sequence idea, although genuine $B$-band versions
of such a galaxy type are not necessarily known. It is possible that
NGC 5470 would be less S0-like in a higher resolution mid-IR image. The
E(d)8 part of the classification strictly recognizes the highly
flattened thick disk with a strong pointed oval shape.

NGC 4634 is a member of the Virgo Cluster and shows a thin planar dust
lane in blue light. The boxiness of its thick disk could be an
environmentally driven product.  Kormendy \& Bender (2012) have
interpreted the broad boxy zone outside the edge-on S0 galaxy NGC 4638,
also a Virgo Cluster member, as possibly being a thick disk
environmentally flared by harassment from cluster encounters.
NGC 4634 is classified in Table 6 as Sd sp / E(b)7-8. Like NGC 5470,
the 3.6$\mu$m image of NGC 4634 shows little or no central
concentration. Also, the thin disk in this case has limited extent.

\subsubsection{Extraplanar Disks}

An extraplanar disk in a galaxy is a catastrophically-acquired disk
where the material lies in a different plane from the disk of the
receiving galaxy. The best-known extraplanar disks are polar rings,
where a small gas-rich companion is disrupted into a polar orbit
around a more massive S0 galaxy (Schweizer, Whitmore, \& Rubin 1983).
An inclined ring is an extraplanar disk where a companion has been
disrupted along a lower inclination orbit. Inclined and polar ring
galaxies are most easily recognized when both the receiving disk and
the extraplanar disk are nearly edge-on (Whitmore et al. 1990).

\begin{figure}
\figurenum{25}
\plotone{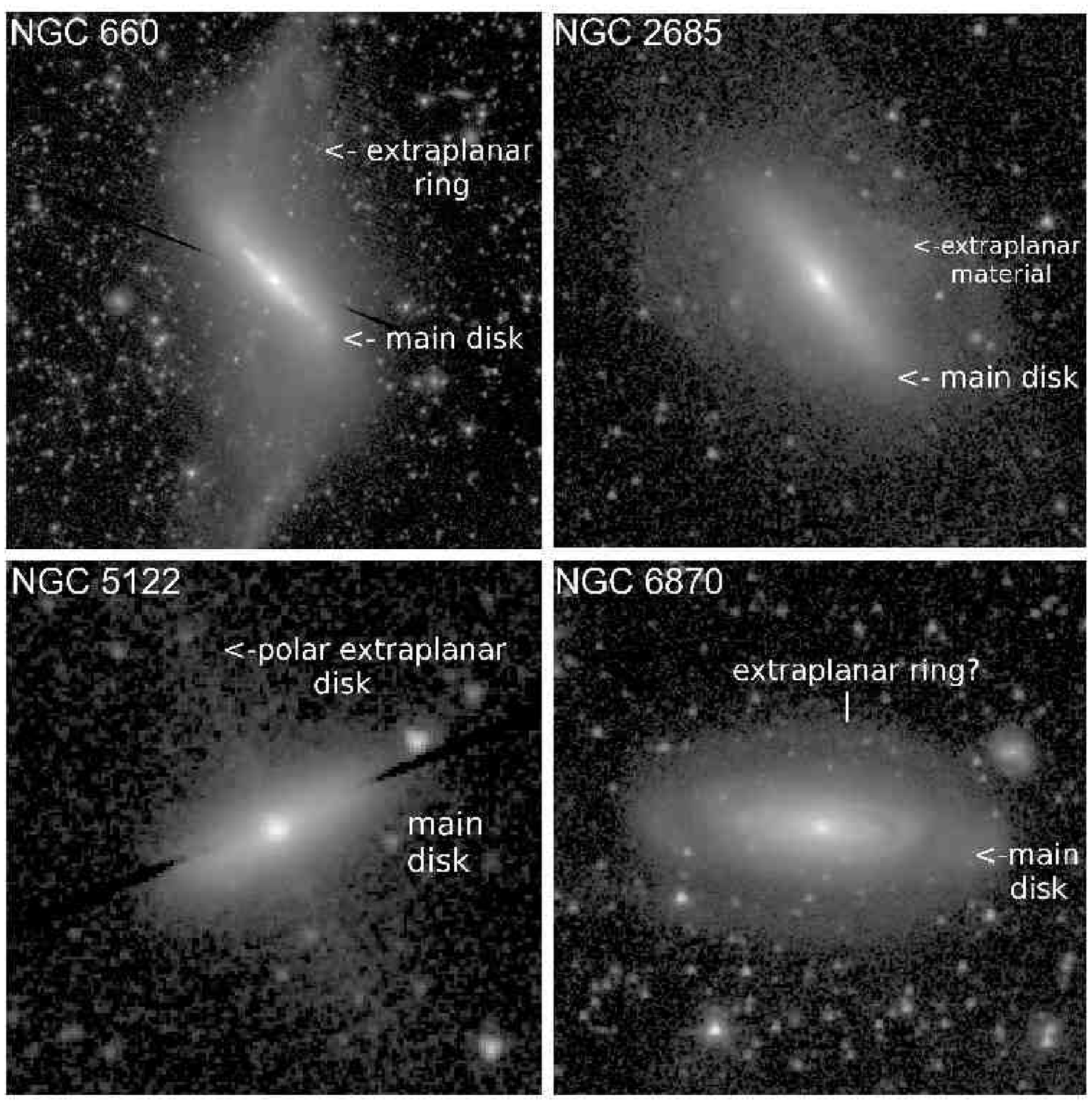}
\caption{Montage showing four S$^4$G galaxies having definite or
possible
extraplanar disk material. The main (receiving) disk and the
extraplanar disk
are indicated.
}
\label{extraplanar-disks}
\end{figure}

Galaxies showing extraplanar material in the form of an inclined disk
or ring have been of great interest for studies of disruptive
encounters, merger histories, and the shapes of dark matter halos
(e.g., Casertano et al. 1991, and most recently Iodice \& Corsini
2013).  Several S$^4$G galaxies show definite or possible extraplanar
disk material.  Four examples are shown in
Figure~\ref{extraplanar-disks}. Three of these (NGC 660, 2685, and
5122) were already known from optical observations, and we have already
described IC 51. Whitmore et al. (1990) show how detection of polar
ring galaxies, where the extraplanar disk is oriented at nearly
90$^{\circ}$ to the main disk (typically an S0 galaxy), depends
strongly on viewing geometry.

NGC 660 is interesting in that in blue light, the main disk shows an
aligned dust lane that is at an angle to a second dust lane. It was
classified as a polar-ring galaxy (PRG) in Phase 1 (paper I), but here
we prefer the term ``inclined ring galaxy" (IRG) since the extraplanar
material in cases like NGC 660 is not necessarily oriented orthogonally
to the main disk (van Driel et al. 1995). The S$^4$G 3.6$\mu$m image of
NGC 660 reveals an interesting characteristic: a broad X pattern
flanked by two bright ansae. This indicates that the main component of
NGC 660 has a bar (see also Luetticke et al. 2004).

The well-known extraplanar material in NGC 2685 is seen in the
3.6$\mu$m image as a small tipped partial ring of knots. This has been
interpreted as part of a ring at a large angle to the main stellar
object. J\'ozsa et al. (2009) argue that there are indeed two disks of
different orientations in NGC 2685, but the second disk (which has the
bulk of the HI) is extremely warped in the inner regions and appears
extraplanar as a result. NGC 2685 is likely not a classical polar ring
galaxy.

NGC 5122 (lower left panel of Figure~\ref{extraplanar-disks}) shows
a clear edge-on polar disk and is the best example of a polar ring
galaxy in the catalog.  Whitmore et al. (1990) argue that for every
case like this in a sample, there should be a few where the orientation
of the two disks is far enough from edge-on that the object could
masquerade as a relatively normal-looking system. NGC 6870, shown
in Figure~\ref{extraplanar-disks}, could be such a case.  The galaxy
shows a bright inner, highly tilted (but not edge-on) disk that
seems to have less inclined material crossing just inside the ends
of the inner disk, giving the galaxy the subtle appearance of a
hat.

\begin{figure}
\figurenum{26}
\plotone{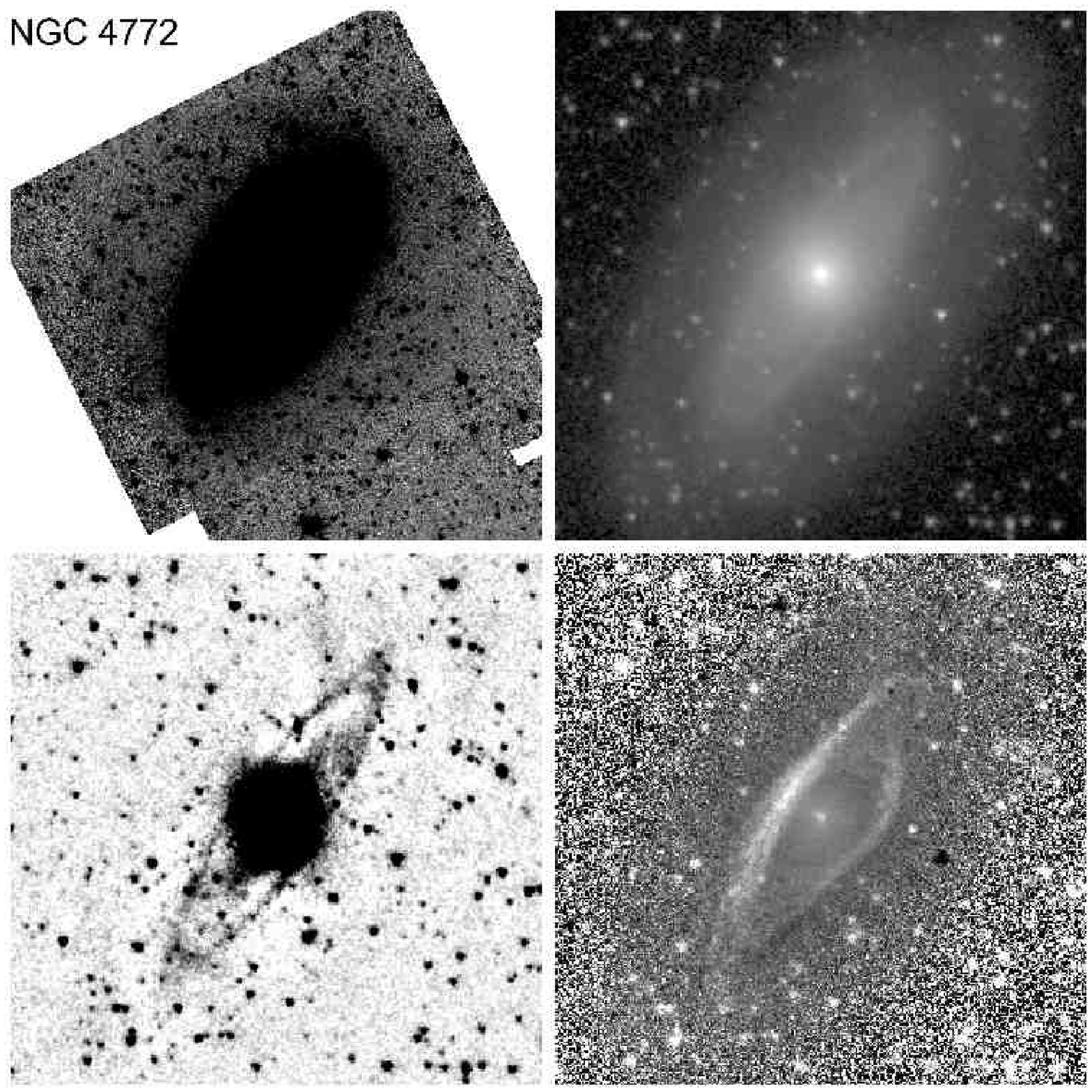}
\caption{A look at S$^4$G galaxy NGC 4772, a candidate for a
masquerading inclined ring galaxy: ({\it top left}) - frame showing
the very faint outer disk light that is round enough to suggest an
inclination of less than 40$^{\circ}$; ({\it upper right}) - the
structure in the overexposed inner oval zone is shown at twice the
scale; this reveals the highly elongated inner ring. {\it lower left} -
this frame is an unsharp-masked version of the upper right frame
showing how thin and well-defined the inner ring is; {\it lower right}
- a $g-$[3.6] color index map of NGC 4772 coded such that blue features
are dark and red features are light. This shows the complex dust and
near-side extinction in the region of the inner ring. The scale of the
last three frames is twice that of the upper left frame. The lower left
frame is in intensity units; the other frames are in units of mag
arcsec$^{-2}$.
}
\label{ngc4772}
\end{figure}

NGC 4772 (Figure~\ref{ngc4772}) is a galaxy that, in blue light,
shows a bright bulge in the center of an odd ring with strong
near-side extinction.  (This can be seen in the lower right panel
of Figure~\ref{ngc4772}, which shows a $g$ $-$ [3.6] color index
map of the galaxy.) The extinction suggests a substantial inclination,
but the faintest outer 3.6$\mu$m isophotes of the galaxy (shown in
the upper left frame of Figure~\ref{ngc4772}) are much rounder than
the ring and imply an inclination of less than 40$^{\circ}$. This
outer light has the morphology of an R$_1^{\prime}$ outer pseudoring
relative to a broad oval which is shown at twice the scale in the
upper right panel of Figure~\ref{ngc4772}. An unsharp-masked version
of this image shows a very thin, regular ring with no hint of excess
star formation around its major axis. If the ring were coplanar
with the broad oval, it would likely show such an excess (e.g.,
Crocker et al. 1996).  The implication is that the R$_1^{\prime}$
outer pseudoring and the larger oval are part of one disk, while
the inner ring could be part of a second disk that is not coplanar
with the first one. The bulge in NGC 4772 is significant and largely
spherical, making it unlikely that the inner ring is intrinsically
oval.

As noted by Whitmore et al. (1990), the only way to prove that a
specific case is a genuine PRG or IRG is to make kinematic
observations.  Haynes et al.  (2000) observed NGC 4772 in HI and found
that the galaxy has two somewhat kinematically decoupled HI rings of
different shapes and position angles (associated with the
R$_1^{\prime}$ outer pseudoring and the inner ring). From the velocity
field, these authors concluded that the galaxy has suffered a minor
merger.

NGC 4772 demonstrates how comparisons between optical and mid-IR images
can lead to candidates for extraplanar disks that we know have to
exist, but which are overlooked due to unfavorable orientations of the
disks.

\subsection{Late-type and Spheroidal Galaxies}

The Virgo Cluster is well-represented in the S$^4$G database. Included
in the sample are objects that were classified by Binggeli, Sandage, \&
Tammann (1985) as dwarf elliptical (dE) and dwarf S0 (dS0)
galaxies. Kormendy (2012) reviews the evidence that dE and dS0 galaxies
are environmentally-modified, bulgeless extreme late-type galaxies
whose existence favors reconsideration of the old van den Bergh (1976)
``parallel sequence" idea (see also Laurikainen et al. 2011; Cappellari
et al. 2011). For historical reasons, Kormendy collectively calls these
objects ``spheroidals" (Sph), although the term is not meant to imply
anything about intrinsic shapes. This conclusion is based on effective
parameter correlations (effective surface brightness $\mu_e$ and
effective radius $r_e$ versus absolute $V$-band magnitude, for example)
which revealed that dE and dS0 galaxies occupy areas where irregular
galaxies and very late-type spirals are found.  Genuine dwarf
ellipticals, like M32, are much rarer than Sph galaxies, and the S$^4$G
sample, in fact, includes no genuine dwarf elliptical galaxies.

\begin{figure}
\figurenum{27}
%\plotone{spheroidals1-new.ps}
\begin{center}
\includegraphics[height=5.0in,angle=0]{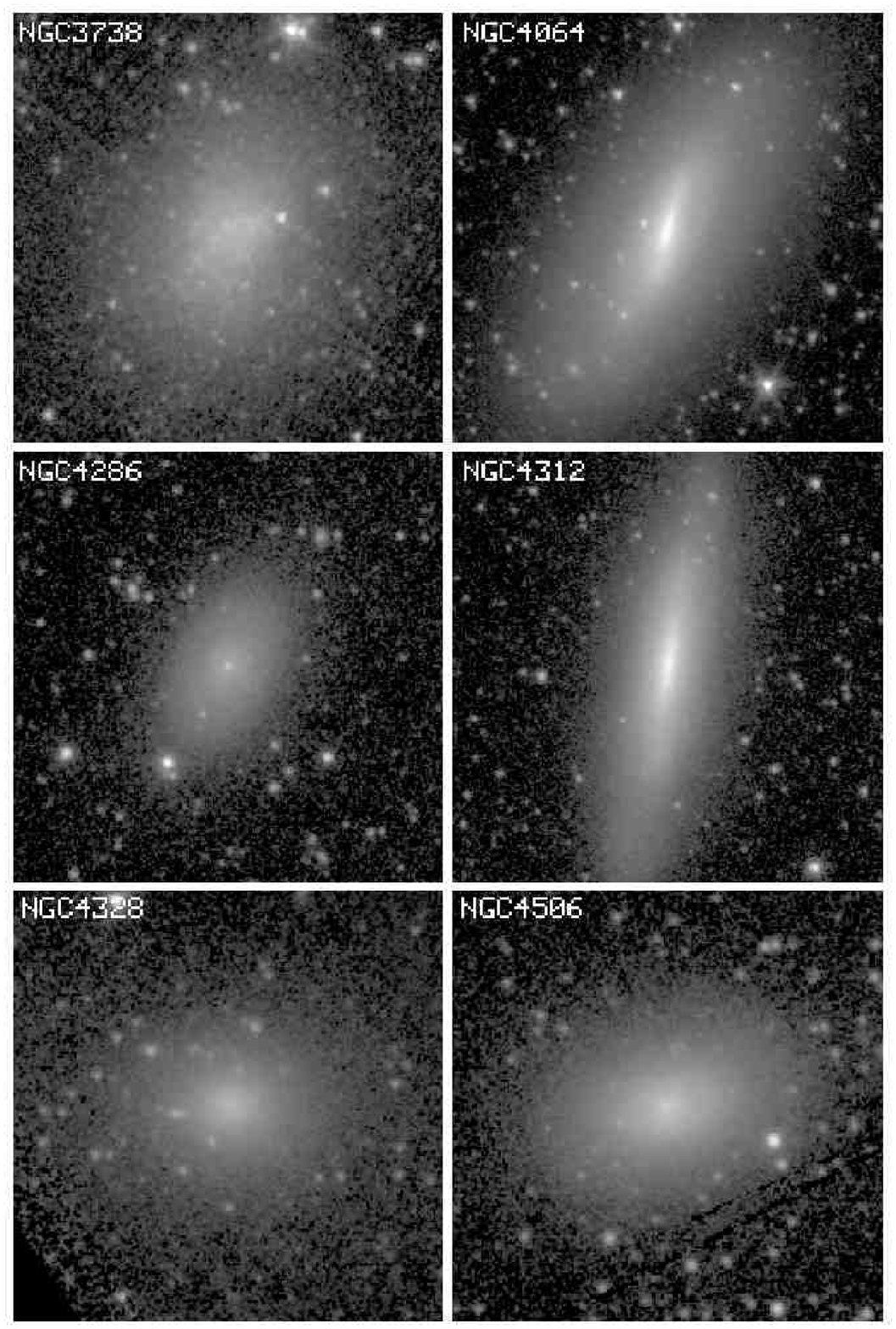}
\end{center}
\caption{Six examples of Kormendy spheroidal galaxies, including two
(NGC 4064 and 4312) that could be interpreted as bulgeless S0 galaxies
in the mid-IR.
}
\label{sph1}
\end{figure}

The connection between irregulars and Sph galaxies seems evident in
S$^4$G images. Because the effects of star formation are reduced, a
$B$-band irregular galaxy can look like a dE or dS0 galaxy at
3.6$\mu$m. An example is NGC 3738, shown in Figure~\ref{sph1}.  In
Table 6, the galaxy is classified as dIm (dE) / Sph, indicating it has
a bright background that resembles a dE galaxy. Such a galaxy, if
environmentally modified, would probably look very much like a Virgo
Cluster dE or dS0. We use ``/ Sph" to indicate a likely connection.
Figure~\ref{sph1} shows several other examples. NGC 4328 shows a clear
inner lens but no bar, while NGC 4506 shows a very faint, low
luminosity spiral, embedded in both cases in an Sph background.

Kormendy (2012) also shows that higher-luminosity Sph galaxies are
bulgeless S0s. Two galaxies likely to be related to these are NGC 4064
and 4312 (Figure~\ref{sph1}). NGC 4312 is a definite member of the
Virgo Cluster while NGC 4064 is a likely member. Both appear
highly-inclined and show a narrower elongation in the inner regions
that could be interpreted as a bar. The appearance of these galaxies at
3.6$\mu$m favors an early CVRHS stage, yet both appear largely
bulge-less. These could be environmentally-modified late-type galaxies.
This result is similar to what was found by Laurikainen et al. (2010),
where some S0s from the NIRS0S were shown to have very small bulges and
were presented as evidence in support of the parallel sequence
classification.

\begin{figure}
\figurenum{28}
%\plotone{spheroidals2-new.ps}
\begin{center}
\includegraphics[height=3.5in,angle=0]{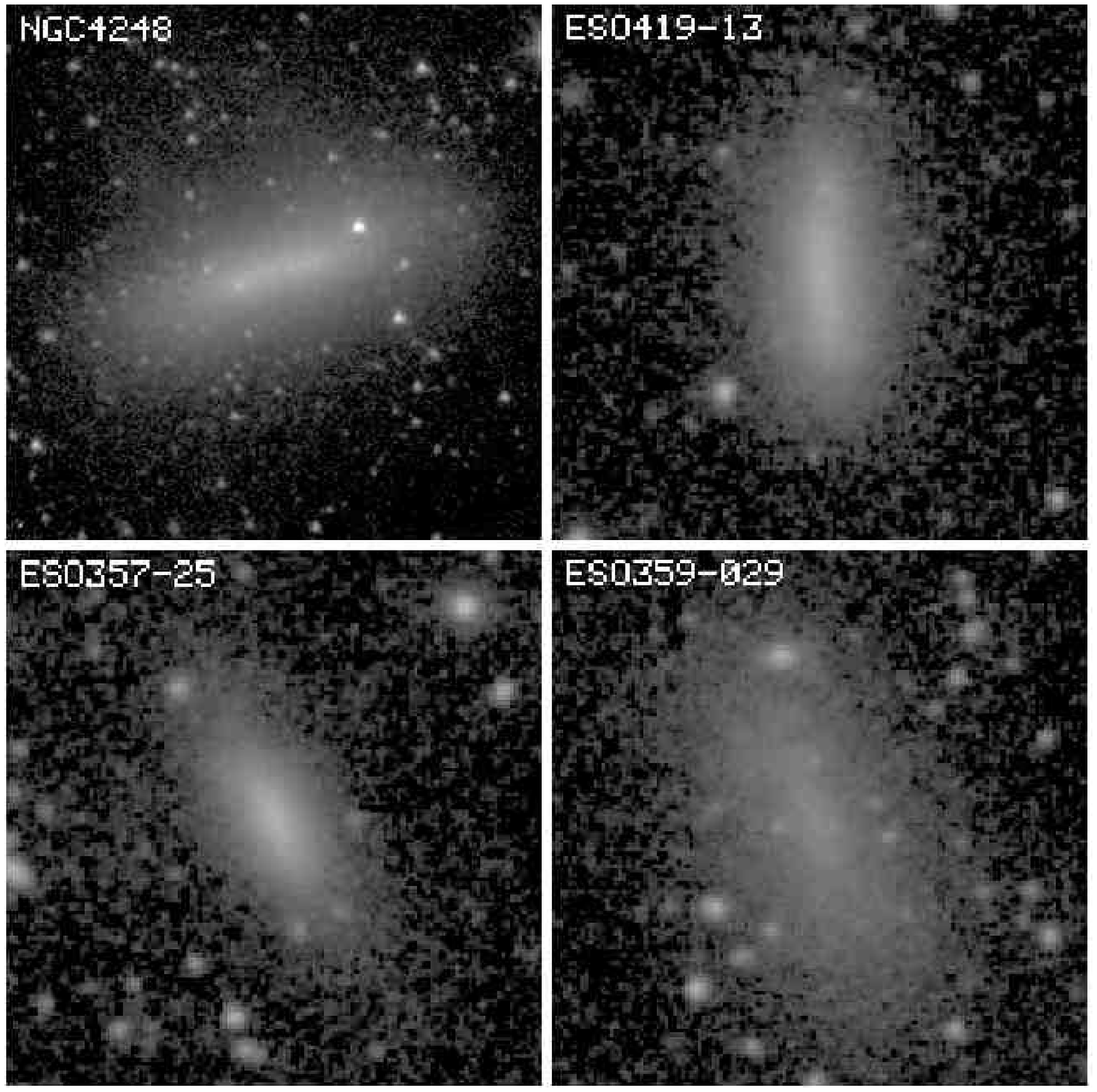}
\end{center}
\caption{Four low-luminosity, likely Sph galaxies showing an inner zone
that is more elongated than the outer isophotes parallel to the major
axis. None is a member of the Virgo Cluster.}
\label{sph2}
\end{figure}

Figure~\ref{sph2} shows four similar low-luminosity non-Virgo Cluster
members, each of which has a similar projected outer isophotal shape
and a narrower brightness enhancement along the major axis. In NGC
4248, the isophotal differences between the inner and outer regions
seem obvious; the same is true for ESO 419$-$13.  In both cases, the
inner elongation is parallel to the background major axis. While the
inner zones could be bars, another interpretation is that these
galaxies are edge-on Im galaxies embedded in a 3D background Sph. ESO
357$-$25 and ESO 359$-$29 appear related to these two objects but the
central ``stripe" is less well-defined.

\begin{figure}
\figurenum{29}
%\plotone{extreme-spirals-names.ps}
\begin{center}
\includegraphics[height=5.0in,angle=0]{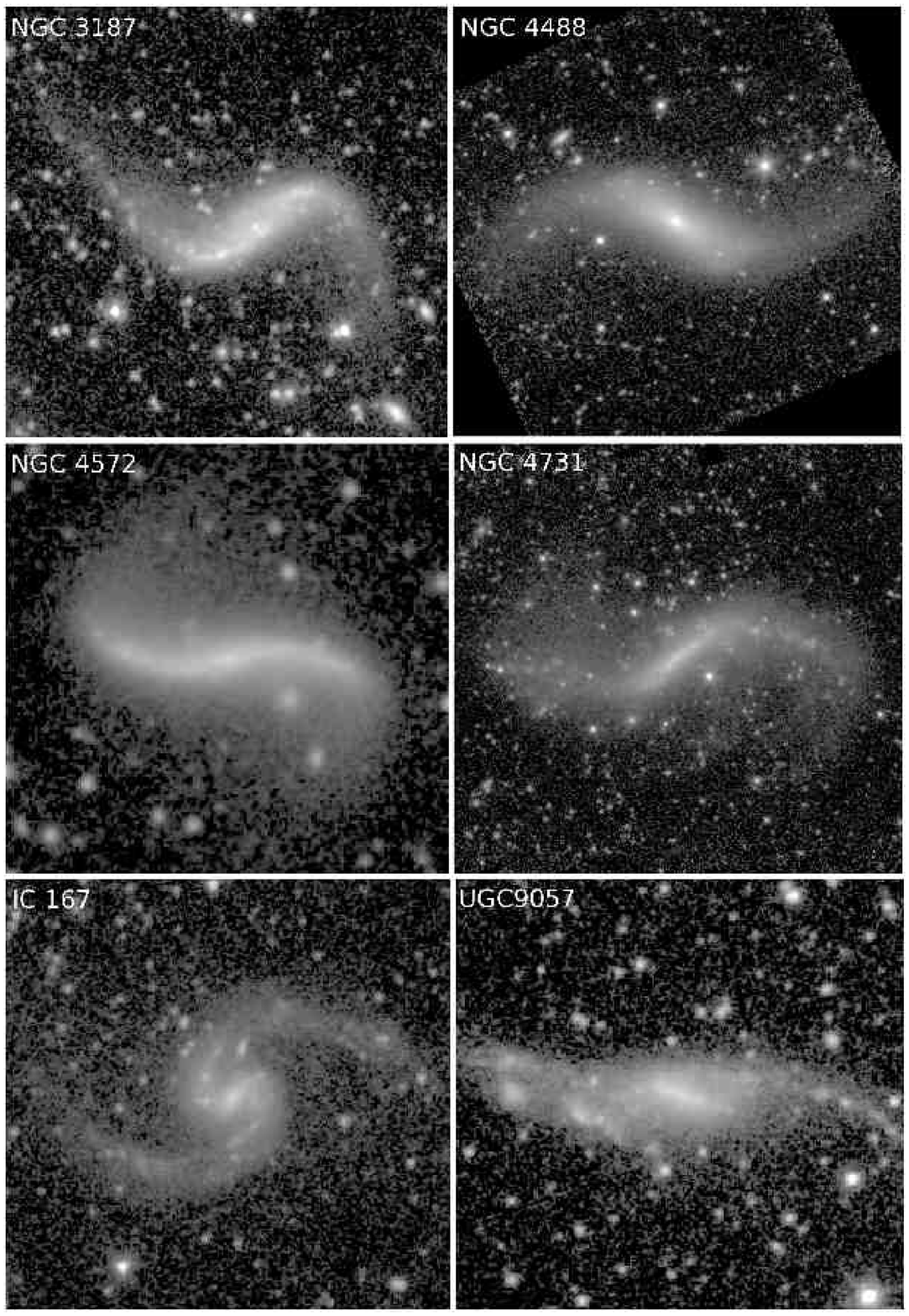}
\end{center}
\caption{Six galaxies having what appears to be extremely open and
extensive
spiral structure. At least one (NGC 4572) could be a case of extreme
warping
(super-spw in Table 1).}
\label{extspirals}
\end{figure}

Kormendy's conclusion concerning dE and dS0 galaxies has been
challenged by Graham (2013 and references therein) as being due to
the misleading nature of non-linear parameter correlations. Plotted
in a different way, regular ellipticals and dE and dS0 galaxies do
not necessarily divert from a smooth connection to more luminous
ellipticals.

\subsection{The Mid-IR Morphology of Spiral Arms}

The morphology of spiral structure in the mid-IR is of great interest
for several reasons: (1) the mid-IR reveals the underlying stellar
density enhancements associated with the arms, allowing a more reliable
judgment of the significance of the spiral to the overall dynamics and
evolution of a given disk galaxy; (2) the minimally-extinguished view of
spiral arms provided by mid-IR imaging allows not only the true
character of the arms (such as pitch angle and relative amplitude) to
be meaured, but also all star-forming regions associated with the arms
will be seen, including those which may have been heavily obscured; and
(3) like the classification of bars, the classification of spiral arms
in the mid-IR should be more reliable than in blue light.

In this Section, we examine the morphology of spiral arms using arm
classifications that are independent of CVRHS classifications, and
also look more closely at a few examples of extreme spiral structure
in the S$^4$G.

\subsubsection{Arm Classifications}

Arm classifications for galaxies are based on the symmetry, continuity,
and number of spiral arms. Arm Classes are distinct from the Hubble and
de Vaucouleurs classifications in that they are independent of the
pitch angle or bulge/disk ratio. Elmegreen (1981) introduced the term
``flocculent" to describe galaxies with short spiral arm pieces, in
contrast to the Lin \& Shu (1966) grand design spirals with long
symmetric spiral arms. Multiple arm galaxies fall in between these
extremes but are more similar to grand design galaxies, with inner
two-arm symmetry branching to many long arms. Elmegreen \& Elmegreen
(1982) devised a 12-point classification system that examined nuances
of these structures. Arm classes from $B$-band images were determined
for over 700 galaxies based on the Palomar Observatory Sky Survey and
high resolution atlas images (Elmegreen \& Elmegreen 1987).

The physical nature of flocculent, multiple arm, and grand design
galaxies is revealed through a comparison of blue and near-IR
($I$-band) spiral arm strengths, in which the contrast between arm and
interarm regions is essentially the same in the $B$- and $I$-bands for
grand design and multiple arm galaxies but not for flocculent
galaxies. Also, the arm amplitude is stronger in grand design galaxies
(Elmegreen \& Elmegreen 1984). The underlying mechanism that explains
global symmetry in the grand design and multiple arm galaxies is a
density wave, which flocculent galaxies appear to lack in blue light.

Previous studies of 2$\mu$m images, which like IRAC images are less
affected by dust extinction than optical images, showed that sometimes
an ``optically flocculent" spiral pattern looked grand design in the
near-IR. Four examples of this, including NGC 5055, were studied
by Thornley (1996), and NGC 253 and 7217 are two more examples
described in the dVA. Paper I compared $B$-band and 3.6$\mu$m images of
the well-known optically-flocculent spiral NGC 2841, showing that it also
has an underlying stellar grand design spiral.

To help elucidate whether underlying disks might contain an obscured
density wave pattern, arm classifications were extended to 3.6$\mu$m by
Elmegreen et al. (2011) in 46 S$^4$G galaxies; most did not change
their Arm Class going from $B$-band to 3.6$\mu$m. Spiral arm properties
were studied through measurements of symmetric components, arm-interarm
amplitudes, and Fourier transforms. As for the $B$-band images, the
3.6$\mu$m images showed stronger Fourier components and stronger arm
amplitudes in grand design and multiple arm galaxies than in flocculent
galaxies.

Arm classifications are now extended to the full sample of S$^4$G
galaxies in Table 9, where flocculent, multi-arm, and grand design
morphologies are symbolized using F, M, and G, respectively. These
classifications are given only for 1114 S$^4$G spiral galaxies that are
not too inclined to make the arm class indistinguishable, and are
listed also in the notes to Table 6. The classifiable subsample contains
50\% flocculent, 32\% multi-arm, and 18\% grand design cases. While
flocculent galaxies tend to be late-type, the arm classes span the
range of CVRHS spiral stages. Of these galaxies, 317 are in common with
galaxies for which a $B$-band arm class was previously published. A
comparison of the classifications showed that 60\% of the galaxies have
the same Arm Class going from $B$-band to 3.6$\mu$m, 28\% changed from
flocculent to multi-arm or multi-arm to grand design, and 9\% changed
from grand design to multi-arm or multi-arm to flocculent. Only 2
galaxies (NGC 3381 and NGC 5383) changed from grand design to
flocculent going to 3.6$\mu$m; 7 galaxies changed from flocculent to
grand design (NGC 1068, 3310, 3982, 4413, 4750, 7714, 7727). In all
cases, the IR images have deeper exposures and better resolution
(and less dust extinction) than the $B$-band images, so more structure
is revealed.

 \begin{deluxetable*}{lclclclclc}
 \tablewidth{0pc}
 \tablenum{9}
 \tablecaption{Arm Classifications
 for 1114 S$^4$G Galaxies\tablenotemark{a}
 }
 \tablehead{
 \colhead{Galaxy} &
 \colhead{AC} & 
 \colhead{Galaxy} & 
 \colhead{AC} & 
 \colhead{Galaxy} & 
 \colhead{AC} & 
 \colhead{Galaxy} & 
 \colhead{AC} & 
 \colhead{Galaxy} & 
 \colhead{AC}  
 \\
 \colhead{1} &
 \colhead{2} &
 \colhead{3} &
 \colhead{4} &
 \colhead{5} &
 \colhead{6} &
 \colhead{7} &
 \colhead{8} &
 \colhead{9} &
 \colhead{10} 
 }
 \startdata
NGC    45       & F & NGC   899       & F & NGC  1365       & G & NGC  2684       & F & NGC  3185       & G \\
NGC    63       & M & NGC   907       & F & NGC  1367       & G & NGC  2701       & F & NGC  3198       & M \\
NGC   115       & F & NGC   908       & M & NGC  1385       & F & NGC  2710       & G & NGC  3206       & F \\
NGC   131       & F & NGC   918       & M & NGC  1398       & M & NGC  2712       & M & NGC  3213       & F \\
NGC   134       & M & NGC   941       & F & NGC  1406       & G & NGC  2715       & F & NGC  3225       & F \\
NGC   150       & G & NGC   986       & G & NGC  1415       & G & NGC  2731       & F & NGC  3227       & G \\
NGC   157       & G & NGC   988       & F & NGC  1421       & G & NGC  2735       & G & NGC  3246       & F \\
NGC   178       & F & NGC   991       & F & NGC  1425       & M & NGC  2742       & M & NGC  3254       & M \\
NGC   210       & G & NGC  1022       & G & NGC  1433       & G & NGC  2743       & F & NGC  3259       & M \\
NGC   247       & F & NGC  1035       & F & NGC  1436       & G & NGC  2748       & M & NGC  3264       & F \\
NGC   253       & G & NGC  1036       & F & NGC  1438       & G & NGC  2750       & M & NGC  3274       & F \\
NGC   255       & F & NGC  1042       & M & NGC  1452       & G & NGC  2770       & M & NGC  3287       & F \\
NGC   275       & F & NGC  1051       & F & NGC  1473       & F & NGC  2776       & M & NGC  3294       & M \\
NGC   289       & M & NGC  1068       & G & NGC  1483       & F & NGC  2780       & G & NGC  3299       & F \\
NGC   298       & F & NGC  1073       & M & NGC  1493       & F & NGC  2805       & M & NGC  3310       & G \\
NGC   300       & M & NGC  1076       & F & NGC  1494       & F & NGC  2841       & M & NGC  3319       & G \\
NGC   337       & F & NGC  1084       & F & NGC  1512       & G & NGC  2854       & G & NGC  3320       & M \\
NGC   337A      & F & NGC  1087       & F & NGC  1515       & G & NGC  2856       & G & NGC  3321       & M \\
NGC   360       & M & NGC  1090       & M & NGC  1518       & F & NGC  2882       & M & NGC  3338       & M \\
NGC   406       & F & NGC  1097       & G & NGC  1519       & F & NGC  2903       & M & NGC  3344       & M \\
NGC   428       & F & NGC  1179       & F & NGC  1532       & G & NGC  2906       & M & NGC  3346       & F \\
NGC   450       & F & NGC  1187       & M & NGC  1546       & G & NGC  2919       & G & NGC  3351       & G \\
NGC   470       & M & NGC  1232       & M & NGC  1559       & F & NGC  2938       & F & NGC  3359       & M \\
NGC   485       & G & NGC  1249       & F & NGC  1566       & G & NGC  2964       & G & NGC  3361       & M \\
NGC   488       & M & NGC  1253       & M & NGC  1637       & M & NGC  2964       & M & NGC  3364       & F \\
NGC   493       & M & NGC  1255       & F & NGC  1640       & M & NGC  2967       & M & NGC  3368       & G \\
NGC   514       & M & NGC  1258       & F & NGC  1672       & G & NGC  2978       & G & NGC  3370       & M \\
NGC   578       & M & NGC  1292       & F & NGC  1679       & F & NGC  2985       & M & NGC  3381       & F \\
NGC   600       & M & NGC  1299       & F & NGC  1688       & M & NGC  3020       & F & NGC  3389       & F \\
NGC   613       & M & NGC  1300       & G & NGC  1703       & M & NGC  3021       & M & NGC  3403       & M \\
NGC   615       & M & NGC  1306       & F & NGC  1792       & M & NGC  3023       & F & NGC  3423       & F \\
NGC   628       & M & NGC  1309       & F & NGC  1808       & G & NGC  3031       & G & NGC  3430       & M \\
NGC   658       & M & NGC  1310       & M & NGC  1879       & F & NGC  3041       & M & NGC  3433       & G \\
NGC   672       & F & NGC  1313       & F & NGC  2104       & F & NGC  3049       & G & NGC  3437       & M \\
NGC   685       & F & NGC  1325       & M & NGC  2460       & M & NGC  3055       & M & NGC  3445       & F \\
NGC   691       & M & NGC  1325A      & F & NGC  2500       & F & NGC  3057       & F & NGC  3455       & F \\
NGC   701       & M & NGC  1337       & F & NGC  2537       & F & NGC  3061       & F & NGC  3485       & M \\
NGC   718       & G & NGC  1338       & M & NGC  2541       & F & NGC  3066       & G & NGC  3486       & M \\
NGC   723       & F & NGC  1341       & F & NGC  2543       & G & NGC  3147       & M & NGC  3488       & F \\
NGC   755       & F & NGC  1345       & F & NGC  2552       & F & NGC  3153       & F & NGC  3495       & M \\
NGC   770       & M & NGC  1347       & F & NGC  2604       & M & NGC  3155       & G & NGC  3504       & G \\
NGC   772       & M & NGC  1350       & G & NGC  2608       & M & NGC  3162       & M & NGC  3507       & G \\
NGC   803       & M & NGC  1353       & M & NGC  2633       & G & NGC  3169       & F & NGC  3511       & M \\
NGC   864       & M & NGC  1357       & M & NGC  2648       & G & NGC  3177       & G & NGC  3512       & M \\
NGC   895       & M & NGC  1359       & F & NGC  2681       & G & NGC  3184       & M & NGC  3513       & G \\
 \enddata
 \end{deluxetable*}
 %\clearpage
 \begin{deluxetable*}{lclclclclc}
 \tablewidth{0pc}
 \tablenum{9 (cont.)}
 \tablecaption{Arm Classifications
 for 1114 S$^4$G Galaxies (cont.)\tablenotemark{a}
 }
 \tablehead{
 \colhead{Galaxy} &
 \colhead{AC} & 
 \colhead{Galaxy} & 
 \colhead{AC} & 
 \colhead{Galaxy} & 
 \colhead{AC} & 
 \colhead{Galaxy} & 
 \colhead{AC} & 
 \colhead{Galaxy} & 
 \colhead{AC}  
 \\
 \colhead{1} &
 \colhead{2} &
 \colhead{3} &
 \colhead{4} &
 \colhead{5} &
 \colhead{6} &
 \colhead{7} &
 \colhead{8} &
 \colhead{9} &
 \colhead{10} 
 }
 \startdata
NGC  3521       & M & NGC  3810       & M & NGC  4102       & M & NGC  4321       & G & NGC  4559       & M \\
NGC  3547       & M & NGC  3813       & M & NGC  4106       & G & NGC  4348       & G & NGC  4561       & F \\
NGC  3549       & G & NGC  3846A      & F & NGC  4108       & F & NGC  4351       & F & NGC  4567       & M \\
NGC  3583       & G & NGC  3850       & F & NGC  4108B      & F & NGC  4353       & F & NGC  4568       & M \\
NGC  3589       & F & NGC  3876       & F & NGC  4116       & F & NGC  4376       & F & NGC  4569       & G \\
NGC  3596       & M & NGC  3887       & M & NGC  4120       & F & NGC  4378       & G & NGC  4571       & F \\
NGC  3614       & M & NGC  3888       & M & NGC  4123       & M & NGC  4380       & M & NGC  4572       & G \\
NGC  3622       & F & NGC  3893       & M & NGC  4127       & M & NGC  4384       & F & NGC  4579       & G \\
NGC  3623       & G & NGC  3896       & F & NGC  4133       & M & NGC  4385       & F & NGC  4580       & M \\
NGC  3625       & G & NGC  3898       & M & NGC  4136       & F & NGC  4390       & F & NGC  4591       & M \\
NGC  3626       & G & NGC  3901       & F & NGC  4141       & F & NGC  4395       & F & NGC  4592       & F \\
NGC  3627       & G & NGC  3906       & F & NGC  4142       & F & NGC  4396       & F & NGC  4593       & G \\
NGC  3629       & F & NGC  3912       & F & NGC  4145       & M & NGC  4409       & F & NGC  4595       & F \\
NGC  3631       & M & NGC  3913       & F & NGC  4152       & M & NGC  4411A      & M & NGC  4597       & F \\
NGC  3652       & G & NGC  3930       & F & NGC  4157       & M & NGC  4411B      & M & NGC  4602       & M \\
NGC  3654       & F & NGC  3938       & M & NGC  4158       & M & NGC  4412       & G & NGC  4618       & F \\
NGC  3655       & M & NGC  3949       & F & NGC  4159       & F & NGC  4413       & G & NGC  4625       & F \\
NGC  3659       & F & NGC  3953       & M & NGC  4162       & M & NGC  4414       & F & NGC  4628       & G \\
NGC  3664       & F & NGC  3955       & G & NGC  4165       & G & NGC  4416       & F & NGC  4630       & F \\
NGC  3672       & M & NGC  3956       & F & NGC  4178       & F & NGC  4428       & M & NGC  4632       & F \\
NGC  3673       & M & NGC  3972       & G & NGC  4180       & G & NGC  4430       & F & NGC  4633       & F \\
NGC  3675       & M & NGC  3976       & M & NGC  4189       & M & NGC  4448       & M & NGC  4635       & F \\
NGC  3681       & M & NGC  3981       & G & NGC  4192       & G & NGC  4450       & G & NGC  4639       & M \\
NGC  3683A      & M & NGC  3982       & G & NGC  4193       & M & NGC  4462       & G & NGC  4642       & M \\
NGC  3684       & M & NGC  3985       & F & NGC  4204       & F & NGC  4470       & F & NGC  4647       & F \\
NGC  3686       & M & NGC  3992       & M & NGC  4210       & M & NGC  4480       & M & NGC  4651       & M \\
NGC  3687       & M & NGC  4020       & F & NGC  4212       & M & NGC  4487       & F & NGC  4653       & M \\
NGC  3689       & G & NGC  4027       & F & NGC  4216       & G & NGC  4490       & F & NGC  4654       & M \\
NGC  3691       & F & NGC  4030       & M & NGC  4234       & F & NGC  4496A      & F & NGC  4658       & F \\
NGC  3701       & M & NGC  4032       & F & NGC  4237       & M & NGC  4498       & F & NGC  4668       & F \\
NGC  3705       & M & NGC  4035       & F & NGC  4238       & F & NGC  4501       & M & NGC  4680       & G \\
NGC  3715       & M & NGC  4037       & F & NGC  4254       & M & NGC  4502       & F & NGC  4682       & M \\
NGC  3718       & G & NGC  4041       & M & NGC  4258       & G & NGC  4504       & M & NGC  4688       & F \\
NGC  3726       & M & NGC  4045       & M & NGC  4260       & G & NGC  4517A      & F & NGC  4689       & M \\
NGC  3730       & M & NGC  4049       & F & NGC  4273       & M & NGC  4519       & F & NGC  4701       & M \\
NGC  3752       & F & NGC  4050       & G & NGC  4276       & F & NGC  4523       & F & NGC  4713       & F \\
NGC  3755       & F & NGC  4051       & M & NGC  4288       & F & NGC  4525       & F & NGC  4722       & M \\
NGC  3756       & M & NGC  4062       & M & NGC  4294       & F & NGC  4527       & G & NGC  4725       & G \\
NGC  3769       & G & NGC  4067       & M & NGC  4298       & F & NGC  4531       & G & NGC  4731       & G \\
NGC  3780       & M & NGC  4080       & F & NGC  4299       & F & NGC  4534       & F & NGC  4750       & G \\
NGC  3782       & F & NGC  4085       & M & NGC  4303       & M & NGC  4535       & M & NGC  4765       & F \\
NGC  3786       & G & NGC  4088       & M & NGC  4303A      & F & NGC  4536       & M & NGC  4771       & M \\
NGC  3788       & G & NGC  4094       & M & NGC  4313       & G & NGC  4540       & F & NGC  4775       & F \\
NGC  3794       & F & NGC  4096       & M & NGC  4314       & G & NGC  4545       & M & NGC  4779       & M \\
NGC  3795A      & F & NGC  4100       & G & NGC  4319       & G & NGC  4548       & G & NGC  4781       & F \\
 \enddata
 \end{deluxetable*}
 %\clearpage
 \begin{deluxetable*}{lclclclclc}
 \tablewidth{0pc}
 \tablenum{9 (cont.)}
 \tablecaption{Arm Classifications
 for 1114 S$^4$G Galaxies (cont.)\tablenotemark{a}
 }
 \tablehead{
 \colhead{Galaxy} &
 \colhead{AC} & 
 \colhead{Galaxy} & 
 \colhead{AC} & 
 \colhead{Galaxy} & 
 \colhead{AC} & 
 \colhead{Galaxy} & 
 \colhead{AC} & 
 \colhead{Galaxy} & 
 \colhead{AC}  
 \\
 \colhead{1} &
 \colhead{2} &
 \colhead{3} &
 \colhead{4} &
 \colhead{5} &
 \colhead{6} &
 \colhead{7} &
 \colhead{8} &
 \colhead{9} &
 \colhead{10} 
 }
 \startdata
NGC  4790       & F & NGC  5236       & M & NGC  5597       & M & NGC  5954       & M & NGC  7163       & G \\
NGC  4793       & M & NGC  5240       & M & NGC  5600       & F & NGC  5956       & M & NGC  7167       & G \\
NGC  4795       & G & NGC  5247       & G & NGC  5604       & M & NGC  5957       & G & NGC  7171       & G \\
NGC  4800       & M & NGC  5248       & G & NGC  5624       & F & NGC  5958       & F & NGC  7184       & M \\
NGC  4806       & M & NGC  5254       & M & NGC  5630       & F & NGC  5962       & F & NGC  7188       & G \\
NGC  4808       & M & NGC  5289       & G & NGC  5633       & M & NGC  5963       & M & NGC  7191       & G \\
NGC  4814       & M & NGC  5297       & M & NGC  5645       & F & NGC  5964       & M & NGC  7205       & M \\
NGC  4826       & M & NGC  5300       & F & NGC  5660       & M & NGC  5970       & M & NGC  7218       & F \\
NGC  4897       & M & NGC  5313       & M & NGC  5661       & G & NGC  5985       & M & NGC  7247       & G \\
NGC  4899       & M & NGC  5320       & M & NGC  5665       & F & NGC  6015       & F & NGC  7254       & G \\
NGC  4900       & F & NGC  5334       & F & NGC  5667       & F & NGC  6063       & M & NGC  7290       & M \\
NGC  4902       & M & NGC  5336       & M & NGC  5668       & F & NGC  6070       & M & NGC  7307       & F \\
NGC  4904       & F & NGC  5339       & G & NGC  5669       & F & NGC  6106       & F & NGC  7314       & M \\
NGC  4928       & F & NGC  5346       & F & NGC  5676       & M & NGC  6118       & M & NGC  7328       & G \\
NGC  4942       & F & NGC  5347       & G & NGC  5678       & F & NGC  6140       & F & NGC  7371       & M \\
NGC  4948A      & F & NGC  5350       & G & NGC  5691       & F & NGC  6155       & F & NGC  7412       & M \\
NGC  4951       & M & NGC  5362       & M & NGC  5693       & F & NGC  6181       & G & NGC  7416       & G \\
NGC  4961       & F & NGC  5364       & M & NGC  5708       & F & NGC  6207       & F & NGC  7418       & M \\
NGC  4965       & F & NGC  5371       & G & NGC  5713       & F & NGC  6217       & M & NGC  7418A      & F \\
NGC  4980       & F & NGC  5375       & M & NGC  5740       & M & NGC  6236       & F & NGC  7421       & M \\
NGC  4981       & M & NGC  5376       & M & NGC  5744       & G & NGC  6237       & F & NGC  7424       & M \\
NGC  4995       & M & NGC  5383       & F & NGC  5756       & G & NGC  6239       & F & NGC  7437       & F \\
NGC  5002       & F & NGC  5426       & M & NGC  5757       & M & NGC  6255       & F & NGC  7448       & M \\
NGC  5005       & M & NGC  5427       & G & NGC  5768       & M & NGC  6267       & F & NGC  7456       & F \\
NGC  5012       & M & NGC  5430       & M & NGC  5774       & F & NGC  6339       & M & NGC  7463       & G \\
NGC  5016       & M & NGC  5443       & G & NGC  5783       & M & NGC  6395       & F & NGC  7479       & G \\
NGC  5033       & M & NGC  5448       & G & NGC  5789       & F & NGC  6412       & M & NGC  7496       & G \\
NGC  5042       & F & NGC  5452       & G & NGC  5798       & F & NGC  6434       & M & NGC  7497       & F \\
NGC  5054       & M & NGC  5457       & M & NGC  5806       & M & NGC  6870       & M & NGC  7513       & G \\
NGC  5055       & M & NGC  5464       & F & NGC  5832       & F & NGC  6887       & M & NGC  7531       & M \\
NGC  5068       & F & NGC  5468       & M & NGC  5850       & G & NGC  6889       & F & NGC  7537       & M \\
NGC  5079       & F & NGC  5474       & F & NGC  5861       & M & NGC  6902       & M & NGC  7541       & M \\
NGC  5085       & M & NGC  5476       & F & NGC  5878       & M & NGC  6902B      & F & NGC  7552       & G \\
NGC  5088       & F & NGC  5480       & M & NGC  5879       & M & NGC  6923       & M & NGC  7582       & G \\
NGC  5105       & F & NGC  5486       & F & NGC  5885       & M & NGC  6925       & M & NGC  7590       & M \\
NGC  5109       & F & NGC  5520       & M & NGC  5892       & M & NGC  7051       & M & NGC  7599       & M \\
NGC  5112       & M & NGC  5534       & G & NGC  5894       & M & NGC  7059       & F & NGC  7606       & M \\
NGC  5116       & G & NGC  5560       & G & NGC  5899       & G & NGC  7070       & F & NGC  7625       & F \\
NGC  5117       & F & NGC  5566       & G & NGC  5915       & F & NGC  7091       & F & NGC  7661       & G \\
NGC  5147       & F & NGC  5569       & F & NGC  5916       & G & NGC  7107       & F & NGC  7689       & M \\
NGC  5169       & M & NGC  5577       & F & NGC  5921       & M & NGC  7140       & M & NGC  7713       & F \\
NGC  5194       & G & NGC  5584       & M & NGC  5937       & M & NGC  7151       & F & NGC  7714       & G \\
NGC  5204       & F & NGC  5585       & F & NGC  5949       & F & NGC  7154       & F & NGC  7716       & M \\
NGC  5205       & G & NGC  5587       & G & NGC  5950       & G & NGC  7162       & M & NGC  7721       & M \\
NGC  5218       & M & NGC  5595       & M & NGC  5951       & F & NGC  7162A      & F & NGC  7723       & M \\
 \enddata
 \end{deluxetable*}
 %\clearpage
 \begin{deluxetable*}{lclclclclc}
 \tablewidth{0pc}
 \tablenum{9 (cont.)}
 \tablecaption{Arm Classifications
 for 1114 S$^4$G Galaxies (cont.)\tablenotemark{a}
 }
 \tablehead{
 \colhead{Galaxy} &
 \colhead{AC} & 
 \colhead{Galaxy} & 
 \colhead{AC} & 
 \colhead{Galaxy} & 
 \colhead{AC} & 
 \colhead{Galaxy} & 
 \colhead{AC} & 
 \colhead{Galaxy} & 
 \colhead{AC}  
 \\
 \colhead{1} &
 \colhead{2} &
 \colhead{3} &
 \colhead{4} &
 \colhead{5} &
 \colhead{6} &
 \colhead{7} &
 \colhead{8} &
 \colhead{9} &
 \colhead{10} 
 }
 \startdata
NGC  7724       & G & IC   2007       & F & UGC  1195       & F & UGC  5897       & M & UGC  7943       & F \\
NGC  7727       & G & IC   2051       & M & UGC  1547       & F & UGC  5922       & M & UGC  7950       & F \\
NGC  7741       & F & IC   2056       & M & UGC  1551       & F & UGC  5934       & F & UGC  8041       & F \\
NGC  7743       & G & IC   2361       & G & UGC  1670       & F & UGC  5976       & F & UGC  8042       & F \\
NGC  7750       & G & IC   2604       & F & UGC  1753       & F & UGC  5989       & F & UGC  8052       & G \\
NGC  7755       & M & IC   2627       & G & UGC  1862       & F & UGC  6023       & M & UGC  8053       & F \\
NGC  7757       & M & IC   2969       & F & UGC  2081       & F & UGC  6104       & M & UGC  8056       & F \\
NGC  7764       & F & IC   2995       & M & UGC  2443       & G & UGC  6157       & F & UGC  8067       & M \\
NGC  7793       & F & IC   3102       & G & UGC  3070       & F & UGC  6162       & F & UGC  8084       & F \\
NGC  7798       & G & IC   3115       & G & UGC  4151       & M & UGC  6169       & G & UGC  8085       & M \\
NGC  7817       & G & IC   3259       & F & UGC  4169       & F & UGC  6194       & G & UGC  8153       & F \\
IC    163       & F & IC   3391       & F & UGC  4238       & F & UGC  6249       & F & UGC  8282       & F \\
IC    167       & G & IC   3392       & G & UGC  4390       & F & UGC  6296       & G & UGC  8385       & F \\
IC    529       & M & IC   3476       & F & UGC  4499       & F & UGC  6309       & G & UGC  8449       & F \\
IC    600       & F & IC   3517       & F & UGC  4543       & F & UGC  6320       & F & UGC  8489       & F \\
IC    718       & F & IC   3742       & F & UGC  4549       & M & UGC  6335       & G & UGC  8516       & F \\
IC    749       & F & IC   4216       & M & UGC  4621       & G & UGC  6345       & F & UGC  8588       & F \\
IC    750       & G & IC   4221       & M & UGC  4623       & M & UGC  6517       & M & UGC  8597       & F \\
IC    758       & F & IC   4237       & M & UGC  4701       & G & UGC  6534       & F & UGC  8658       & G \\
IC    764       & G & IC   4351       & M & UGC  4714       & F & UGC  6575       & F & UGC  8733       & F \\
IC    769       & G & IC   4407       & F & UGC  4787       & F & UGC  6670       & F & UGC  8795       & G \\
IC    776       & F & IC   4468       & G & UGC  4824       & M & UGC  6713       & F & UGC  8851       & F \\
IC    797       & F & IC   4536       & F & UGC  4841       & M & UGC  6773       & F & UGC  8877       & F \\
IC    800       & F & IC   4901       & M & UGC  4845       & G & UGC  6816       & F & UGC  8892       & F \\
IC    851       & F & IC   4986       & F & UGC  4867       & G & UGC  6818       & F & UGC  8909       & F \\
IC    863       & M & IC   5007       & F & UGC  4871       & F & UGC  6849       & F & UGC  8995       & F \\
IC   1014       & F & IC   5039       & M & UGC  4922       & M & UGC  6879       & F & UGC  9126       & F \\
IC   1055       & M & IC   5069       & M & UGC  4982       & M & UGC  6900       & F & UGC  9215       & F \\
IC   1066       & F & IC   5078       & M & UGC  5114       & F & UGC  6903       & G & UGC  9274       & F \\
IC   1067       & M & IC   5156       & M & UGC  5172       & F & UGC  6917       & M & UGC  9291       & M \\
IC   1125       & F & IC   5201       & M & UGC  5228       & M & UGC  6930       & F & UGC  9299       & F \\
IC   1151       & G & IC   5267       & M & UGC  5238       & F & UGC  6983       & M & UGC  9310       & F \\
IC   1158       & M & IC   5269A      & F & UGC  5358       & G & UGC  7009       & F & UGC  9569       & F \\
IC   1251       & F & IC   5269B      & M & UGC  5391       & F & UGC  7133       & F & UGC  9601       & F \\
IC   1447       & M & IC   5269C      & F & UGC  5478       & F & UGC  7143       & M & UGC  9661       & F \\
IC   1532       & F & IC   5270       & F & UGC  5522       & F & UGC  7184       & M & UGC  9663       & F \\
IC   1870       & F & IC   5271       & M & UGC  5612       & F & UGC  7189       & F & UGC  9730       & G \\
IC   1892       & F & IC   5273       & M & UGC  5646       & M & UGC  7239       & F & UGC  9746       & G \\
IC   1914       & F & IC   5321       & F & UGC  5676       & F & UGC  7271       & F & UGC  9837       & M \\
IC   1933       & F & IC   5325       & M & UGC  5688       & F & UGC  7590       & F & UGC  9858       & G \\
IC   1952       & G & IC   5332       & M & UGC  5695       & F & UGC  7690       & F & UGC  9875       & F \\
IC   1953       & F & UGC    99       & F & UGC  5707       & F & UGC  7699       & F & UGC  9936       & F \\
IC   1954       & M & UGC   313       & G & UGC  5740       & F & UGC  7700       & F & UGC  9951       & F \\
IC   1986       & F & UGC   891       & F & UGC  5832       & F & UGC  7848       & M & UGC 10020       & F \\
IC   1993       & M & UGC  1014       & F & UGC  5841       & M & UGC  7911       & F & UGC 10041       & F \\
 \enddata
 \end{deluxetable*}
 %\clearpage
 \begin{deluxetable*}{lclclclclc}
 \tablewidth{0pc}
 \tablenum{9 (cont.)}
 \tablecaption{Arm Classifications
 for 1114 S$^4$G Galaxies (cont.)\tablenotemark{a}
 }
 \tablehead{
 \colhead{Galaxy} &
 \colhead{AC} & 
 \colhead{Galaxy} & 
 \colhead{AC} & 
 \colhead{Galaxy} & 
 \colhead{AC} & 
 \colhead{Galaxy} & 
 \colhead{AC} & 
 \colhead{Galaxy} & 
 \colhead{AC}  
 \\
 \colhead{1} &
 \colhead{2} &
 \colhead{3} &
 \colhead{4} &
 \colhead{5} &
 \colhead{6} &
 \colhead{7} &
 \colhead{8} &
 \colhead{9} &
 \colhead{10} 
 }
 \startdata
UGC 10054       & F & ESO  287- 37    & F & ESO  441- 17    & F & ESO  550- 24    & F & PGC 31979       & F \\
UGC 10290       & F & ESO  288- 13    & F & ESO  442- 13    & G & ESO  551- 31    & F & PGC 32091       & M \\
UGC 10310       & F & ESO  289- 26    & G & ESO  443- 69    & M & ESO  572- 18    & F & PGC 35705       & F \\
UGC 10413       & G & ESO  298- 15    & F & ESO  443- 80    & F & ESO  572- 30    & F & PGC 36217       & F \\
UGC 10437       & F & ESO  298- 23    & F & ESO  443- 85    & M & ESO  576-  1    & G & PGC 36274       & F \\
UGC 10445       & F & ESO  300- 14    & F & ESO  445- 89    & F & ESO  576-  3    & M & PGC 36551       & F \\
UGC 10721       & M & ESO  305-  9    & F & ESO  479-  4    & F & ESO  576- 17    & F & PGC 38250       & F \\
UGC 10736       & F & ESO  340- 42    & F & ESO  480- 25    & F & ESO  576- 32    & F & PGC 41965       & F \\
UGC 10791       & F & ESO  341- 32    & F & ESO  481- 18    & M & ESO  576- 50    & M & PGC 42868       & F \\
UGC 10803       & M & ESO  342- 13    & F & ESO  482- 35    & M & ESO  576- 59    & F & PGC 43020       & F \\
UGC 10806       & F & ESO  342- 50    & M & ESO  485- 21    & G & ESO  580- 22    & F & PGC 43345       & M \\
UGC 10854       & F & ESO  345- 46    & F & ESO  502- 16    & F & ESO  580- 30    & M & PGC 43458       & F \\
UGC 12151       & F & ESO  347-  8    & F & ESO  502- 20    & F & ESO  580- 41    & F & PGC 44735       & F \\
UGC 12178       & F & ESO  355- 26    & M & ESO  503- 22    & F & ESO  601- 31    & F & PGC 44952       & F \\
UGC 12681       & F & ESO  357- 12    & F & ESO  504- 10    & F & ESO  602- 30    & F & PGC 44954       & G \\
UGC 12682       & F & ESO  358-  5    & F & ESO  504- 24    & F & PGC  2492       & F & PGC 45195       & F \\
UGC 12707       & F & ESO  358- 15    & F & ESO  504- 28    & F & PGC  3853       & M & PGC 45257       & F \\
UGC 12709       & F & ESO  358- 20    & F & ESO  506- 29    & F & PGC  6228       & M & PGC 45650       & M \\
UGC 12732       & F & ESO  358- 54    & F & ESO  507- 65    & F & PGC  6244       & F & PGC 45824       & F \\
UGC 12846       & F & ESO  362-  9    & F & ESO  508-  7    & F & PGC  6626       & F & PGC 45877       & M \\
UGC 12856       & F & ESO  400- 25    & F & ESO  508- 11    & F & PGC  6667       & F & PGC 45958       & M \\
ESO   12- 10    & F & ESO  402- 26    & G & ESO  508- 19    & F & PGC  6667       & F & PGC 47721       & M \\
ESO   13- 16    & F & ESO  403- 24    & F & ESO  508- 24    & G & PGC  6898       & F & PGC 48087       & M \\
ESO   26-  1    & M & ESO  403- 31    & F & ESO  508- 51    & F & PGC  7900       & F & PGC 48179       & F \\
ESO   27-  1    & M & ESO  404-  3    & M & ESO  509- 26    & G & PGC  8295       & F & PGC 49521       & F \\
ESO   27-  8    & G & ESO  404- 12    & M & ESO  509- 74    & M & PGC  9559       & F & PGC 51523       & F \\
ESO   54- 21    & F & ESO  404- 27    & G & ESO  510- 58    & F & PGC 11248       & G & PGC 52460       & F \\
ESO   79-  5    & F & ESO  407-  9    & M & ESO  510- 59    & G & PGC 11367       & M & PGC 52853       & F \\
ESO   79-  7    & F & ESO  407- 14    & M & ESO  532- 22    & F & PGC 11744       & F & PGC 53093       & M \\
ESO   85- 14    & F & ESO  408- 12    & F & ESO  533- 28    & M & PGC 12068       & F & PGC 53134       & F \\
ESO   85- 30    & F & ESO  410- 18    & F & ESO  539-  7    & F & PGC 12608       & F & PGC 53568       & F \\
ESO  116- 12    & F & ESO  411- 26    & F & ESO  541-  4    & M & PGC 12633       & M & PGC 53779       & F \\
ESO  145- 25    & F & ESO  418-  8    & F & ESO  541-  5    & F & PGC 12664       & F & PGC 53796       & G \\
ESO  150-  5    & F & ESO  418-  9    & F & ESO  544- 30    & F & PGC 12981       & F & PGC 54944       & F \\
ESO  187- 35    & F & ESO  420-  9    & F & ESO  545-  2    & F & PGC 13684       & F & PGC 66242       & F \\
ESO  187- 51    & F & ESO  421- 19    & F & ESO  545- 16    & F & PGC 13716       & M & PGC 66559       & F \\
ESO  202- 41    & F & ESO  422-  5    & F & ESO  547-  5    & F & PGC 14487       & F & PGC 68061       & F \\
ESO  234- 43    & F & ESO  422- 41    & F & ESO  548-  5    & F & PGC 15869       & F & PGC 68771       & F \\
ESO  234- 49    & F & ESO  423-  2    & F & ESO  548- 25    & G & PGC 16784       & F & PGC 69293       & F \\
ESO  237- 52    & F & ESO  438- 17    & M & ESO  548- 32    & F & PGC 27616       & F & PGC 69448       & M \\
ESO  238- 18    & F & ESO  440-  4    & F & ESO  548- 82    & F & PGC 27810       & M & PGC 72252       & F \\
ESO  248-  2    & F & ESO  440- 11    & M & ESO  549- 18    & G & PGC 27833       & F & PGC1063216      & F \\
ESO  285- 48    & F & ESO  440- 46    & F & ESO  549- 35    & F & PGC 28380       & G &  &  \\
\enddata
\tablenotetext{a}{Arm classes: F=flocculent, M=multi-armed, G=grand design; all
classifications are due to D. Elmegreen}
\end{deluxetable*}

\subsubsection{Extreme Spirals}

Arm classifications are a useful way of distinguishing different
classes of spiral patterns, but like any aspect of galaxy structure,
there are extremes even among these classes.  In a large survey of
galaxy images, there is a high probability of finding unusually
strong examples of some aspects of galaxy morphology. One is the
extremely oval inner rings recognized by Buta (2014) in the EFIGI
SDSS optical galaxy sample (Baillard et al. 2011). These are SB
rings which have an intrinsic minor-to-major axis ratio of $\leq$0.5
compared to the average of 0.81$\pm$0.06 (Buta 1995). Here we bring
attention to ``extreme spirals."

An ``extreme spiral" is a grand design spiral having apparently
large amplitude spiral arms of high pitch angle, and often showing
so little background disk light that it is difficult to determine
the orientation parameters of the system. Given the depth of S$^4$G
3.6$\mu$m images, such systems are worthy of further investigation
to determine how they form and how their properties differ from
more normal spirals; most may be tidal in origin. We do not have a 
quantitative formal definition of such galaxies. They are merely
recognized as being unusually strong and open compared to more
average grand-design spirals.

Figure~\ref{extspirals} shows six prominent examples from the S$^4$G
database:

NGC 3187 - This galaxy is a member of a small group including NGC 3185,
3190, and 3193. One arm remains perpendicular to the apparent bar to a
very large radius. The arms could be driven by an interaction, or,
more likely, are due to the bar but perturbed by an interaction.

NGC 4488 - This Virgo Cluster galaxy has a strong and highly unusual
boxy bar-like zone from which two smooth and open arms emerge. It is
not clear whether the arms and the apparent bar are coplanar features.
Graham et al. (2012) have also noted a ``bow-tie" aspect of the boxy
zone of NGC 4488, and suggest the apparent arms may be due to an
interaction.

NGC 4572 - This galaxy is a strong candidate for an extremely warped
disk with considerable extraplanar material. The galaxy has a small
nearby companion, and also has a similar redshift to that of a bright
nearby elliptical galaxy, NGC 4589. Alternatively, it could be simply
a strong tidal spiral, due to its possible local interaction. The
Table 6 classification, S$\underline{\rm c}$d sp-superw?, brings
attention to the warped disk possibility. 

NGC 4731 - This appears to be an extreme late-type barred spiral. The
bar is thin and likely is also flat (see also Figure~\ref{lt-bars}). The
spiral arms, however, are very open and show a strong right angle turn
at large radius. The behavior is similar to NGC 4572, although we are
not suggesting that NGC 4731 is another case of extreme warping.

IC 167 - an exceptional case with a short bar and very strong spiral
structure. At the NED distance of 41.6 Mpc, the diameter of the galaxy
is $\approx$38 kpc.

UGC 9057 - An extreme late-type spiral with very open spiral arms, but
not an extreme case of size.

\begin{figure}
\figurenum{30}
\plotone{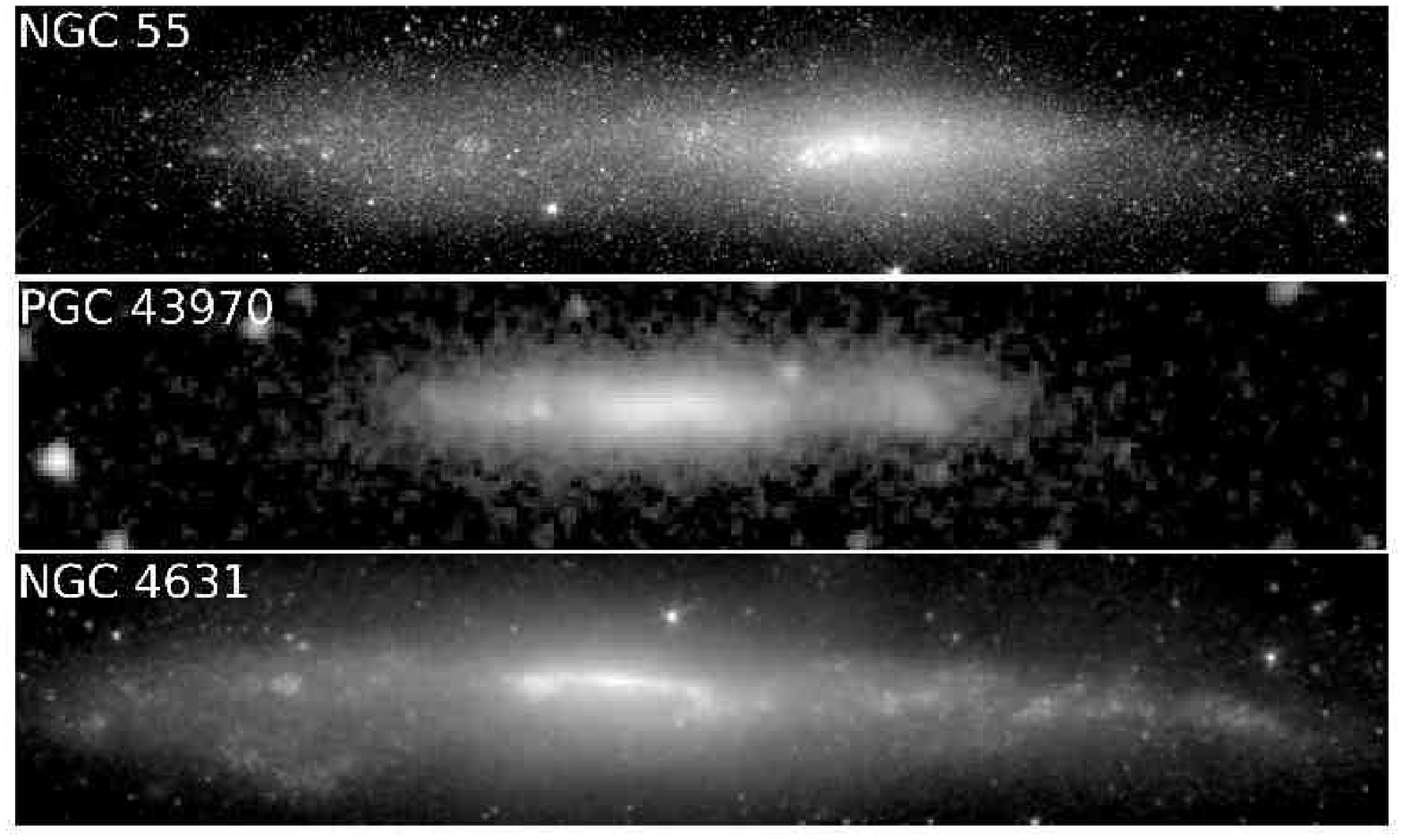}
\caption{({\it top two panels}) Two highly-inclined late-type barred
spirals showing the characteristic asymmetry of stage Sm (Magellanic).
{\it bottom} de Vaucouleurs used the appearance of NGC 4631 in blue
light to deduce that it was a spiral of type SB(s)d, even though the
bar was not detectable in that band. At 3.6$\mu$m, the bar is
prominent.}
\label{ngc4631}
\end{figure}

\subsection{Recognizing edge-on Magellanic Barred Spirals and
the Special Case of NGC 4631}

The classification of NGC 4631 (de Vaucouleurs \& de Vaucouleurs 1964)
sparked an interesting historical disagreement between Gerard de
Vaucouleurs and Geoffrey Burbidge. De Vaucouleurs was well-known for
classifying some edge-on galaxies with what appeared to be
overly-detailed types given what could actually be discerned in an
image. The appearance of NGC 4631 led de Vaucouleurs to classify it as
type SB(s)d, but the ``B" and the ``(s)" were nothing more than
inferences based on the asymmetric shape. Burbidge, Burbidge, \&
Prendergast (1964) argued that no more than a type of Sc could be
justified. Now, however, as we show in Figure~\ref{ngc4631}, there is
very likely a bar in NGC 4631 that is vertically thin. The feature
was not clearly distinguishable in blue light. In Table 6, the
adopted mid-IR  classification is SB(s)d sp pec, confirming de
Vaucouleurs's original interpretation.

The morphological aspects de Vaucouleurs used to assess the type of NGC
4631 can be used for other galaxies. For example, NGC 55
(Figure~\ref{ngc4631}, top frame) is an almost edge-on late-type spiral
classified as type SB(s)m sp in RC3. The galaxy is almost exactly
edge-on, yet there is little doubt as to its classification.  The
brightness enhancement on the west side is likely to be a bar, and the
asymmetric extension from this area is likely to be the single main
spiral arm that is characteristic of type SB(s)m (de Vaucouleurs \&
Freeman 1972). (The mean mid-IR classification for the galaxy in Table
6 is SA$\underline{\rm B}$(s)m sp.) In Figure~\ref{ngc4631}, PGC 43970
is another example of an edge-on galaxy that is likely to be a highly
inclined SB(s)m Magellanic spiral. Many SAB/SBdm and SAB/SBm S$^4$G
galaxies have been recognized in this manner. When recognized in the
edge-on view, there is an indication that Sm,Im galaxies may be less
flattened than, for example, Sd galaxies. The S$^4$G provides an
excellent database for re-investigating the intrinsic flattenings of
galaxies of different types.

\section{Summary}

The S$^4$G has provided a remarkable and valuable view of galaxy
morphology in the mid-IR. Our main results from this study are as
follows:

\noindent
1. Using the precepts of CVRHS morphology, we give detailed
classifications for more than 2400 nearby galaxies in the mid-IR.
Together with paper I, this provides the first such examination of
its kind in this wavelength domain. The classifications are based on
two independent examinations of logarithmic, sky-subtracted images in
units of mag arcsec$^{-2}$. These are numerically combined to provide
final mean classifications. Together with the paper I study, we also
provide an atlas of images of all of the galaxies classified.

\noindent
2. The dust-penetrated nature of mid-IR morphology allows us to see the
full complement of stellar structures in galaxies, as per the original
goal of the survey (Sheth et al.  2010). The main structures of CVRHS
morphology are summarized in Table 1, and many are illustrated either
in the montages or the atlas images accompanying this article. Other
structures are described in the host of papers referenced in Section
1.

\noindent
3. Figures~\ref{comp1} - ~\ref{comp2} show that that CVRHS
classifications of S$^4$G galaxies have both good internal and external
consistency. This consistency, however, does not preclude the existence
of resolution and especially inclination biases in the catalog,
independent of the sample bias against gas-poor galaxies. We have shown
that the high-inclination half of the sample has higher frequencies of
later type and barred galaxies than does the lower-inclination half.
For histograms and estimates of fractions of types, families, and
varieties, we have adopted an inclination limit of
$\approx$60$^{\circ}$, based on the measured 3.6$\mu$m isophotal axis
ratio at $\mu_{AB}$ = 25.5 mag arcsec$^{-2}$.

\noindent
4. Within this inclination limit, 48.5$\pm$1.4\% of the galaxies in the S$^4$G
sample are what we call ``extreme late-type galaxies," meaning galaxies
in the type range Scd to Im. Scd-Sdm galaxies are essentially pure disk
galaxies while some Sm and Im galaxies are also likely flat. The
emphasis on these galaxies stems partly from the volume-limited nature
of the survey, and partly on the fact that an HI radial velocity was
used as a selection criterion.

\noindent
5. Also within the adopted inclination limit, the barred CVRHS
classification fraction, $f (F\geq 0.5)$, in the S$^4$G depends on
mid-IR type. For spiral galaxies in the mid-IR type range Scd-Sm, the
fraction is $f (F\geq 0.5)$ = 81.0$\pm$2.0\%, while over the type range
S0/a to Sc, the fraction is $f (F\geq 0.5)$ = 55.0$\pm$ 2.2\%  When
examined further on a type by type basis, it appears that much of the
drop for the S0/a to Sc subsample occurs for stages S$\underline{\rm
b}$c-Sc. This drop does not appear to be present when $B$-band RC3
classifications are used for the same galaxies, and may be linked in
part to the ``earlier effect" in mid-IR classification where S0/a
to Sc galaxies in blue light shift by $\approx$ one stage interval
on average when classified in the IR, while later $B$-band types do
not.

The bars in mid-IR S0/a to Sc galaxies are typically not the same kinds
of features as those in Scd-Sm galaxies. It is mainly the earlier type
galaxies that show bars having 3D box/peanut/X patterns, stellar ansae,
or ``barlenses." The bars in the later type spirals are often linear
chains of star-forming regions that are both azimuthally and vertically
thin. This dichotomy in bar character has been known for a long time
(e.g. Elmegreen \& Elmegreen 1985), and because the distinction is
important for estimates of the cosmologically significant bar fraction
(Sheth et al.  2008, 2014a) we defer the main discussion of the actual
S$^4$G bar fraction (as opposed to the visual barred family
classification fraction) to Sheth et al. (2014b).

\noindent
6. (s)-variety spirals average at a much later mid-IR stage (Scd) than
do ring-lens variety spirals (S0$^+$ to Sa). Outer pseudorings are seen
on average in Sb-Sbc galaxies compared to S0$^+$ to S0/a for outer
rings and lenses. The inner ring-lens fraction drops rapidly with
advancing mid-IR stage.

\noindent
7. Sph galaxies are well-represented in the S$^4$G, although not all
may have been recognized as such because these galaxies are
best-defined using parameter correlations. Dwarf irregulars can show a
regular elliptical background at 3.6$\mu$m, and some early-type disk
galaxies are actually bulgeless. These galaxies have been found by
Kormendy \& Bender (2012) to have a natural ``home' in a parallel
sequence among extreme late-type galaxies. S$^4$G morphology supports
this interpretation.

\noindent
8. Grand design patterns are found in less than 20\% of the
classifiable spiral galaxies in the S$^4$G catalogue. The most common
arm type is flocculent, owing to the sample emphasis on Scd-Sm stages.

\noindent
9. The S$^4$G sample includes many galaxies and morphological
features worthy of follow-up studies. By ``follow-up studies," we
mean more detailed investigations of structure, such as using mass
maps to approximate gravitational potentials, obtaining kinematic
data, and correlating mid-IR structure with galaxy structure in
optical and ultraviolet wavebands. Many follow-up studies have been
made by the S$^4$G team as listed in Section 1. We have highlighted
special cases of interest, and have in particular emphasized the
numerous examples of apparently embedded disks in 3D early-type
systems.

We thank the anonymous referee for many helpful comments that greatly
improved this paper. RB acknowledges the support of a grant from the
Research Grants Committee of the University of Alabama. KS acknowledges
the support of the National Radio Astronomy Observatory. The National
Radio Astronomy Observatory is a facility of the National Science
Foundation operated under cooperative agreement by Associated
Universities, Inc. EA and AB acknowledge financial support to the DAGAL
network from the People Programme (Marie Curie Actions) of the European
Union's Seventh Framework Programme FP7/2007-2013/ under REA grant
agreement number PITN-GA-2011-289313. EA and AB also acknowledge
financial support from the CNES (Centre National d'Etudes Spatiales -
France). LCH acknowledges support from the Kavli Foundation, Peking
University, and the Chinese Academy of Science through grant No.
XDB09030100 (Emergence of Cosmological Structures) from the Strategic
Priority Research Program.  This research has made use of the NASA/IPAC
Extragalactic Database (NED), which is operated by the Jet Propulsion
Laboratory, California Institute of Technology, under contract with the
National Aeronautics and Space Administration.

\centerline{Appendix: Partial Phase 3 Analysis}

\begin{figure}
\figurenum{31}
\plotone{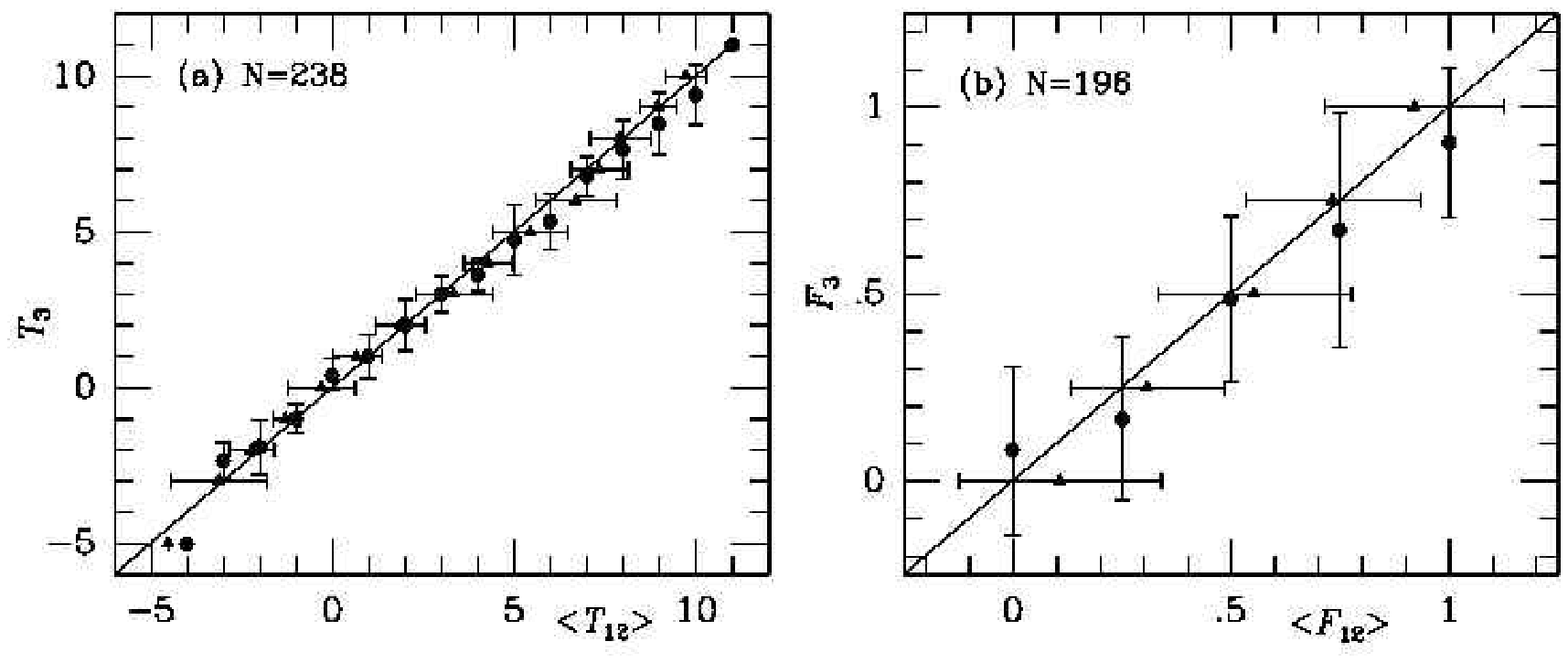}
\caption{}
\label{phase3}
\end{figure}

The referee has suggested that we further assess the significance of
the underline classifications in the mean stage, family, and variety
classifications in Table 6 by carrying out a partial Phase 3 analysis
for a random subset of 10\% of the S$^4$G sample. For this purpose, we
reclassified every 10th galaxy in Table 6; the results are collected in
Table 10. Figure~\ref{phase3} compares these classifications with those
in Table 6. For all four comparisons, the results are very similar to
those found between the Phase 1 and 2 classifications
(Figures~\ref{comp1} and ~\ref{ivov}). 

 \begin{deluxetable*}{llll}
 \tablewidth{0pc}
 \tablenum{10}
 \tablecaption{Phase 3 CVRHS Classifications
 for 241 S$^4$G Galaxies\tablenotemark{a}
 }
 \tablehead{
 \colhead{Galaxy} &
 \colhead{Type} & 
 \colhead{Galaxy} & 
 \colhead{Type} 
 \\
 \colhead{1} &
 \colhead{2} &
 \colhead{3} &
 \colhead{4} 
 }
 \startdata
NGC    14       & (L)IAB(s)m                                                                                 & UGC  3070      &S$\underline{\rm A}$B(s)dm                                                                 \\
UGC   156       & SAB(s)m                                                                                    & PGC 15869      &SB(r$^{\prime}$l,s)d                                                                       \\
NGC   131       & SAB(r$\underline{\rm s}$)dm                                                                & NGC  1703      &SA(s,nl)c                                                                                  \\
NGC   157       & SA(s)bc                                                                                    & ESO  486- 21   &IAm / Sph                                                                                  \\
PGC  2689       & Im                                                                                         & ESO  422- 41   &S$\underline{\rm A}$B(l)d                                                                  \\
ESO   79-  7    & SB(rs)d                                                                                    & IC   2135      &Scd sp / E8                                                                                \\
ESO  541-  5    & SAB(s)m                                                                                    & UGC  4148      &SAB(s)m                                                                                    \\
UGC   711       & Scd sp / E(d)9                                                                             & UGC  4306      &SA(r)0$^+$                                                                                 \\
UGC   891       & SAB(s)m                                                                                    & NGC  2604      &SB(r$\underline{\rm s}$)cd                                                                 \\
UGC   964       & SAB(s)d sp / E(d)6                                                                         & UGC  4551      &SA(r)0$^+$                                                                                 \\
UGC  1020       & SAB$_a$0$^{\circ}$:                                                                        & NGC  2683      &(RL)SA$\underline{\rm B}$$_{ax}$($\underline {\rm r}$s)a                                   \\
UGC  1133       & dIm                                                                                        & NGC  2712      &(R$^{\prime}$)SAB$_a$(s,nl)b                                                               \\
NGC   658       & SA(rs)b                                                                                    & NGC  2726      &(RL)SA(rs,rs)a                                                                             \\
NGC   672       & (R$^{\prime}$)SB(s)d                                                                       & NGC  2775      &SA(l,$\underline {\rm r}$s)0/a                                                             \\
NGC   680       & SA(l)0$^{\circ}$ pec                                                                       & UGC  4871      &SAB(s)dm                                                                                   \\
NGC   723       & SA(s)c                                                                                     & NGC  2787      &SA$\underline{\rm B}$$_a$($\underline {\rm r}$s,bl)0$^+$                                   \\
PGC  7682       & Sdm sp / E(d)6                                                                             & UGC  4988      &SAB(s)m                                                                                    \\
UGC  1670       & S$\underline{\rm A}$B(s)m                                                                  & NGC  2894      &SA(s:)0/a                                                                                  \\
ESO  545-  3    & Sc sp                                                                                      & UGC  5139      &dIm                                                                                        \\
UGC  1862       & SA(s)d                                                                                     & UGC  5179      &SA0$^{\circ}$                                                                              \\
UGC  1981       & dIm                                                                                        & NGC  2976      &SA(s)d / Sph                                                                               \\
UGC  2082       & Scd sp / E(d)8                                                                             & NGC  3024      &SB(s)m sp                                                                                  \\
NGC  1051       & S$\underline{\rm A}$B(s)dm                                                                 & NGC  3027      &SB(s)dm                                                                                    \\
NGC  1097       & (R$^{\prime}$)SB($\underline {\rm r}$s,nr)ab pec                                           & UGC  5354      &(R$^{\prime}$)SAB(s)d                                                                      \\
ESO  356- 18    & SAB(s)m sp                                                                                 & UGC  5401      &SA$\underline{\rm B}$(s)m                                                                  \\
NGC  1187       & SA$\underline{\rm B}$(rs)bc                                                                & NGC  3057      &SB(r$\underline{\rm s}$)d                                                                  \\
ESO  300- 14    & S$\underline{\rm A}$B(r$\underline{\rm s}$)cd                                              & UGC  5478      &SA$\underline{\rm B}$(s)dm                                                                 \\
NGC  1258       & (R$^{\prime}$R$^{\prime}$)SAB(s)b pec                                                      & NGC  3177      &SA(rs,r$\underline{\rm s}$)b                                                               \\
PGC 12439       & Sc sp / E(d)8                                                                              & NGC  3182      &SA(r)0$^+$                                                                                 \\
PGC 12608       & SAB(s)dm                                                                                   & UGC  5612      &(R$^{\prime}$)SB(r$\underline{\rm s}$)m                                                    \\
NGC  1316C      & SAB$_a$(s)0$^+$                                                                            & NGC  3248      &(RL)SA(l)0$^{\circ}$                                                                       \\
NGC  1341       & (L)SB(rs)dm / Sph                                                                          & UGC  5708      &SB(s)d sp                                                                                  \\
NGC  1347       & SAB$_a$(rs)d                                                                               & NGC  3252      &Sc sp / E7                                                                                 \\
NGC  1357       & (R$^{\prime}$L)SA(r$^{\prime}$l,$\underline {\rm r}$s)0/a                                  & NGC  3321      &SA$\underline{\rm B}$(rs)cd                                                                \\
IC   1962       & SAB(s)m sp                                                                                 & UGC  5832      &(R$^{\prime}$)SB(s)m                                                                       \\
NGC  1398       & (R$^{\prime}$R)SB($\underline {\rm r}$s,bl)a                                               & UGC  5841      &SB(s)ab                                                                                    \\
NGC  1433       & (R$_1^{\prime}$)SB(p,r,bl,nr,nb)a                                                          & UGC  5898      &SAB(s)dm sp                                                                                \\
ESO   15-  1    & SB(s)m                                                                                     & NGC  3395      &SAB(s)c pec                                                                                \\
PGC 90694       & SBd sp / E8                                                                                & NGC  3424      &SA$\underline{\rm B}$$_x$(l)a                                                              \\
IC   2051       & SAB($\underline {\rm r}$s,nd)b                                                             & NGC  3447      &SA$\underline{\rm B}$(s)dm pec                                                             \\
ESO  549- 35    & SAB(s)d                                                                                    & NGC  3468      &SA(l)0$^{\circ}$ pec                                                                       \\
NGC  1511A      & SAc sp / E(d)6-7                                                                           & UGC  6112      &S$\underline{\rm A}$B(s)d                                                                  \\
NGC  1518       & SB(s)dm                                                                                    & ESO  502- 16   &SAB(s)m                                                                                    \\
PGC 14626       & SB(s)dm                                                                                    & ESO  502- 20   &S$\underline{\rm A}$B(s)cd                                                                 \\
IC   2056       & SA(s)b                                                                                     & UGC  6271      &SA(l)0$^+$ [c]                                                                             \\
 \enddata
 \end{deluxetable*}
 %\clearpage
 \begin{deluxetable*}{llll}
 \tablewidth{0pc}
 \tablenum{10 (cont.)}
 \tablecaption{Phase 3 CVRHS Classifications
 for the S$^4$G Sample (cont.) \tablenotemark{a}}
 \tablehead{
 \colhead{Galaxy} &
 \colhead{Type} & 
 \colhead{Galaxy} & 
 \colhead{Type} 
 \\
 \colhead{1} &
 \colhead{2} &
 \colhead{3} &
 \colhead{4} 
 }
 \startdata
UGC  6309       & (R$_{12}^{\prime}$)SB(s)bc                                                                 & NGC  4411B     &SA(s)c                                                                                     \\
UGC  6345       & SB(s)dm                                                                                    & UGC  7557      &SAB(r$\underline{\rm s}$)d                                                                 \\
UGC  6390       & Sd sp                                                                                      & UGC  7579      &SB(s)m sp / Sph                                                                            \\
NGC  3666       & SA(s)bc                                                                                    & UGC  7596      &dIm (dE) / Sph                                                                             \\
IC   2764       & SA(r,nl)0$^{\circ}$                                                                        & NGC  4461      &(R$^{\prime}$L)SA($\underline {\rm r}$s,l)0$^+$                                            \\
NGC  3701       & SA($\underline {\rm r}$s)bc                                                                & NGC  4489      &SA(rl,nl)0$^{\circ}$                                                                       \\
UGC  6534       & SA(s)cd                                                                                    & NGC  4502      &SA$\underline{\rm B}$$_a$(s)cd                                                             \\
UGC  6570       & SA(l)0$^-$ + PR disk?                                                                      & IC   3474      &Sdm sp                                                                                     \\
NGC  3773       & SA0$^-$ pec                                                                                & NGC  4523      &SA$\underline{\rm B}$(s)m                                                                  \\
NGC  3795       & SA(s)cd sp                                                                                 & NGC  4535      &SAB(s)c                                                                                    \\
UGC  6682       & SA$\underline{\rm B}$(s)m                                                                  & NGC  4548      &(R$^{\prime}$)SB(rs,bl)a                                                                   \\
NGC  3870       & SB$_a$(r$^{\prime}$l)0$^{\circ}$ [cd]                                                      & UGC  7774      &SB(s)dm sp                                                                                 \\
NGC  3893       & SA(s)c                                                                                     & ESO  442- 13   &SAB(s)cd                                                                                   \\
IC   2963       & Sdm sp / E(d)7                                                                             & ESO  506- 29   &SB(s)dm                                                                                    \\
NGC  3930       & SAB(r$\underline{\rm s}$)cd                                                                & UGC  7848      &SAB$_a$(rs)cd                                                                              \\
ESO  440- 27    & Sc sp / E(d)7-8                                                                            & NGC  4625      &SAB(r$\underline{\rm s}$)m                                                                 \\
PGC 37373       & SAB(s)cd                                                                                   & NGC  4636      &E2                                                                                         \\
NGC  3981       & SAB(s)bc pec                                                                               & NGC  4654      &SB(r$\underline{\rm s}$)cd                                                                 \\
UGC  6956       & SA$\underline{\rm B}$(s)m                                                                  & IC   3742      &SB(s)m                                                                                     \\
UGC  6978       & SB(s)m                                                                                     & NGC  4684      &SA(l)0$^-$ sp                                                                              \\
NGC  4037       & SA$\underline{\rm B}$(rs)b                                                                 & NGC  4700      &SB(s)m sp                                                                                  \\
NGC  4051       & SAB(rs,AGN)bc                                                                              & UGC  7991      &Sd spw / E(d)8                                                                             \\
NGC  4067       & SB(rs)b                                                                                    & NGC  4753      &ETG pec / S0$^{\circ}$ pec                                                                 \\
IC   2996       & SA(rs)bc                                                                                   & NGC  4779      &SA$\underline{\rm B}$(rs)b                                                                 \\
NGC  4108       & (R$^{\prime}$)S$\underline{\rm A}$B(rs)bc                                                  & NGC  4810      &IBm                                                                                        \\
UGC  7125       & SA$\underline{\rm B}$(s)m sp                                                               & NGC  4808      &SAB(rs)cd                                                                                  \\
ESO  505- 23    & SB(s)d sp                                                                                  & PGC 44358      &Sd sp                                                                                      \\
UGC  7170       & Sd spw / E(d)8-9                                                                           & PGC 44735      &SAB(r$\underline{\rm s}$)dm                                                                \\
UGC  7184       & SB(rs)dm                                                                                   & PGC 44952      &SA(s)cd                                                                                    \\
IC    769       & (R$^{\prime}$)SAB(s)bc                                                                     & NGC  4941      &(RL)SA($\underline {\rm r}$s)a                                                             \\
UGC  7242       & dIm                                                                                        & IC   4182      &SA(s)m                                                                                     \\
IC   3062       & SA(s)bc                                                                                    & NGC  4984      &(R$^{\prime}$,R)SAB$_a$(l,bl,nl)a                                                          \\
NGC  4216       & (R$_2^{\prime}$)SA$\underline{\rm B}$$_{xa}$(r,nl)ab sp                                    & ESO  576-  3   &SB(r$\underline{\rm s}$)dm                                                                 \\
NGC  4235       & SAB$_x$(0/a sp                                                                             & NGC  5016      &S$\underline{\rm A}$B($\underline {\rm r}$s)c                                              \\
NGC  4248       & dIm sp / Sph pec                                                                           & PGC 45958      &SA(s)c                                                                                     \\
NGC  4266       & S0$^{\circ}$ sp / E6                                                                       & NGC  5054      &(R$^{\prime}$)SA(s,nr)bc                                                                   \\
NGC  4276       & SA$\underline{\rm B}$(r$\underline{\rm s}$)d                                               & NGC  5073      &SB$_x$(s)0/a sp                                                                            \\
NGC  4286       & dSA(l)0$^{\circ}$,N / Sph                                                                  & NGC  5088      &(R$^{\prime}$)SA(s)cd                                                                      \\
NGC  4303       & SAB(rs,nl)bc                                                                               & NGC  5117      &SA$\underline{\rm B}$(s)cd                                                                 \\
NGC  4316       & S(r:)b: sp / E(d)7                                                                         & NGC  5147      &SAB(s)c                                                                                    \\
NGC  4344       & SA(r)0$^+$ [c]                                                                             & NGC  5170      &(R$^{\prime}$)SAB$_x$(l)a sp / E(d)8-9                                                     \\
UGC  7490       & SA(s)dm                                                                                    & PGC 47721      &(R$^{\prime}$,R)SA(rs)ab                                                                   \\
IC   3298       & SB(s)m sp                                                                                  & ESO  444- 78   &dIm                                                                                        \\
NGC  4389       & SB(rs)ab [d]                                                                               & UGC  8642      &SB(s)m sp                                                                                  \\
NGC  4396       & SA(s:)c sp                                                                                 & UGC  8688      &Im                                                                                         \\
 \enddata
 \end{deluxetable*}
 %\clearpage
 \begin{deluxetable*}{llll}
 \tablewidth{0pc}
 \tablenum{10 (cont.)}
 \tablecaption{Phase 3 CVRHS Classifications
 for the S$^4$G Sample (cont.) \tablenotemark{a}}
 \tablehead{
 \colhead{Galaxy} &
 \colhead{Type} & 
 \colhead{Galaxy} & 
 \colhead{Type} 
 \\
 \colhead{1} &
 \colhead{2} &
 \colhead{3} &
 \colhead{4} 
 }
 \startdata
ESO  577- 38    & dIm sp                                                                                     & NGC  6181      &S$\underline{\rm A}$B(rs)bc                                                                \\
NGC  5337       & (L)SB(rs)0/a                                                                               & NGC  6267      &SB(rs)b                                                                                    \\
NGC  5339       & SAB(rs,bl)ab                                                                               & NGC  6339      &SB(s)cd                                                                                    \\
NGC  5360       & Im sp / E6 / Sph                                                                           & IC   4901      &(R$^{\prime}$,R$^{\prime}$)SB($\underline {\rm r}$s)bc                                     \\
ESO  510- 26    & SAB(s)m                                                                                    & ESO  340- 17   &SB(s)d [0/a] pec                                                                           \\
NGC  5443       & (R$^{\prime}$)SA$\underline{\rm B}$$_x$($\underline {\rm r}$s)a                            & ESO  234- 49   &SA(r)cd                                                                                    \\
NGC  5474       & SA(s)m                                                                                     & ESO  341- 32   &SA$\underline{\rm B}$(r$\underline{\rm s}$)dm                                              \\
NGC  5476       & S$\underline{\rm A}$B(s)c                                                                  & ESO  402- 26   &(R$_1^{\prime}$R$_2^{\prime}$)S$\underline{\rm A}$B($\underline {\rm r}$s,nl)b             \\
NGC  5526       & Sc sp / E7-8                                                                               & ESO  287- 37   &SA$\underline{\rm B}$$_a$(s)dm                                                             \\
NGC  5560       & SB(s)d spw                                                                                 & ESO  288- 13   &SA$\underline{\rm B}$(s)dm                                                                 \\
UGC  9206       & Im sp / Sph                                                                                & NGC  7162      &S$\underline{\rm A}$B($\underline {\rm r}$s)b                                              \\
UGC  9245       & SB(r$\underline{\rm s}$)dm                                                                 & ESO  601-  7   &SB(s)d sp                                                                                  \\
UGC  9291       & SAB(rs)c                                                                                   & PGC 68061      &Sb sp pec / E7:                                                                            \\
NGC  5673       & SB(s)d sp                                                                                  & ESO  532- 32   &dIm sp                                                                                     \\
NGC  5668       & S$\underline{\rm A}$B(rs)cd                                                                & ESO  602-  3   &SB(s)m sp                                                                                  \\
NGC  5691       & SB(s)dm [0/a]                                                                              & PGC 68771      &SB(s)dm                                                                                    \\
NGC  5719       & (R$^{\prime}$)SA(l)0/a / E5-6                                                              & NGC  7314      &SAB(rs)bc                                                                                  \\
ESO  580- 22    & SB(s)dm                                                                                    & UGC 12151      &SB(rs)d                                                                                    \\
PGC 52853       & SA(s)cd                                                                                    & ESO  346-  1   &Sd sp / E(d)8                                                                              \\
IC   1066       & SA(s)b                                                                                     & IC   5267      &(R$^{\prime}$)SA($\underline {\rm r}$s,r$^{\prime}$l)0/a                                   \\
NGC  5798       & SA(s)d                                                                                     & IC   5269C     &SAB(s)dm pec                                                                               \\
UGC  9661       & SB(s)dm                                                                                    & ESO  469-  8   &SAB(s)m sp                                                                                 \\
PGC 91413       & SB(s)dm sp                                                                                 & NGC  7537      &SA($\underline {\rm r}$s)b                                                                 \\
NGC  5866B      & dSA(s)0/a [d] / Sph                                                                        & NGC  7625      &(L)SAB(s)0/a [dm]                                                                          \\
UGC  9815       & SB(s)dm sp                                                                                 & ESO  240-  4   &SB(s)d sp                                                                                  \\
NGC  5929       & E0 pec                                                                                     & IC   5334      &(R$^{\prime}$)SA(l)0/a / E(d)6-7                                                           \\
NGC  5963       & S$\underline{\rm A}$B[rs=(R$_2^{\prime}$)SAB(s)c]b                                         & ESO  605- 15   &dIm sp                                                                                     \\
NGC  5981       & S0$^+$ spw / E(d)7                                                                         & NGC  7731      &(R$_1$L,R$_2^{\prime}$)S$\underline{\rm A}$B$_a$(r,nl)a                                    \\
UGC  9991       & Sdm                                                                                        & NGC  7755      &(R$^{\prime}$)SA$\underline{\rm B}$(rs,nr)bc                                               \\
NGC  6010       & S0$^-$ sp / E(d)7                                                                          & NGC  7793      &SA(s)c                                                                                     \\
UGC 10290       & SB(s)dm                                                                                    &
......... & .......... \\
 \enddata
 \tablenotetext{a}{Subsample is selected as 
 every 10th galaxy from Table 6}
 \end{deluxetable*}

The partial Phase 3 analysis was used as follows. Table 5 gives
$\sigma_{<1,2>3} = \sigma (\Delta T_{<1,2>3}) = {\sqrt{{\Sigma
(T_3-<T>)^2} \over {N-1}}}$, where $T_3$ is the Phase 3 stage index and
$<T>$ is the mean of the Phase 1 and 2 stages in Table 6. The idea is
to derive $\sigma (T_3)$ and compare it with $\sigma (T)$ from the
Phase 1 and 2 analysis alone (Col. 2 of Table 5); we should find that
$\sigma(T_3)$ $\approx$ $\sigma (T)$ if the classifications are
consistent. For this purpose, we derive $\sigma(T_3) =
\sqrt{\sigma(\Delta T_{<1,2>3})^2 - \sigma(<T>)^2}$. This gives $\sigma
(T_3)$ = 0.69 of a stage interval, very similar to $\sigma (T)$ = 0.74
from the Phase 1 and 2 analysis alone.  This supports our Phase 1,2
estimate of $\sigma (<T>)$ = 0.5 stage interval and suggests that while
the half-step underline stage classifications are only marginally
significant, it is still worth retaining them.

The partial Phase 3 analysis also supports our Phase 1,2 assessment of
the significance of the underline family classifications S$\underline
{\rm A}$B and SA$\underline{\rm B}$. Setting $\Delta F_{<1,2>3} = F_3 -
<F>$, we derive $\sigma (F_3)$ = 0.79 of a family interval compared to
0.68 of a family interval for Phases 1 and 2 alone (Table 5). This
supports our Phase 1,2 estimate of $\sigma (<F>)$ = 0.5 family
interval, indicating that the CVRHS sequence SA, S$\underline {\rm
A}$B, SAB, SA$\underline {\rm B}$, and SB is more internally
significant than half-step stages.

For both inner and outer varieties, the internal standard deviations
for the partial Phase 3 analysis are similar to those of the mean Phase
1,2 classifications. Table 5 gives $\sigma (IV_3)$ = 1.19 variety
intervals compared to $\sigma (IV)$ = 0.82 of a variety interval from
Phases 1 and 2 alone. This is reasonable agreement considering the
size of the Phase 3 sample compared to the Phase 1,2 sample. As for
stages, the underline inner varieties also have marginal significance
on the whole, but are still worth retaining in Table 6 because of the
very good correlation for the standard inner varieties (s) to (r).

The same approach to outer varieties gives $\sigma (OV_3)$ = 1.29 outer
variety intervals, which agrees well with $\sigma (OV)$ = 1.27 from the
Phase 1,2 analysis in spite of the much smaller sample. Consistent
classification of outer features appears to be more difficult than for
stage, family, and inner variety, but nevertheless $\sigma (<OV>)$ =
0.90 (Table 5) supports the marginal significance of the multiple
categories used.

\begin{figure}
\figurenum{31 (cont.)}
\plotone{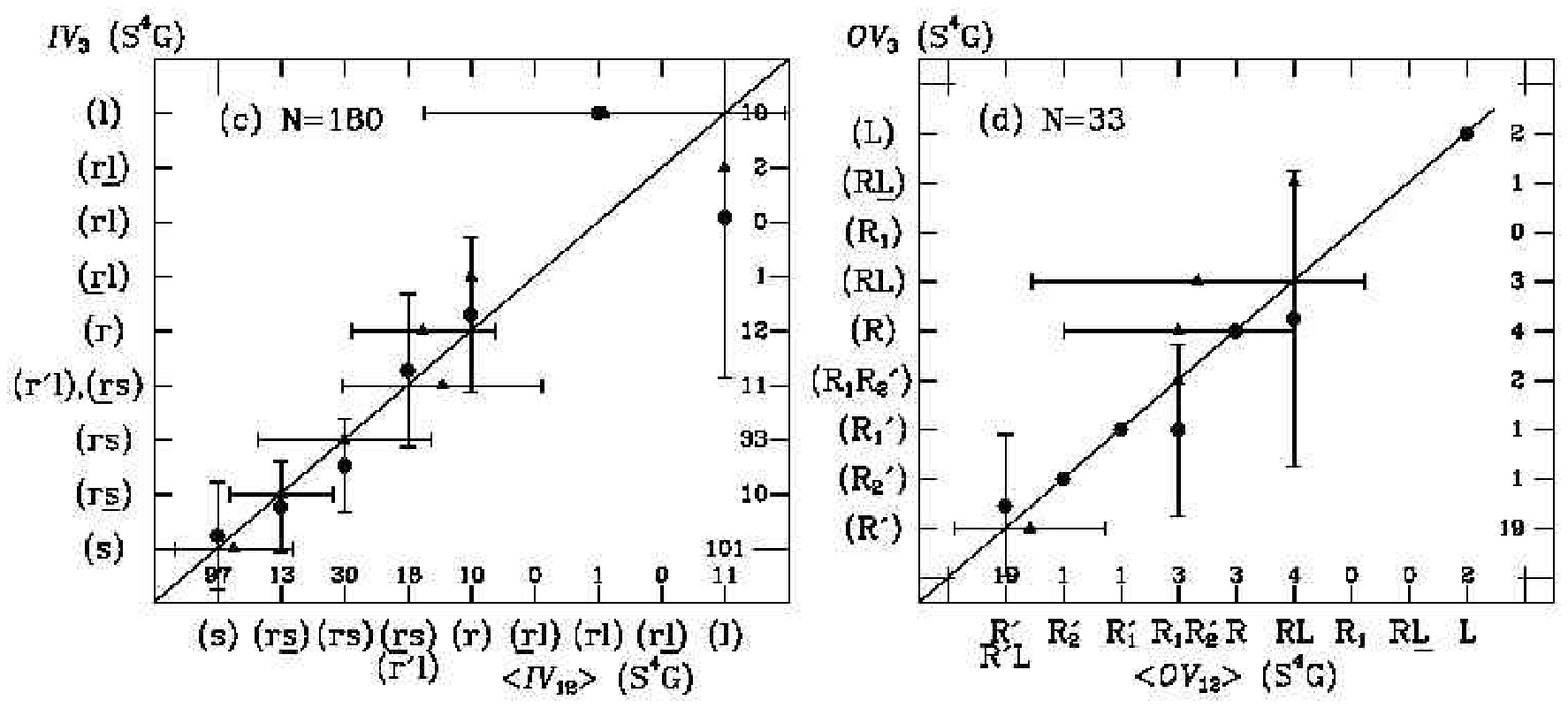}
\caption{Comparisons between Phase 3 classifications (Table 10) and the
mean classifications from Phases 1 and 2 given in Table 6. In (a) and
(b), the comparisons use the numerical codes listed in Table 3. In (c)
and (d), the comparisons use the letter classifications instead. Error
bars are standard deviations in the system of the Table 3 codes in each
case. The number N in each panel is the number of features and
includes multiple features in some cases. The number in each category
is also given to show that the dominant classifications are ``(s)" for
inner variety and ``none" for outer variety.}
\end{figure}

\newpage
\noindent
\centerline{REFERENCES}

\noindent
Athanassoula, E. 2005, MNRAS, 358, 1477

\noindent
Athanassoula, E. 2012, in Secular Evolution of Galaxies, XXIII Canary Islands Winter School
of Astrophysics, J. Falcon-Barroso \& J. H. Knapen, eds., Cambridge, Cambridge University
Press, p. 305

\noindent
Athanassoula, E. \& Bureau, M. 1999,  ApJ, 522, 699

\noindent
Athanassoula, E., Romero-G\'omez, M., \& Masdemont, J.
J. 2009a, \mnras, 394, 67

\noindent
Athanassoula, E., Romero-G\'omez, M., Bosma, A., \& Masdemont, J.
J. 2009b, \mnras, 400, 1706

\noindent
Athanassoula, E., Laurikainen, E., Salo, H., Bosma, A., 2014, arXiv 1405.6726

\noindent
Aguerri, J. A. L., M\'endez-Abreu, J., \& Corsini, E. M. 2009, A\&A, 495, 491

\noindent
Baillard, A. et al. 2011, A \& A, 532, 75

\noindent
Barazza, F., Jogee, S., \& Marinova, I. 2008, ApJ, 675, 1194

\noindent
Binggeli, B., Sandage, A., \& Tammann, G. A. 1985, \aj, 90, 1681

\noindent
Binney, J. \& Tremaine, S. 2008, Galactic Dynamics (Second ed.)

\noindent
Block, D. L. \& Puerari, I. 1999, A\&A, 342, 627

\noindent
Briggs, F. H. 1990, ApJ, 352, 15

\noindent
Burbidge, E. M., Burbidge, G. R. \& Prendergast, K. H. 1964, ApJ, 140, 1620

\noindent Bureau, M. \& Freeman, K. C. 1999, \aj, 118, 126

\noindent Bureau, M., Athanassoula, E., Chung, A., and Aronica, G. 2004, in Penetrating Bars Through Masks of
Cosmic Dust, D. L. Block, et al., eds., Dordrecht, Springer, p. 139

\noindent
Bureau, M., Aronica, G., Athanassoula, E., Dettmar, R.-J., Bosma, A., Freeman, K. C. 2006, \mnras, 370, 753

\noindent
Buta, R. 1984, Publ. Astron. Soc. Australia, 5, 472

\noindent
Buta, R. 1995, ApJS, 96, 39

\noindent
Buta, R. 2012, in Secular Evolution of Galaxies, XXIII Canary Islands Winter School
of Astrophysics, J. Falcon-Barroso \& J. H. Knapen, eds., Cambridge, Cambridge University
Press, p. 155

\noindent
Buta, R. 2013, in Planets, Stars, and Stellar Systems, Volume 6:
Extragalactic Astronomy and Cosmology, T. D. Oswalt, W.
C. Keel, eds., Springer Science+Business Media, Dordrecht, p. 1

\noindent
Buta, R. 2014, in The Structure and Dynamics of Disk Galaxies, ASP Conference Series Vol. 480, M. Seigar 
\& P. Treuthardt, eds., p. 53

\noindent
Buta, R. \& Combes, F. 1996, Galactic Rings, Fund. of Cosmic Physics, 17, 95

\noindent
Buta R. \& Crocker D. A. 1991, \aj, 102, 1715

\noindent
Buta R. \& Crocker D. A. 1993, \aj, 106, 939

\noindent
Buta, R. J., Corwin, H. G., \& Odewahn, S. C. 2007, The de Vaucouleurs Atlas of Galaxies, Cambridge: Cambridge U. Press (dVA)

\noindent
Buta, R., Laurikainen, E., Salo, H., Block, D. L., \& Knapen, J. H. 2006, \aj, 132, 1859

\noindent
Buta, R. et al. 2010a, \apjs, 190, 147 (paper I)

\noindent
Buta, R., Laurikainen, E., Salo, H., \& Knapen, J. H. 2010b, \apj, 721, 259

\noindent
Buta, R., Mitra, S., de Vaucouleurs, G., \& Corwin, H. G. 1994, \aj, 107, 118

\noindent
Buta, R., Ryder, S. D., Madsen, G. J., Wesson, K., Crocker, D. A., \& Combes, F.
2001, AJ, 121, 225

\noindent
Cappellari, M., Emsellem, E., Krajnovi\'c, D. et al. 2011, MNRAS, 416, 1680

\noindent
Casertano, S., Sackett, P. D., \& Briggs, F. 1991, Warped Disks and Inclined Rings
Around Galaxies, Cambridge, Cambridge University Press

\noindent
Campbell, L. A. et al. 2014, arXiv 1406.4867

\noindent
Cisternas, M. et al. 2013, ApJ, 776, 50

\noindent
Coe, D., Ben\'itez, N., S\'anchez, S. F., Jee, M., Bouwens, R., \& Ford, H. 2006, AJ, 132, 926

\noindent
Comer\'on, S. 2013, A\&A, 551, L4

\noindent
Comer\'on, S., Knapen, J. H., Beckman, J. E., Laurikainen, E., Salo, H., Martinez-Valpuesta, I., \& Buta, R. J. 2010, \mnras, 402, 2462

\noindent
Comer\'on, S. et al. 2011a, ApJ, 741, 28

\noindent
Comer\'on, S. et al. 2011b, ApJ, 738, L17

\noindent
Comer\'on, S. et al. 2011c, ApJ, 729, 18

\noindent
Comer\'on, S. et al. 2012, ApJ, 759, 98

\noindent
Comer\'on, S. et al. 2014, A \& A, 562, 121

\noindent
Contopoulos, G. \& Grosbol, P. 1989, A\&A Reviews 1, 261

\noindent
Crocker, D. A., Baugus, P. D., \& Buta, R. J. 1996, AoJS, 105, 353

\noindent
Danby, J. M. A. 1965, AJ, 70, 501

\noindent
de Looze, I. et al. 2012, MNRAS, 427, 2797

\noindent
de Vaucouleurs, G. 1959, Handbuch der Physik, 53, 275

\noindent
de Vaucouleurs, G. 1963, \apjs, 8, 31

\noindent
de Vaucouleurs, G. and de Vaucouleurs, A. 1964, Reference Catalogue of Bright Galaxies, Austin University of Texas Monographs in Astronomy No. 1 (RC1)

\noindent
de Vaucouleurs, G., de Vaucouleurs, A., and Corwin, H. G. 1976, Second Reference Catalogue of Bright Galaxies, Austim University of Texas Monographs in
Astronomy No. 2 (RC2)

\noindent
de Vaucouleurs, G., de Vaucouleurs, A., Corwin, H. G., Buta, R., Paturel, G., \& Fouque\', P. 1991, Third Reference Catalogue of Bright Galaxies, New York, Springer (RC3)

\noindent
de Vaucouleurs, G. \& Freeman, K. C. 1972, Vistas in Astronomy 14, 163

\noindent
de Zeeuw, P. T. et al. 2002, \mnras, 329, 513

\noindent
Elmegreen, B. G. \& Elmegreen, D. M. 1985, AoJ, 288, 438

\noindent
Elmegreen, D. M. 1981, ApJS, 47, 229

\noindent
Elmegreen, D. M., \& Elmegreen, B. G. 1982, MNRAS, 201, 1021

\noindent
Elmegreen, D. M., \& Elmegreen, B. G. 1984, ApJS, 54, 127

\noindent
Elmegreen, D. M., \& Elmegreen, B. G. 1987, ApJ, 314, 3

\noindent
Elmegreen, D. M., Elmegreen, B. G., Marcus, M., Shahinyan, K., Yau, M., \& Petersen, M. 2009, ApJ, 701, 306

\noindent
Elmegreen, D. M. et al. 2011, ApJ, 737, 32

\noindent
Elmegreen, D. M. et al. 2014, ApJ, 780, 32

\noindent
Erwin, P. 2004, A\&A, 415, 941

\noindent
Erwin, P. \& Debattista, V. P. 2013, MNRAS, 431, 3060

\noindent
Eskridge, P. B. et al. 2000, \aj, 119, 536

\noindent
Eskridge, P. B. et al. 2002, \apjs, 143, 73

\noindent
Fazio, G. G. et al. 2004, \apjs, 154, 10

\noindent
Graham, A. W. 2013, in Planets, Stars, and Stellar Systems, Volume 6:
Extragalactic Astronomy and Cosmology, T. D. Oswalt, W.
C. Keel, eds., Springer Science+Business Media, Dordrecht, p. 91

\noindent
Graham, A. W., Spitler, L. R., Forbes, D. A., Lisker, T., Moore, B., \& Janz, J. 2012, ApJ, 750, 121

\noindent
Haynes, M. P., Jore, K. P., Barrett, E. A., Broeils, A., \& Murray,
B. M. 2000, AJ, 120, 703

\noindent
Helou, G. et al. 2004, ApJS, 154, 253

\noindent
Huertas-Company, M., Rouan, D., Tasa, L., Soucail, G., \& Le F\`evre, O.
2008, A\&A, 478, 971

\noindent
Holwerda, B. et al. 2014, ApJ, 781, 12

\noindent
Hubble, E. 1926, \apj, 64, 321

\noindent
Hubble, E. 1936, {\it The Realm of the Nebulae}, Yale Univ.  Press, Yale.

\noindent
Ilyina, M. \& Sil'chenko, O. K. 2011, AstL, 37, 589

\noindent
Ilyina, M., Sil'chenko, O. K., \& Afanasiev, V. L. 2014, MNRAS, 439, 334

\noindent
Iodice, E. \& Corsini, E. M. 2013, Multi-Spin Galaxies, ASP Conference Series Vol.
486

\noindent
Jedrzejewski, R. 1987, MNRAS, 226, 747

\noindent
Jord\'an, A. et al. 2007, ApJS, 169, 213

\noindent
J\'ozsa, G. I. G., Oosterloo, T. A., Morganti, R., Klein, U., \& Erben. T. 2009, A\&A, 494, 489

\noindent
Kendall, S., Kennicutt, R. C., Clarke, C., \& Thornley, M. 2008, MNRAS, 387, 1007

\noindent
Kim, T. et al. 2012, ApJ, 753, 43

\noindent
Kim, T. et al. 2014, ApJ, 782, 64

\noindent
Knapen, J. H. 2010, in Galaxies and their Masks, D. L. Block, K. C. Freeman, \& I. Puerari,
Springer, New York, p. 201

\noindent
Knapen, J. H., Erroz-Ferrer, S., Roa, J., Bakos, J., Cisternas, M.,
Leaman, R., \& Szymanek, N. 2014, A \& A, accepted (arXiv 1406.4107) 

\noindent
Kormendy, J. 1979, \apj, 227, 714

\noindent
Kormendy, J. 1984, ApJ, 286, 116

\noindent
Kormendy, J. 2012, in Secular Evolution of Galaxies, XXIII Canary Islands Winter School
of Astrophysics, J. Falcon-Barroso \& J. H. Knapen, eds., Cambridge, Cambridge University
Press, p. 1

\noindent
Kormendy, J. \& Bender, R. 1996, \apj, 464, L119

\noindent
Kormendy, J. \& Bender, R. 2012, \apjs, 198, 2

\noindent
Kormendy, J. \& Kennicutt, R. C. 2004, \araa, 42, 603 

\noindent
Kormendy, J., Fisher, D. B., Cornell, M. E., \& Bender, R. 2009, ApJS, 182, 216

\noindent
Laine, S., Shlosman, I., Knapen, J. H., \& Peletier, R. F. 2002, ApJ, 567, 97

\noindent
Laine, J. et al. 2014, MNRAS, accepted (arXiv 1404.0559)

\noindent
Laine, S. et al. 2014, MNRAS, submitted

\noindent
Laurikainen, E., Salo, H., Buta, R., \& Knapen, J. 2009, \apj, 392, L34

\noindent
Laurikainen, E., Salo, H., Buta, R., Knapen, J., and Comer\'on, S. 2010, \mnras, 405, 1089

\noindent
Laurikainen, E., Salo, H., Buta, R., \& Knapen, J. 2011, \mnras, 418, 1452

\noindent
Laurikainen, E., Salo, H., Athanassoula, E., Bosma, A., Buta, R., \& Janz, J. 2013, MNRAS, 430, 3489

\noindent
Laurikainen, E., Salo, H., Athanassoula, E., Bosma, A., Herrera-Endoqui, M. 2014, MNRAS, L80

\noindent
Lee, B. et al. 2013, ApJ, 774, 47

\noindent
Lin, C. C., \& Shu, F. H. 1966, PNAS, 55, 229

\noindent
Luetticke, R., Dettmar, R.-J., \& Pohlen, M. 2000a, A\&AS, 145, 405

\noindent
Luetticke, R., Dettmar, R.-J., \& Pohlen, M. 2000b, A\&A, 362, 435

\noindent
Luetticke, R., Pohlen, M., \& Dettmar, R.-J. 2004, A\&A, 527, 539

\noindent
Madore, B. F., Nelson, E., \& Petrillo, K. 2009, \apjs, 181, 572

\noindent
Martin, P. 1996, in ``Barred Galaxies," ASP conf. Ser. 91, R. Buta, D.
A. Crocker, \& B. G. Elmegreen, eds., San Francisco, ASP, p. 70

\noindent
Matthews, L. \& de Grijs, R. 2004, AJ, 128, 137

\noindent
Marinova, I. \& Jogee, S. 2007, ApJ, 659, 1176

\noindent
Martinez-Valpuesta, I., Knapen, J. H., \&  Buta, R. J. 2007, AJ, 134, 1863

\noindent
Meidt, S. et al. 2012, ApJ, 744, 17

\noindent
Meidt, S. et al. 2014, ApJ, accepted (arXiv 1402-5210) 

\noindent
Munoz-Mateos, J. et al. 2013, ApJ, 771, 59

\noindent
Munoz-Mateos, J. et al. 2014, submitted

\noindent
Nair, P. \& Abraham, R. G. 2010, ApJ, 714, L260

\noindent
Pahre, M., Ashby, M. L. N., Fazio, G. G., \& Willner, S. P. 2004, \apjs, 154, 235

\noindent
Paturel, G. et al. 2003, A \& A, 412, 45

\noindent
Querajeta, M. et al. 2014, in preparation 

\noindent
Radburn-Smith, D. J., de Jong, R. S., Streich, D., Bell, E. F., Dalcanton, J. J., Dolphin, A. E.,
Stilp, A. M., Monachesi, A., Holwerda, B., \& Bailin, J. 2014, ApJ, 780, 105

\noindent
Regan, M. \& Teuben, P. 2004, \apj, 600, 595

\noindent
Romero-G\'omez, M., Masdemont, J. J., Athanassoula, E., \&
Garc\'ia-G\'omez, C. 2006, \aap, 453, 39

\noindent
Romero-G\'omez, M., Athanassoual, E., Masdemont, J. J., \&
Garc\'ia-G\'omez, C. 2007, \aap, 472, 63

\noindent
Saha, K., de Jong, R., \& Holwerda, B. 2009, MNRAS, 396, 409

\noindent
Salo, H. et al. 2014, in preparation

Sandage, A. 2005, ARAA, 43, 581

\noindent
Sandage, A. 1961, The Hubble Atlas of Galaxies, Carnegie Inst. of Wash. Publ. No. 618

\noindent
Sandage, A. \& Bedke, J. 1994, The Carnegie Atlas of Galaxies, Carnegie Inst. of Wash. Pub. No. 638

\noindent
Sandage, A. \& Tammann, G. A. 1981, A Revised Shapley-Ames Catalog, Carnegie Inst. of Wash. Publ. No. 635

\noindent
Schweizer, F., Whitmore, B. C., \& Rubin, V. C. 1983, AJ, 88, 909

\noindent
Schweizer, F., Ford, W. K., Jedrzejewski, R., \ Giovanelli, R. 1987, ApJ, 320, 454

\noindent
Sellwood, J. 2013, in Planets, Stars, and Stellar Systems, Volume 5:
Extragalactic Astronomy and Cosmology, T. D. Oswalt, G.
Gilmore, eds., Springer Science+Business Media, Dordrecht, p. 923

\noindent
Sheth, K. et al. 2008, ApJ, 675, 1141

\noindent
Sheth, K. et al. 2010, \pasp, 122, 1397

\noindent
Sheth, K. et al. 2013, Spitzer Proposal ID \#10043

\noindent
Sheth, K. et al. 2014a, AAS 223, 205.02

\noindent
Sheth, K. et al. 2014b, in preparation

\noindent
Theys, J. C. \& Spiegel, J. C. 1976, \apj, 208, 650

\noindent
Thornley, M. 1996, ApJ, 469, L54

\noindent
van den Bergh, S. 1976, \apj, 206, 883

\noindent
van den Bergh, S. 2009, \apj, 694, L120

\noindent
van Driel, W. et al. 1995, AJ, 109, 942

\noindent
Werner, M. et al. 2004, ApJS, 154, 1

\noindent
Whitmore, B. C., Lucas, R. A., McElroy, D. B., Steiman-Cameron, T. Y., Sackett, P. D., Olling, R. P.
1990, AJ, 100, 1489

\noindent
Williams, R. E. et al. 1996, AJ, 112, 1335

\noindent
Zaritsky, D. et al. 2013, ApJ, 772, 135

\noindent
Zaritsky, D. et al. 2014, AJ, 147, 134

\end{document}